\documentclass[12pt]{article}
\usepackage{amssymb}
\usepackage{graphics}
\usepackage{psfrag}

\DeclareFontFamily{U}{rsf}{}
\DeclareFontShape{U}{rsf}{m}{n}{
  <5> <6> rsfs5 <7> <8> <9> rsfs7 <10-> rsfs10}{}
\DeclareMathAlphabet\Scr{U}{rsf}{m}{n}

\catcode`\@=11
\def\@citex[#1]#2{%
\if@filesw \immediate \write \@auxout {\string \citation {#2}}\fi
\@tempcntb\m@ne \let\@h@ld\relax \def\@citea{}%
\@cite{%
  \@for \@citeb:=#2\do {%
    \@ifundefined {b@\@citeb}%
      {\@h@ld\@citea\@tempcntb\m@ne{\bf ?}%
      \@warning {Citation `\@citeb ' on page \thepage \space undefined}}%
      {\@tempcnta\@tempcntb \advance\@tempcnta\@ne%
      \@tempcntb\number\csname b@\@citeb \endcsname \relax%
      \ifnum\@tempcnta=\@tempcntb 
        \ifx\@h@ld\relax%
          \edef \@h@ld{\@citea\csname b@\@citeb\endcsname}%
        \else%
          \edef\@h@ld{\ifmmode{-}\else--\fi\csname b@\@citeb\endcsname}%
        \fi%
      \else
        \@h@ld\@citea\csname b@\@citeb \endcsname%
        \let\@h@ld\relax%
      \fi}%
    \def\@citea{,\penalty\@highpenalty\,}%
  }\@h@ld
}{#1}}

\def\@citeb#1#2{{[#1]\if@tempswa , #2\fi}}
\def\@citeu#1#2{{$^{#1}$\if@tempswa , #2\fi }}
\def\@citep#1#2{{#1\if@tempswa , #2\fi}}

\def\bcites{         
        \catcode`\@=11
        \let\@cite=\@citeb
        \catcode`\@=12
}

\def\upcites{         
        \catcode`\@=11
        \let\@cite=\@citeu
        \catcode`\@=12
}

\def\plaincites{      
        \catcode`\@=11
        \let\@cite=\@citep
        \catcode`\@=12
}

\newcount\hour
\newcount\minute
\newtoks\amorpm
\hour=\time\divide\hour by 60
\minute=\time{\multiply\hour by 60 \global\advance\minute by-\hour}
\edef\standardtime{{\ifnum\hour<12 \global\amorpm={am}%
        \else\global\amorpm={pm}\advance\hour by-12 \fi
        \ifnum\hour=0 \hour=12 \fi
        \number\hour:\ifnum\minute<10 0\fi\number\minute\the\amorpm}}
\edef\militarytime{\number\hour:\ifnum\minute<10 0\fi\number\minute}

\def\draftlabel#1{{\@bsphack\if@filesw {\let\thepage\relax
   \xdef\@gtempa{\write\@auxout{\string
      \newlabel{#1}{{\@currentlabel}{\thepage}}}}}\@gtempa
   \if@nobreak \ifvmode\nobreak\fi\fi\fi\@esphack}
        \gdef\@eqnlabel{#1}}
\def\@eqnlabel{}
\def\@vacuum{}
\def\marginnote#1{}
\def\draftmarginnote#1{\marginpar{\raggedright\scriptsize\tt#1}}
\def\draft{
        \pagestyle{plain}
        \overfullrule=2pt
        \oddsidemargin -.5truein
        \def\@oddhead{\sl \phantom{\today\quad\militarytime} \hfil
        \smash{\Large\sl DRAFT} \hfil \today\quad\militarytime}
        \let\@evenhead\@oddhead
        \let\label=\draftlabel
        \let\marginnote=\draftmarginnote
        \def\ps@empty{\let\@mkboth\@gobbletwo
        \def\@oddfoot{\hfil \smash{\Large\sl DRAFT} \hfil}
        \let\@evenfoot\@oddhead}
        \def\@eqnnum{(\theequation)\rlap{\kern\marginparsep\tt\@eqnlabel}%
        \global\let\@eqnlabel\@vacuum}  }

\def\section{\@startsection {section}{1}{\z@}{3.ex plus 1ex minus
 .2ex}{2.ex plus .2ex}{\large\bf}}
\def\subsection{\@startsection{subsection}{2}{\z@}{2.75ex plus 1ex minus
 .2ex}{1.5ex plus .2ex}{\bf}}        

\def\appendix{{\newpage\section*{Appendix}}\let\appendix\section%
        {\setcounter{section}{0}
        \gdef\thesection{\Alph{section}}}\section}

\def\abstract{\if@twocolumn
\section*{Abstract}
\else 
\begin{center}
{\bf Abstract\vspace{-.5em}\vspace{0pt}}
\end{center}
\quotation
\fi}

\catcode`\@=12

\newcommand{\be}{\begin{equation}}
\newcommand{\ee}{\end{equation}}
\newcommand{\bea}{\begin{eqnarray}}
\newcommand{\eea}{\end{eqnarray}}

\def\to{\rightarrow}

\def\lae{\mathrel{\mathop{\smash{\lower .5 ex \hbox{$\stackrel<\sim$}}}}}
\def\lae{\mathrel{\mathop{\smash{\lower .5 ex \hbox{$\stackrel>\sim$}}}}}

\def\l:{\mathopen{:}\,}
\def\r:{\,\mathclose{:}}

\catcode`\@=11
\def\theequation{\arabic{equation}}
\catcode`\@=12

\bcites

\catcode`\@=11
\def\theequation{\thesection.\arabic{equation}}
\@addtoreset{equation}{section}
\@addtoreset{footnote}{section}
\@addtoreset{footnote}{subsection}
\catcode`\@=12

\newcommand{\wtx}{\widetilde{x}}

\newcommand{\Ost}{O_+}
\newcommand{\Ons}{O_-}

\newcommand{\nn}{\nonumber}

\newcommand{\beq}{\begin{equation}}
\newcommand{\eeq}{\end{equation}}
\newcommand{\beqa}{\begin{eqnarray}}
\newcommand{\eeqa}{\end{eqnarray}}
\newcommand{\dd}{{\rm d}}

\newcommand{\Z}{{\mathbb Z}}
\newcommand{\ZZ}{{\mathbb Z}}

\newcommand{\R}{{\mathbb R}}

\newcommand{\C}{{\mathbb C}}

\newcommand{\PP}{{\mathbb P}}
\newcommand{\e}{\,{\rm e}}
\newcommand{\CP}{\mathbb{CP}}
\newcommand{\wh}{\widehat}
\newcommand{\wt}{\widetilde}
\newcommand{\sigmaU}{\sigma_0}
\newcommand{\Sym}{{\rm Sym}}

\begin{document}

\begin{titlepage}

\begin{center}

\hfill\today\\
\hfill TUW-13-12

\vskip 2.5 cm
{\large \bf Linear Sigma Models With
Strongly Coupled Phases\\[0.2cm]
--- One Parameter Models ---}
\vskip 1 cm

{Kentaro Hori${}^{{}^u}$~ and ~Johanna Knapp${}^{{}^v}$}\\
\vskip 0.5cm {\sl ${}^{{}^u}$Kavli IPMU, The University of Tokyo, Kashiwa,
Japan,\\[0.1cm] and ${}^{{}^v}$TU Vienna, Wiedner
  Hauptstrasse 8-10, 1040 Vienna, Austria}

\end{center}

\vskip 0.5 cm
\begin{abstract}
We systematically construct a class of two-dimensional $(2,2)$ supersymmetric
gauged linear sigma models with phases in which a continuous subgroup of
the gauge group is totally unbroken.
We study some of their properties by employing a recently developed
technique.
The focus of the present work is on models with one K\"ahler parameter.
The models include those corresponding to Calabi-Yau threefolds,
extending three examples found earlier by a few more,
as well as Calabi-Yau manifolds of other dimensions and non-Calabi-Yau
manifolds.
The construction leads to predictions of
equivalences of D-brane categories, systematically extending
earlier examples.
There is another type of surprise.
Two distinct superconformal field theories
corresponding to Calabi-Yau threefolds
with different Hodge numbers, $h^{2,1}=23$ versus $h^{2,1}=59$,
have exactly the same quantum K\"ahler moduli space.
The strong-weak duality plays a crucial r\^ole in confirming
this, and also is useful in the actual computation
of the metric on the moduli space.

\end{abstract}

\end{titlepage}

\newpage
\section{Introduction}

$(2,2)$ superconformal field theories in $1+1$ dimensions are 
interesting systems of study.
Those with $c=9$ and integral R-charges are of special importance
for supersymmetric string compactifications.
There have been a lot of surprises and our understanding has
increased considerably, but there is a lot to be understood,
both in individual systems and in the grasp of the whole picture.
We expect to encounter new surprises ahead and we look forward to face them
with sincerity.


A driving force of the progress in our understanding
has been  the linear sigma models \cite{Witten:1993yc}.
The earlier study was mostly centered on models with Abelian gauge groups,
in which there are as many Fayet-Iliopoulos (FI) parameters as the dimension
of the gauge groups. The FI parameter space is decomposed into ``phases'',
and the gauge symmetry is broken to a finite subgroup in each of them. 
There, the classical analysis can be used to find the nature of the low
energy theory. On the other hand, in theories with non-Abelian gauge groups,
a continuous subgroup of the gauge group is typically unbroken in some of the
phases. In such a phase, we need to have some understanding of the
low energy dynamics of the strongly coupled systems with unbroken gauge symmetry.
This makes the analysis difficult but at the same time interesting.
In some cases, it is possible to understand the low energy dynamics
to the extent that we can find the nature of the theory
\cite{Hori:2006dk,Hori:2011pd}, but that is not always the case.
Recently, some new techniques to study $(2,2)$ supersymmetric gauge theories
 have been developed
\cite{Benini:2012ui,Doroud:2012xw,Jockers:2012dk,Gomis:2012wy,Hori:2013ika,Sugishita:2013jca,Honda:2013uca,Benini:2013nda,Benini:2013xpa,Grassi:2007va,Gadde:2013dda},
and a natural idea would be to apply them to the strongly coupled phases.

In this paper, we systematically construct a class of linear sigma models
with strongly coupled phases, extending the examples found
in \cite{Hori:2006dk,Hori:2011pd}, and then employ the recent
techniques to study some of their properties.
We find two series of theories, (A) and (S).
The gauge group $G$ is of the form
\beq
G\,\,\sim\,\, U(1)\times H
\eeq
where $H$ is (A) a symplectic group or (S) an orthogonal group, and the matter
consists of a number of $H$-singlets $p^1,p^2,\ldots$ and
a number of $H$-fundamentals $x_1, x_2,\ldots$ .
The superpotential is of the form
\beq
\mbox{(A)}:~~\sum_{i,j}A^{ij}(p)[x_ix_j],\quad
\mbox{(S)}:~~\sum_{i,j}S^{ij}(p)(x_ix_j),
\eeq
where $[x_ix_j]$ or $(x_ix_j)$ are respectively the symplectic or orthogonal
bilinear invariants and $A^{ij}(p)$ or $S^{ij}(p)$ 
are antisymmetric or symmetric matrices with polynomial entries.
The system has two phases, $r\gg 0$ and $r\ll 0$, where $r$ is
the FI parameter of the $U(1)$ factor.
$r\ll 0$ is the strongly coupled phase where $H$ factor is totally unbroken,
while $r\gg 0$ is generically weakly coupled.
Each system has a dual description \cite{Hori:2011pd},
in which $r\ll 0$ is generically weakly coupled and $r\gg 0$ is strongly
coupled.

The models include those in which the understanding of
low energy dynamics \cite{Hori:2006dk,Hori:2011pd} can be used to find
the nature of the theory at the strongly coupled phase.
In particular, there are models in which the $r\ll 0$ phase corresponds to
a smooth compact Calabi-Yau threefolds, as in the following list:
\beq
\begin{array}{|c|cccc|c|}
\hline
\mbox{model}&h^{1,1}&h^{2,1}&\mbox{$r\ll 0$ phase}&\mbox{$r\gg 0$ phase}&
\mbox{ref}\\
\hline
({\rm A}^2_{(-2)^7,1^7})&1&50&\mbox{Pfaffian in $\PP^6$}&
\mbox{Int. in  $G(2,7)$}&\cite{rodland98,Hori:2006dk}\\
\hline
({\rm S}^{8,0}_{(-2)^{32},1^8})&1&65&\mbox{Sym-det in  $\PP^{31}$}
&\mbox{$\Z_2$ quad(8) on $\PP^3$}&
\cite{Caldararu:2007tc,Hori:2011pd},*\\
\hline
({\rm S}^{2,+}_{(-2)^5,1^5})&1&26&\mbox{Double sym-det in $\PP^4$}&
\mbox{Reye congruence}&\cite{hosonotakagi11,Hori:2011pd}\\
\hline
({\rm A}^2_{(-1)^4,(-2)^3,1^5})&1&51&\mbox{Pfaffian in $\PP^6_{1111222}$}&
\mbox{$\Z_2$ quad(4) on $B_5$}&\cite{kanazawa10},*\\
\hline
({\rm A}^2_{(-1)^6,(-2),1^4,0})&1&59&\mbox{Pfaffian in $\PP^6_{1111112}$}&
\mbox{$\Z_2$ quad(4) on $B_4$}&\cite{kanazawa10},*\\
\hline
({\rm A}^2_{(-2)^7,3,1^4})&1&61&\mbox{Pfaffian in $\PP^6$}&
\mbox{$\Z_2$ quad(4) $\PP^3$}&\cite{Tonoli,kanazawa10},*\\
\hline
({\rm A}^2_{(-2)^5,(-4)^2,3^2,1^3})&1&61&\mbox{Pfaffian in $\PP^6_{1111122}$}&
\mbox{Pseudo Hybrid}&\cite{kanazawa10},*\\
\hline
({\rm S}^{2,+}_{(-1)^2,(-2)^3,1^4})&1&23&
\mbox{Double sym-det in $\PP^4_{11222}$}&
\mbox{$\Z_2$ quad(2) on $B/\tau$}&*\\
\hline
\end{array}
\eeq
Let us briefly explain the ingredients of the table.
The name of each model is shown as (A$^k_{q}$) or (S$^{k,\bullet}_q$): 
the superscript $k$ encodes the rank
of the group $H$ and the subscript $q$ shows
the $U(1)$ charges of the matters ($\bullet=\pm, 0$ labels the type
of the group, see the main text).
$h^{1,1}$ and $h^{2,1}$ are the numbers of exactly marginal twisted chiral
and chiral operators respectively.
``Pfaffian'' or ``Sym-det'' (for ``symmetric determinantal variety'')
is the locus of $p$ in which the rank of $A(p)$ or $S(p)$ is restricted.
``Double sym-det'' is a certain double cover of the symmetric determinantal
variety.
``$\Z_2$ quad(n) on $M$'' means a hybrid model which is a fibration over $M$
of a $\Z_2$ Landau-Ginzburg orbifold of $n$ variables with quadratic
superpotential. $B_5$, $B_4$ and $B$ are certain Fano threefolds
($B_5$ is the intersection of three hyperplanes in $G(2,5)$,
$B_4$ is a hyperplane in $G(2,4)$ and $B$ is the intersection of three 
symmetric bilinears in $\CP^3\times \CP^3$; $\tau$ is the exchange of
the two $\CP^3$'s). The references show where the Calabi-Yau manifold
is first constructed and studied, or where the linear
sigma model is first constructed. ``$*$'' stands for the present work.

The other phase $r\gg 0$ of some of these models is not
a simple non-linear sigma model but a hybrid model, as was
just mentioned.
Semiclassical analysis can give us some hint concerning the
nature of the model \cite{Caldararu:2007tc,Hori:2011pd}
but that cannot access the actual content of
the quantum theory. In this situation, we computed the two sphere
partition functions \cite{Benini:2012ui,Doroud:2012xw} (with a
correction \cite{Hori:2013ika,Honda:2013uca}), 
and learned of the asymptotic behavior of
the metric on the K\"ahler moduli space, assuming the conjecture of 
\cite{Jockers:2012dk}.
We found that the metrics for all hybrids of the type
``$\Z_2$ quad(2m) on $M$'' 
behave the same way as in the geometric phase,
where the limiting point is a cusp singularity at infinite distance.
One hybrid which is not of that type behaves differently and the limiting point
is at a finite distance.
This is the pattern found in \cite{Aspinwall:2009qy}
where the former is called the {\it true hybrid}
and the latter is called the {\it pseudo-hybrid}.
(The former is a special case of {\it good hybrid} in the sense of
\cite{Bertolini:2013xga}.)

A surprise was waiting for us after
the careful computation of the metric on the K\"ahler moduli space.
Two of the above models, (A$^2_{(-1)^6,(-2),1^4,0}$) 
and (S$^{2,+}_{(-1)^2,(-2)^3,1^4}$), have exactly the same moduli space as
a K\"ahler manifold!  The parameters are related by
$t_A= -t_S+$ constant:
the geometric phase of one model corresponds to the true hybrid phase
of the other.
The two models correspond to completely different
superconformal field theories as the numbers of exactly marginal
chiral operators are different, $59$ versus $23$, and yet they have exactly
the same K\"ahler metric on the twisted chiral moduli space.
It would be interesting to see if the equivalence extends to more refined
structure, for example by
examining the hemisphere partition function
\cite{Hori:2013ika,Sugishita:2013jca,Honda:2013uca}.

We also study models in which $A(p)$ and $S(p)$ are linear in $p$, more
systematically than in \cite{Hori:2011pd}.
We find that the dual model can be simplified in such a case,
and the two phases can be described in a symmetric manner.
The model with the data $A:\C^M\hookrightarrow \wedge^2 V^*$ or 
$S:\C^M\hookrightarrow \Sym^2V^*$ is dual to
the model with the orthogonal
data $A^{\vee}:\C^{M^{\vee}}\hookrightarrow \wedge^2V$ or 
$S^{\vee}:\C^{M^{\vee}}\hookrightarrow \Sym^2V$.
This clearly shows that the gauge theory duality has a close connection to
projective duality. 
Our construction implies the equivalences of D-brane categories
between the two phases, and they seem to correspond to
the equivalences of derived categories in the framework of
``homological projective duality'' \cite{KuznetsovHPD}. 
Such a relation had been observed in sporadic cases
\cite{Hori:2006dk,BorisovCaldararu,kuznetsovRodland,Caldararu:2007tc}.
The present work extend these examples systematically.

{\bf Acknowledgments:} 
We would like to thank
Alexey Bondal,
Richard Eager,
David Favero,
Sergey Galkin,
Daniel Halpern-Leistner,
Shinobu Hosono,
Ludmil Katzarkov,
Alexander Kuznetsov,
Dave Morrison,
Tony Pantev,
Mauricio Romo,
Hiromichi Takagi
and Duco van Straten
for discussions, instructions and encouragement.
J.K. would like to thank Hans Jockers, Wolfgang
Lerche, Emanuel Scheidegger, Eric Sharpe and Harald
Skarke for helpful discussions.
This work was supported by WPI Initiative, MEXT, Japan at Kavli IPMU,
the University of Tokyo.
The research of KH was also supported by JSPS Grant-in-Aid for
Scientific Research No. 21340109.

\section{The linear sigma models (A) and (S)}\label{sec:AandS}

\subsection{Low energy 
dynamics of symplectic and orthogonal gauge theories}
\label{subsec:low}

Here we describe the essential part of the low energy dynamics of
symplectic and orthogonal gauge theories with fundamental matter fields,
claimed in \cite{Hori:2011pd} with some non-trivial evidence.
We shall say that a theory is {\it regular} when there is no non-compact
quantum Coulomb branch. This is a condition to have a well behaved theory
with discrete spectrum when the Higgs branch is lifted or
compactified by the superpotential.

Let us first consider the $USp(k)$ gauge theory with $N$ massless
fundamentals, $x_1,\ldots, x_N$, and vanishing superpotential
($k=2,4,6,\ldots$ and $N=0,1,2,3,\ldots$).
It is regular if and only if $N$ is odd.
For $N\leq k$, there is no zero energy state --- supersymmetry is broken.
For $N=k+1$, the low energy theory is the free conformal field theory
of the mesons, $[x_jx_j]=\sum_{a,b=1}^kx_i^aJ_{ab}x_j^b$,
where $J_{ab}$ is the symplectic structure defining the gauge group.
For higher odd $N\geq k+3$, there is a duality: the theory flows to the same
fixed point as the theory with the dual gauge group
\beq
~~~USp(k)~~\longleftrightarrow~~ USp(N-k-1),
\eeq
with $N$ fundamentals $\wtx^1,\ldots,\wtx^N$ and
${N(N-1)\over 2}$ singlets $a_{ij}=-a_{ji}$ with the superpotential
\beq
W=\sum_{i,j=1}^Na_{ij}[\wtx^i\wtx^j].
\label{USpduality}
\eeq
The mesons in the original theory
correspond to the singlets in the dual, $[x_ix_j]=a_{ij}$.

Next, let us consider the $O(k)$ or $SO(k)$ gauge theory with
$N$ massless fields in the fundamental
representation, $x_1,\ldots, x_N$, and vanishing superpotential
($k=1,2,3,\ldots$ and $N=0,1,2,3,\ldots,$).
For $k\geq 3$, we have a mod 2 theta angle 
associated to the $\Z_2$ fundamental group of the gauge group.
An $O(k)$ theory can be regarded as a $\Z_2$ orbifold of
an $SO(k)$ theory, and there are two possibilities,
denoted by $O_+(k)$ and $O_-(k)$, from the choice of orbifold group action.
The theory with $k\geq 2$ is regular
when $N-k$ is odd and the mod 2 theta angle is turned off,
or when $N-k$ is even and the mod 2 theta angle is turned on.
For $N\leq k-2$, there is no zero energy state --- supersymmetry is broken.
The rest applies only to regular theories.
For $N=k-1$, the $SO(k)$ and $O_-(k)$ theories flow in the infra-red limit to
the free theory of the mesons,
$(x_ix_j)=\sum_{a=1}^kx_i^ax_j^a$.
The $O_+(k)$ theory flows to two copies of such a free theory.
For $N\geq k$, there is a duality:
the theory flows to the same infra-red fixed point as
the theory with the dual gauge group
\beqa
~~~O_+(k)&\longleftrightarrow& SO(N-k+1)\nn\\
~~~SO(k)&\longleftrightarrow& O_+(N-k+1)\\
~~~O_-(k)&\longleftrightarrow& O_-(N-k+1),\nn
\eeqa
with $N$ vectors
$\wtx^1,\ldots,\wtx^N$ and
${N(N+1)\over 2}$ singlets $s_{ij}=s_{ji}$ with the superpotential
\beq
W=\sum_{i,j=1}^Ns_{ij}(\wtx^i\wtx^j).
\label{Oduality}
\eeq
The mesons in the original theory correspond to the singlets in
the dual, $(x_ix_j)=s_{ij}$.
The symmetry $\Z_2=O(k)/SO(k)$ in the $SO(k)$ theory corresponds to
the quantum $\Z_2$ symmetry of the dual $O_+(N-k+1)$ theory (regarded as
a $\Z_2$ orbifold).
In particular, the baryons
$[x_{i_1}\cdots x_{i_k}]=\det(x^a_{i_b})$ in the $SO(k)$ theory
correspond to twist operators in
the dual $O_+(N-k+1)$ theory.
There are similar correspondences between $\Z_2$ symmetries
in the other dual pairs.

\subsection{Description of the model}

We now introduce the linear sigma model which we will consider
in this paper. The gauge group is
\beq 
G={U(1)\times H\over\Gamma}\quad\mbox{with
$\,H=USp(k)$, $O_{\pm}(k)$, or $SO(k)$},
\label{G}
\eeq and the matter contents are $M$ fields, $p^1,\ldots,p^M$, which
are charged only under the $U(1)$ factor, and $N$ fields, $x_1,\ldots,
x_N$, which are in the fundamental representation of $H$ but can also
carry $U(1)$ charges.  $\Gamma$ is a finite normal subgroup of
$U(1)\times H$ to make the matter representation faithful: When all
the $U(1)$ charges of the $p^i$'s are even and all the $U(1)$ charges
of the $x_j$'s are odd, and if $H\ne SO(k)$ with odd $k$, then
$\Gamma=\{(\pm 1,\pm {\bf 1}_k)\}\cong\Z_2$.  Otherwise, $\Gamma$ is
trivial.  We assume that the $U(1)$ charges are negative for the $p^i$'s
and non-negative for the $x_i$'s.  We consider a superpotential of the
following form \beq W=\left\{\begin{array}{ll} \displaystyle
\sum_{i,j=1}^NA^{ij}(p)[x_ix_j]&~({\rm symplectic}),\\[0.5cm]
\displaystyle \sum_{i,j=1}^NS^{ij}(p)(x_ix_j)&~({\rm orthogonal}),
\end{array}\right.
\label{W}
\eeq
where $A^{ij}(p)$ and $S^{ij}(p)$ are respectively antisymmetric and symmetric
matrices with polynomial entries of degree at least one.
Of course we need $W$ to be gauge invariant --- the $U(1)$ charges
must cancel  between
$A^{ij}(p)$ and $[x_ix_j]$ or between $S^{ij}(p)$ and $(x_ix_j)$. 
Finally and most importantly, we require regularity.
We shall describe the criterion momentarily.
Let us call this model (A$^k_q$) or
(S$^{k,\pm,0}_q)$ if the group $H$ is 
symplectic or orthogonal respectively.
The subscript $q$ shows the $U(1)$ charge assignment, $q_{p^i}$ to $p^i$ and
$q_{x_j}$ to $x_j$. The superscript ``$\pm, 0$'' in (S$^{k,\pm,0}_q$)
 refers to the type
$O_{\pm}(k),SO(k)$ of the orthogonal gauge group.

The model has a dual description, with the gauge group
\beq 
\wt{G}={U(1)\times \wt{H}\over\wt{\Gamma}},
\label{dualG}
\eeq
where $\wt{H}$ is found via (\ref{USpduality}) or (\ref{Oduality}),
and the matter contents are
$M$ fields $p^1,\ldots, p^M$, with the same charge as in the original
theory, $N$ fields, $\wt{x}^1,\ldots,\wt{x}^N$, in the 
fundamental of $\wt{H}$ and with the opposite $U(1)$ charges compared to
$x_1,\ldots, x_N$, and fields $a_{ij}=-a_{ji}$ or
$s_{ij}=s_{ji}$ with the same $U(1)$ charges as
$[x_ix_j]$ or $(x_ix_j)$. 
$\wt{\Gamma}$ is a finite normal subgroup of $U(1)\times \wt{H}$
to make the matter representation faithful:
it is either trivial or isomorphic to $\Z_2$.
The superpotential is
\beq
W=\left\{\begin{array}{ll}
\displaystyle
\sum_{i,j=1}^N\left([\wt{x}^i\wt{x}^j]+A^{ij}(p)\right)a_{ij}
&~({\rm symplectic}),\\[0.5cm]
\displaystyle
\sum_{i,j=1}^N\left((\wt{x}^i\wt{x}^j)+S^{ij}(p)\right)s_{ij}
&~({\rm orthogonal}).
\end{array}\right.
\label{dualW}
\eeq
There is a subtle rule concerning the correspondence
$G\leftrightarrow \wt{G}$, as described below.

\subsection*{Regularity}

The criterion of regularity is:
\beq
\begin{array}{c|c|c|c}
\mbox{gauge group}&k&\mbox{$(N-k)$ odd}&\mbox{$(N-k)$ even}\\
\hline
U(1)\times USp(k)&\mbox{(even)}&\mbox{regular}&\mbox{not regular}\\
\hline
U(1)\times O_{\pm}(k)&\mbox{any}&\mbox{regular iff $\theta_D=0$}&
\mbox{regular iff $\theta_D=\pi$}\\
\hline
U(1)\times SO(k)&\mbox{any}&\mbox{regular iff $\theta_D=0$}&
\mbox{regular iff $\theta_D=\pi$}\\
\hline
{U(1)\times USp(k)\over \{(\pm 1,\pm{\bf 1}_k)\}}
&\mbox{(even)}&\mbox{regular}&\mbox{not regular}\\
\hline
{U(1)\times O_{\pm}(k)\over\{(\pm 1,\pm{\bf 1}_k)\}}&
\mbox{even}&\mbox{regular}&
\mbox{not regular}\\
\hline
{U(1)\times SO(k)\over \{(\pm 1,\pm{\bf 1}_k)\}}&
\mbox{even}&\mbox{regular iff $\theta_D=0$}&
\mbox{regular iff $\theta_D=\pi$}
\end{array}
\label{criterion}
\eeq
In this table, $\theta_D\in \{0,\pi\}$ is the mod 2 discrete theta angle.
The regularity criterion for the group $G=U(1)\times H$
is copied from the previous subsection.
The one for the group $G=(U(1)\times H)/\{(\pm 1,\pm{\bf 1}_k)\}$
will be explained in the next subsection. 
Note that $(U(1)\times O(k))/\{(\pm 1,\pm{\bf 1}_k)\}$ with $k$ odd
is isomorphic to $U(1)\times SO(k)$ which is included in the table.
Also, $(U(1)\times SO(k))/\{(\pm 1,\pm{\bf 1}_k)\}$ with $k$ odd
does not make sense because $-{\bf 1}_k$ is not an element of $SO(k)$.

\subsection*{The rule of $G\leftrightarrow \wt{G}$}

The above description of $G\leftrightarrow \wt{G}$ includes some ambiguity.
The ambiguity exists only
if all the $U(1)$ charges of the $p^i$ are even and all the $U(1)$ charges
of the $x_j$ are odd. Otherwise, there is no ambiguity: we have
$U(1)\times H\leftrightarrow U(1)\times \wt{H}$, with
$H\leftrightarrow \wt{H}$ given by (\ref{USpduality}) or (\ref{Oduality}).
For the group $G=(U(1)\times H)/\{(\pm 1,\pm{\bf 1}_k)\}$ with $k$ even,
there is no ambiguity either: the dual group is
$\wt{G}=(U(1)\times \wt{H})/\{(\pm 1,\pm{\bf 1}_{k^{\vee}})\}$
where $H\leftrightarrow \wt{H}$ is
given by (\ref{USpduality}) or (\ref{Oduality}),
and $k^{\vee}=N-k\mp 1$.
Note that the nonsensical group 
$(U(1)\times SO(k^{\vee})/\{(\pm 1,\pm{\bf 1}_{k^{\vee}})\}$
with $k^{\vee}$ odd does not appear as the dual group since
the theory with gauge group 
$(U(1)\times O_+(k))/\{(\pm 1,\pm{\bf 1}_k)\}$
with $k$ and $(N-k)$ both even is not regular.
The ambiguity arises when $k$ is odd because of the group isomorphism
\beq
{U(1)\times O(k)\over
\{(\pm 1,\pm{\bf 1}_k)\}}\,\,\cong\,\, U(1)\times SO(k).
\eeq
There are three ways to interpret the gauge group:
the left hand side with $O(k)=O_+(k)$, 
the left hand side with $O(k)=O_-(k)$
and the right hand side. A different choice corresponds to a
different dual. We claim that only one choice is acceptable and 
yields the correct dual. The right choice and the dual is
\beqa
\mbox{$N$ even}:&&~{U(1)\times O_+(k)\over
\{(\pm 1,\pm{\bf 1}_k)\}}~~\longleftrightarrow~~ 
{U(1)\times SO(N-k+1)\over
\{(\pm 1,\pm{\bf 1}_{N-k+1})\}},\label{S+k}\\
\mbox{$N$ odd}:&&~{U(1)\times O_-(k)\over
\{(\pm 1,\pm{\bf 1}_k)\}}~~\longleftrightarrow~~ 
{U(1)\times O_-(N-k+1)\over
\{(\pm 1,\pm{\bf 1}_{N-k+1})\}}.\label{S-k}
\eeqa
If we had chosen $O_-(k)$ for $N$ even, we would have a non-regular dual
theory. If we had chosen $O_+(k)$ for $N$ odd, we would have
a nonsensical gauge group in the dual. If we had chosen
$U(1)\times SO(k)$, the matter representation of the dual group
would be non-faithful.
We shall take (S$^{k,+}_q$) and (S$^{k,-}_q$) as the names
of the the systems (\ref{S+k}) and (\ref{S-k}) respectively.

\subsection{Theta angle and regularity}
\label{subsec:theta}

In general, the theta angle specifies a way to sum over different
topological sectors in the path integral.
In two dimensions, the topological type of principal $G$-bundles
 is labeled by an element of the fundamental group of $G$.
Therefore, the theta angle assigns a phase to each element of $\pi_1(G)$.
For locality, it is natural to require the assignment to be a group
homomorphism $\pi_1(G)\to U(1)$.
We also require that it is invariant under the adjoint action,
a condition which is non-trivial only for elements which are not
in the identity component.
Therefore, the theta angle is a parameter of the group
\beq
\Theta_G:={\rm Hom}(\pi_1(G),U(1))^{\pi_0(G)}.
\eeq
For example, for $G=U(1)$, we have $\pi_1(G)\cong\Z$ and
$\Theta_G\cong U(1)$. Its element $\e^{i\theta}\in \Theta_G$ assigns a
phase $\e^{i\theta n}$ to a $U(1)$ bundle over the spacetime $\Sigma$
with magnetic flux $n$.
It can be placed as a factor
$\exp\left(i\theta\int_{\Sigma}{i\over 2\pi}F_A\right)$
in the path-integral ($F_A$ is the $i\R$ valued
fieldstrength of a $U(1)$ gauge potential $A$).
For the product group $G=U(1)\times H$, the theta angle is just a combination
of the ones for the two factors, $\Theta_G\cong \Theta_{U(1)}\times \Theta_H$.
Note that $\Theta_{USp(k)}=\{1\}$ since $USp(k)$ is simply connected.
$\Theta_{SO(k)}=\Theta_{O(k)}$ is isomorphic to $\Z_2$ (see \cite{Hori:2011pd}
for example) and this is what we meant by the discrete mod 2 theta angle,
$\e^{i\theta_D}\in \{\pm 1\}$.
For $G=(U(1)\times H)/\{(\pm 1, \pm {\bf 1}_k)\}$, we must look into it
more carefully:

Let us start with $H=USp(k)$. We note that the group $G$ can be written as
\beq
G={U(1)\times USp(k)\over \{(\pm 1,\pm {\bf 1}_k)\}}
\cong {\R\times USp(k)\over \langle (\pi,-{\bf 1}_k)\rangle},
\eeq
where $\langle (\pi,-{\bf 1}_k)\rangle$ is the free Abelian group
generated by the element $(\pi,-{\bf 1}_k)$
(we consider $U(1)$ as $\R/2\pi\Z$ here).
 Therefore $\pi_1(G)\cong \Z$ and hence
$\Theta_G\cong U(1)$. We define $\e^{i\theta}\in \Theta_G$ to be the one
that assigns $\e^{i\theta}\in U(1)$ to the generator $(\pi,-{\bf 1}_k)$.

Let us next consider $H=SO(k)$, $k$ even. We have
\beq
G={U(1)\times SO(k)\over  \{(\pm 1,\pm {\bf 1}_k)\}}
\cong {\R\times Spin(k)\over \langle (\pi,\gamma_k), (0,-1)\rangle},
\eeq
where $-1\in Spin(k)$ projects to ${\bf 1}_k\in SO(k)$ and
$\pm \gamma_k\in Spin(k)$ projects to $-{\bf 1}_k\in SO(k)$.
$(\pi,\gamma_k)$ and $(0,-1)$ commute with each other and
$(0,-1)$ is of order two. Thus, $\pi_1(G)\cong \Z\times \Z_2$
and $\Theta_G\cong U(1)\times \Z_2$. The element
$(\e^{i\theta},\e^{i\theta_D})\in\Theta_G$ assigns $\e^{i\theta}\in U(1)$ 
and $\e^{i\theta_D}\in \{\pm 1\}$ to $(\pi,\gamma_k)$ and $(0,-1)$
respectively.

Finally, we consider $H=O(k)$, $k$ even.
We only have to impose the $\pi_0(G)\cong\Z_2$
invariance to the theta angle for the case of $H=SO(k)$.
Note that the non-trivial element of $\pi_0(G)$ acts on 
$\pi_1(G)$ as $(\pi,\gamma_k)\mapsto (\pi,-\gamma_k)$
and $(0,-1)\mapsto (0,-1)$. Invariance under this imposes the constraint
$\e^{i\theta}\e^{i\theta_D}=\e^{i\theta}$, that is,
$\e^{i\theta_D}=1$.
Thus, we have $\Theta_G\cong U(1)$. The discrete theta angle does not
survive.

The theta angle enters into the twisted superpotential on the Coulomb
branch. For example, in a $U(1)$ theory it enters as
$\wt{W}_{\Theta}=i\theta\sigma$, where $\sigma$ is the 
fieldstrength superfield. Note that $2\pi i\sigma$ in
$\wt{W}$ can be regarded as zero as long as $\sigma$ is normalized
in the correct way. (The F-component of $\sigma$ has a fieldstrength
and obeys a quantization condition.)
In this way, the periodicity $\theta\equiv \theta+2\pi$ is realized.
For the group
$G=U(1)\times H$, we parametrize the Coulomb branch by
$\sigmaU,\sigma_1,\ldots,\sigma_l$ 
where 
 $\sigmaU$ is the $U(1)$ component and
the latter $l:=[{k\over 2}]$ are the components for the maximal torus of
$H$ given by
\beqa
\sigma_{\mathfrak{h}}
&:=&\left\{\begin{array}{ll}
{\rm diag}(\sigma_1,\ldots,\sigma_l,-\sigma_1,\ldots,-\sigma_l)
&~~
\mathfrak{h}=\mathfrak{usp}(2l),\\
{\rm diag}(R(\sigma_1),\ldots, R(\sigma_l))&~~\mathfrak{h}=\mathfrak{so}(2l)\\
{\rm diag}(R(\sigma_1),\ldots, R(\sigma_l),0)&~~\mathfrak{h}=\mathfrak{so}(2l+1)
\end{array}\right.\nn\\
&&R(\sigma):=\left(\begin{array}{cc}
0&-i\sigma\\
i\sigma&0
\end{array}\right).\nn
\eeqa
These are correctly normalized --- 
$2\pi i$ times any of these coordinates is equivalent to zero.
The above can also be used for the group
$G=(U(1)\times H)/\{(\pm 1,\pm {\bf 1}_k)\}$ but they are not correctly
normalized. The correct normalization can be found from the
isomorphism
\beq
\begin{array}{ccc}
\displaystyle {U(1)\times U(1)^l\over \{(\pm 1,(\pm 1)^l)\}}&\cong&
U(1)\times U(1)^l\\
(g,h_1,\ldots, h_l)&\mapsto&(g^2,gh_1,\ldots, gh_l)
\end{array}
\eeq
The fieldstrength fields $\sigma'_0,\sigma'_1,\ldots,\sigma'_l$ 
on the right hand side are the correctly normalized ones.
They are related to the above by
\beq
\sigma'_0=2\sigmaU,\quad\,\,\,
\sigma'_a=\sigmaU+\sigma_a,\quad a=1,\ldots,l.
\eeq
Let us now write down the theta angle term in each theory.
\beqa
U(1)\times USp(k)&&\wt{W}_{\Theta}=i\theta\sigmaU,\\
U(1)\times SO(k),~~U(1)\times O(k)&&
\wt{W}_{\Theta}=i\theta\sigmaU
+i\theta_D(\sigma_1+\cdots+\sigma_l),\label{thDD}\\
{U(1)\times USp(k)\over \{(\pm 1,\pm {\bf 1}_k)\}}&&
\wt{W}_{\Theta}=i\theta 2\sigmaU,\\
{U(1)\times SO(k)\over  \{(\pm 1,\pm {\bf 1}_k)\}}~\mbox{($k$ even)}
&&\wt{W}_{\Theta}=i\theta 2\sigmaU
+i\theta_D\{(\sigma_1-\sigmaU)+\cdots+(\sigma_l-\sigmaU)\},~~~~~~
\label{thD}\\
{U(1)\times O(k)\over  \{(\pm 1,\pm {\bf 1}_k)\}}~\mbox{($k$ even)}
&&\wt{W}_{\Theta}=i\theta 2\sigmaU.
\eeqa
We have $i\theta 2\sigmaU$ in the latter three cases
and we have the combination $(\sigma_a-\sigmaU)$ in (\ref{thD})
since $(\pi, \gamma_k)\in \pi_1(G)$ can be realized as a connection of
the maximal torus in which each $U(1)$ factor of the numerator of
$(U(1)\times U(1)^l)/\{(\pm 1,(\pm 1)^l)\}$ has
magnetic flux ${1\over 2}$.
The sum $i\theta_D(\sigma_1+\cdots+\sigma_l)$ is in (\ref{thDD}) and
(\ref{thD}) because $-1\in \pi_1(SO(k))$ can be realized as
a connection where one $U(1)$ factor of the maximal torus
has magnetic flux $1$.
Note that each of the above $\wt{W}_{\Theta}$ is Weyl 
invariant. For example, the Weyl group for $SO(k)$ with $k$ even
includes the sign flip of an even number of $\sigma_a$'s, 
say $\sigma_1$ and $\sigma_2$.
This changes (\ref{thD}) by
\beq
\Delta\wt{W}_{\Theta}=-2i\theta_D(\sigma_1+\sigma_2)
=-2i\theta_D(\sigma_1'+\sigma_2'-\sigma'_0)\equiv 0
\eeq
since $\theta_D\in\{0,\pi\}$. Note that this $\wt{W}_{\Theta}$
would not have been invariant under the
sign flip of an odd number of $\sigma_a$'s, say just $\sigma_1$, because then
the change is
\beq
\Delta\wt{W}_{\Theta}=-2i\theta_D\sigma_1
=-i\theta_D(2\sigma_1'-\sigma'_0)\equiv i\theta_D\sigma_0'\not\equiv 0.
\eeq
Thus, it cannot be $O(k)$ Weyl invariant.
This is another way to see that the mod 2 theta angle is gone for the
group $(U(1)\times O(k))/\{(\pm 1,\pm{\bf 1}_k)\}$.
On the other hand, (\ref{thDD}) is fine for
$U(1)\times O(k)$ because $\sigma_a$'s are correctly normalized for that group.
This also means the following for the theory with gauge group
$(U(1)\times SO(k))/\{(\pm 1,\pm{\bf 1}_k)\}$ ($k$ even): 
It has a $\Z_2$ symmetry associated with
$O(k)/SO(k)\cong \Z_2$, but that shifts $\theta$ by $\pi$.
This is related to an ambiguity in the choice of $\gamma_k$:
we could have taken $(\pi,-\gamma_k)$ instead of $(\pi,\gamma_k)$
in the definition of $\e^{i\theta}$.
In fact, (\ref{thD}) corresponds to one choice of $\gamma_k$.
The swap will change it, say, as
$(\sigma_1-\sigmaU)\to (\sigma_1+\sigmaU)$ in one of the terms,
and that results in the $\pi$ shift of $\theta$.

To examine the regularity,
let us write down the full effective twisted superpotential on
the Coulomb branch.
First, the Fayet-Iliopoulos (FI) parameter $r$ 
is combined with the theta angle $\theta$ into a complex parameter
$t=r-i\theta$. The tree level twisted superpotential
is therefore given by
\beq
\wt{W}_{\it tree}(\sigma)=
\left\{\begin{array}{ll}
\displaystyle  -r\sigmaU+\wt{W}_{\Theta}(\sigma)
&~{\rm if}~~G=U(1)\times H,\\[0.2cm]
\displaystyle  -2r\sigmaU+\wt{W}_{\Theta}(\sigma)
&~{\rm if}~~
\displaystyle G={U(1)\times H\over \{(\pm 1,\pm {\bf 1}_k)\}},
\end{array}\right.
\label{Wtree}
\eeq
The effective twisted superpotential is
\beq
\wt{W}_{\it eff}(\sigma)
=\wt{W}_{\it tree}(\sigma)
+2\pi i \rho(\sigma)
-\sum_{\chi}\chi(\sigma)\left(\log(\chi(\sigma))-1\right).
\label{Weff}
\eeq
The term $2\pi i\rho(\sigma)$ comes from the W-boson integral
\cite{Hori:2011pd}, where $\rho$ is half the sum of positive roots of $G$.
The last sum is over the weights of the matter representation.
Let us write down $\rho(\sigma)$ more explicitly.
For $G=U(1)\times H$, it is
\beq
2\pi i\rho(\sigma)=\left\{\begin{array}{ll}
\displaystyle
0&
~{\rm if}~\,\mathfrak{h}=\mathfrak{usp}(2l),\,\mathfrak{so}(2l),\\[0.2cm]
\displaystyle
\pi i(\sigma_1+\cdots+\sigma_l)&
~{\rm if}~\,\mathfrak{h}=\mathfrak{so}(2l+1),
\end{array}\right.
\label{rhocomp}
\eeq
For $G=(U(1)\times H)/\{(\pm 1,\pm{\bf 1}_k)\}$, $k=2l$, it is
\beq
2\pi i\rho(\sigma)=\left\{\begin{array}{ll}
\displaystyle
-{l(l+1)\over 2}\pi i \sigma_0'&
~{\rm if}~\,\mathfrak{h}=\mathfrak{usp}(2l),\\[0.2cm]
\displaystyle
-{l(l-1)\over 2}\pi i\sigma_0'&
~{\rm if}~\,\mathfrak{h}=\mathfrak{so}(2l),
\end{array}\right.
\eeq
To find the regularity in the $H$ sector, we only need to look at the
region in the Coulomb branch where $\sigma_a$'s are much bigger than 
$\sigmaU$. In this limit, the matter contribution is
\beq
-\sum_{\chi}\chi(\sigma)\left(\log\chi(\sigma)-1\right)
~\sim~N\pi i (\sigma_1+\cdots+\sigma_l).
\eeq
The theory is regular if and only if
$\partial_{\sigma_a}\wt{W}_{\it eff}$ (for $G=U(1)\times H$)
or $\partial_{\sigma'_a}\wt{W}_{\it eff}$ (for $G=(U(1)\times H)/\{(\pm 1.
\pm{\bf 1}_k)\}$) does not vanish modulo $2\pi i\Z$, $a=1,\ldots, l$.
This leads to the criterion 
(\ref{criterion}).

\subsection{Strongly coupled phase}\label{subsec:strong}

This theory has two phases, $r\gg 0$ and $r\ll 0$.
In the phase $r\gg 0$, the gauge group $G$ is 
 broken to a finite subgroup at a generic solution to the D-term equation,
while in the $r\ll 0$ phase the $H$ factor of the gauge group is completely
unbroken at any solution of the D-term and F-term equations.
In this sense, the $r\gg 0$ phase is generically weakly coupled while
the $r\ll 0$ phase is strongly coupled.
The situation is opposite in the dual description:
$r\gg 0$ is a strongly coupled phase while
$r\ll 0$ is generically weakly coupled.

Let us look at the $r\ll 0$ phase of the model
(A$^k_q$) ({\it resp.} (S$^{k,\pm,0}_q$)) in some detail.
In this phase, $p=(p^1,\ldots, p^M)$
acquires non-zero values by the D-term equation,
breaking the $U(1)$ factor of $G$ to a finite subgroup,
and spans a weighted projective space of dimension $(M-1)$,
which we denote by $\PP$. 
On the other hand, the $x$ fields all vanish at any
solution to the D-term and F-term equations. (See 
Appendix~\ref{app:linearalgebra} concerning this point.)
Therefore, the $H$ part of the gauge group is completely unbroken,
and the classical analysis is totally invalid.
We may regard the theory as a fibration over $\PP$
of the theory of gauge group $H$ with $N$ fundamentals.
The latter is of the type discussed in Section~\ref{subsec:low},
and we may employ some understanding of
its low energy dynamics which is summarized there.

The fundamentals have mass matrix $A(p)=(A^{ij}(p))$ 
({\it resp}. $S(p)=(S^{ij}(p))$)
by the superpotential (\ref{W}),
and the number of massless fundamentals
is determined by its rank.
If the rank is too big, the number is too small for the theory
to have a zero energy state.
There must be at least $(k+1)$ ({\it resp}. $(k-1)$) massless fundamentals.
Therefore, the low energy dynamics localizes near
\beqa
Y_A&=&\Bigl\{\,p\in\PP\,\,\Bigr|\,\,
{\rm rank}\, A(p)\,\leq \,N-k-1\,\,\Bigr\},
\label{YA}\\[0.2cm]
\Biggl({\it resp}.~~~Y_S&=&\Bigl\{\,p\in \PP\,\,\Bigr|\,\,
{\rm rank}\,S(p)\,\leq \,N-k+1\,\,\Bigr\}.\Biggr)
\label{YS}
\eeqa
$Y=Y_A$ ({\it resp}. $Y_S$) is called the {\it Pfaffian variety}
({\it resp}. {\it symmetric determinantal variety}) and
has codimension $(k+1)k\over 2$ ({\it resp}. $(k-1)k\over 2$) in $\PP$.
Suppose there is no point $p$ with ${\rm rank}\,A(p)<N-k-1$
({\it resp}. ${\rm rank} \, S(p)<N-k+1$). Then
the variety $Y$ is smooth except at the intersection
with the orbifold singularity of the ambient space $\PP$.
In that case, the mass matrix has constant rank
so that the $H$ theory has $(k+1)$ ({\it resp}. $(k-1)$)
massless fundamentals, $x^L_i$, everywhere on $Y$.
As described above, the $H$-theory flows to the free theory of 
$(k+1)k\over 2$ mesons $[x^L_ix^L_j]$ 
({\it resp}. $(k-1)k\over 2$ mesons $(x^L_ix^L_j)$).
The superpotential (\ref{W}) is a perfect pairing between these
mesons and the coordinates transverse to $Y$ in $\PP$.
Therefore, the low energy theory is simply the non-linear sigma model
with $Y$ as the target, except for
the theory with $H=O_+(k)$ where the target is an unramified double cover
$\wt{Y}_S$ of $Y=Y_S$. Suppose now that there is a locus with
${\rm rank}\,A(p)<N-k-1$ ({\it resp}. ${\rm rank} \, S(p)<N-k+1$).
Then, the variety $Y$ has a singularity there in addition to the orbifold
singularity from $\PP$.
In that case, the number of massless fundamentals
jumps at that locus. Unfortunately, we do not always know the nature of the
low energy dynamics of such a theory. 

Let us look at the dual description in the corresponding phase, which is
$\wt{r}\ll 0$.
A part of the F-term equations is
\beq
[\wt{x}^i\wt{x}^j]+A^{ij}(p)=0\quad\forall i,j\quad\,\,
\Bigl({\it resp}.\quad
(\wt{x}^i\wt{x}^j)+S^{ij}(p)=0\quad
\forall i,j\Bigr).
\label{Fdual}
\eeq
The field $p$ is forbidden to vanish by these equations and the
D-term equations. It therefore breaks the $U(1)$ factor
to a finite subgroup and spans $\PP$ again.
Since $\wt{x}^i$'s take values in $\C^{N-k-1}$ ({\it resp}. $\C^{N-k+1}$),
the equations (\ref{Fdual}) can be solved only if
${\rm rank}\, A(p)\leq N-k-1$, i.e. $p\in Y_A$
({\it resp}. ${\rm rank}\, S(p)\leq N-k+1$, i.e. $p\in Y_S$).
If the inequality is saturated everywhere,
then the $\wt{H}$ factor of the gauge group is broken everywhere.
In particular, the gauge group $\wt{G}$ is broken to a finite subgroup.
It is a weakly coupled phase and we can trust the classical analysis.
We find the non-linear sigma model whose target space is the vacuum manifold.
Let us see what that manifold is.
For a given $p\in  Y$, the equation (\ref{Fdual})
has a unique solution for $\wt{x}$ up to gauge, except in the
case $\wt{H}=SO(N-k+1)$ where there are {\it two} solutions
distinguished by the sign of the baryons $\det(\wt{x}^i_{a_j})$.
Thus, the target space is $Y$ in (\ref{YA}) ({\it resp}. (\ref{YS}))
or a double cover of $Y_S$.
We reproduce the same low energy theory as in the original.
In the model (S$^{k,+}_q$),
this specifies the unramified double cover
\beq
\wt{Y}_S=\Bigl\{\,(p,\wt{x})\,\,\Bigr|\,\,
\mbox{stability},\,\,\,
(\wt{x}^i\wt{x}^j)+S^{ij}(p)=0
\,\,\Bigr\}\mbox{\Large $/$}\wt{G}_{\C}
\label{wtYS}
\eeq
of $Y_S$ which was not even constructed in the original model.
What if the rank of $A(p)$ ({\it resp}. $S(p)$) can go lower at some locus?
For Model (A), it is lowered by an even number since $A(p)$ is antisymmetric,
and a continuous subgroup of $\wt{H}$ will be unbroken at such a locus.
Therefore, a classical analysis is not valid.
On the other hand, in Model (S), the rank can drop just by one.
In such a case, the unbroken part of $\wt{H}$ is not continuous, and the
classical analysis can be employed.
We have the non-linear sigma model on the quotient
$\wt{Y}_S$ given by (\ref{wtYS}), which in this case is a double cover
of $Y_S$ ramified along the rank $(N-k)$ locus.
The dual description has an advantage over the original in these cases.

In general, neither the original nor the dual description
can be used to find the nature of the $r\ll 0$ phase.
However,
this does not mean that the theory is sick. Rather, we know that
the theory is regular in the sense that there is no Coulomb branch.
We simply do know not how to directly analyze such a strongly coupled
system at the moment. It would be interesting to employ
recently developed techniques 
\cite{Benini:2012ui,Doroud:2012xw,Jockers:2012dk,Gomis:2012wy,Hori:2013ika,Sugishita:2013jca,Honda:2013uca,Benini:2013nda,Benini:2013xpa,Grassi:2007va,Gadde:2013dda}
to study such linear sigma models.

\subsection{Examples}

Some of the models have been studied in the past. 
\begin{description}
\item[\rm (A$^2_{(-2)^7,1^7}$)] This was introduced in \cite{Hori:2006dk}
to understand R\o dland's work \cite{rodland98}
by a linear sigma model. The $r\gg 0$
phase is the geometric phase corresponding to
the complete intersection $X_A$ of seven hypersurfaces
in the Grassmannian $G(2,7)$. It predicted equivalences of the derived
category, $D^b(X_A)\cong D^b(Y_A)$, despite the fact that $X_A$ and $Y_A$
are not birationally equivalent. The equivalence was later proved by
Borisov-Caldararu and Kuznetsov
\cite{BorisovCaldararu,kuznetsovRodland}.
\item[\rm (S$^{1,+}_{(-2)^4,1^8}$)] This was studied in \cite{Caldararu:2007tc}
as an Abelian analog of \cite{Hori:2006dk}.
 The $r\gg 0$ phase is the geometric phase corresponding to the
complete intersection $X_S$ of four quadrics
 in $\CP^7$. $Y_S$ is singular because of the rank two degeneration
 of $S(p)$. 
 The category of B-branes can be described as a noncommutative resolution
 of $Y_S$, and is shown to be equivalent to $D^b(X_S)$.
\item[\rm (S$^{2,+}_{(-2)^5,1^5}$)] This was introduced in \cite{Hori:2011pd}
 to put the work of Hosono and Takagi \cite{hosonotakagi11} in the
framework of the linear sigma model. The $r\gg 0$ phase is 
the geometric phase corresponding to the Reye congruence $X_S$
which is not birationally equivalent to $Y_S$.
It predicted the derived equivalences,
$D^b(X_S)\cong D^b(Y_S)$, which was later proved by Hosono-Takagi
themselves \cite{hosonotakagiproof}.
\end{description}
All these examples involve Calabi-Yau threefolds, and
$A(p)$ and $S(p)$ are all linear in the $p$'s.

In this paper 
we would like to find and study other examples in which the variety
$Y_A$, $Y_S$ or $\wt{Y}_S$ is a smooth Calabi-Yau threefold.
The Calabi-Yau condition amounts to
\beq
\sum_{i=1}^Mq_{p^i}+k\sum_{j=1}^Nq_{x_j}=0.
\eeq
The condition on the dimension is
\beq
M-1-{k(k\pm 1)\over 2}=3.
\eeq
If we exclude those which are equal to or dual to models
with Abelian gauge groups, one can only find
surprisingly small number of examples.
We found five models:
$$
\mbox{(A$^2_{(-1)^4,(-2)^3,1^5}$),~
(A$^2_{(-1)^6,(-2),1^4,0}$),~
(A$^2_{(-2)^7,3,1^4}$),~
(A$^2_{(-2)^5,(-4)^2,3^2,1^3}$),~
(S$^{2,+}_{(-1)^2,(-2)^3,1^4}$)}
$$
The Pfaffian varieties $Y_A$ in the first four models 
had been studied in Kanazawa's paper \cite{kanazawa10}
which aimed at generalization of R\o dland's example. 
(The fourth had been constructed earlier by Tonoli \cite{Tonoli}.)
They correspond
respectively to $X_5$, $X_{10}$, $X_{13}$ and $X_7$ in the notation of
\cite{kanazawa10}.
The double cover $\wt{Y}_S$ of the symmetric determinantal variety
in the last model is
a new Calabi-Yau threefold
which nobody has studied, as far as we are aware of.
In the next two sections, we study these examples in detail,
with the focus on the other phase $r\gg 0$.

To be precise, the search also found three other models,
$$
\mbox{(A$^4_{(-2)^{14},1^7}$),~(S$^{8,0}_{(-2)^{32},1^8}$),
~(S$^{4,0}_{(-2)^{10},1^5}$)}
$$
in which the Pfaffian $Y_A$ and the symmetric determinants
$Y_S$ are smooth Calabi-Yau threefolds.
However, it turns out they are not new.
These are nothing but the $X_{A^{\vee}}$, $X_{S^{\vee}}$, $X_{S^{\vee}}$
respectively
of the above three models,
i.e. the intersection of seven hyperplanes in $G(2,7)$,
the intersection of four quadrics in $\CP^7$,
and the Reye congruence, where $A^{\vee}$ and $S^{\vee}$ are
the orthogonals of $A$ and $S$ in a certain sense.
This has led us to notice that,
when $A(p)$ and $S(p)$ are linear in $p$,
 the dual model can be simplified to take the same form as the original,
but with different number of $p$ fields and of course with the dual
gauge group.
This will be discussed in Section~\ref{sec:linear} in a more general
context, including models not limiting to threefolds nor Calabi-Yaus.
The simplification allows us to describe the $r\gg 0$ and
the $r\ll 0$ phases symmetrically.

\section{Pfaffian Calabi-Yau threefolds from (A) series}
\label{sec:(A)}

In this section, we study the four models in (A) series in detail.
As mentioned above, the Pfaffian Calabi-Yau threefolds
$Y_A$ that appear in the $r\ll 0$ phase
are those studied by Kanazawa \cite{kanazawa10}.
The main focus here is the nature of the $r\gg 0$ phase.
We shall study it by looking at the classical Lagrangian
and also by looking at the metric on the moduli space, found by 
computing the two sphere partition function.
The classical analysis shows that they are all in hybrid phase.
To be precise, in each of the first three models, it may be regarded as
a fibration over a Fano threefold of a four-variable
$\Z_2$ Landau-Ginzburg orbifold with quadratic superpotential,
while the last model has no such interpretation.
The metric of the moduli space in the first three models
behave in the same way as the geometric phase, where $r\to+\infty$
is at infinite distance, while the one of the last model
is different, where $r\to+\infty$ is at finite distance.
This relation between the classical picture and quantum K\"ahler
metric was observed in a work of Aspinwall and Plesser
\cite{Aspinwall:2009qy}. In their language, the first three are
in true hybrid phase while the last one is in pseudo-hybrid.

Before studying the individual cases, let us describe what can be said
uniformly in all four models. The gauge group is $G=U(1)\times SU(2)$
or ${U(1)\times SU(2)\over\{(\pm 1,\pm {\bf 1}_2)\}}\cong U(2)$, and
the matter consists of seven $SU(2)$ singlets $p^1,\ldots, p^7$
which are negatively charged under $U(1)$,
and five $SU(2)$ doublets $x_1,\ldots, x_5$
which are mostly positively charged under $U(1)$.
The D-term equations are
\beqa
&&\sum_{i=1}^7q_{p^i}|p^i|^2+\sum_{j=1}^5q_{x_j}||x_j||^2
=r_{\mathfrak{u}(1)},\label{D1}\\
&&xx^{\dag}={1\over 2}||x||^2{\bf 1}_2,\label{D2}
\eeqa
and the F-term equation are
\beqa
&&\sum_{j=1}^5A^{ij}(p)x^1_j=\sum_{j=1}^5A^{ij}(p)x^2_j=0,\quad
i=1,\ldots, 5
\label{Nx}\\
&&\sum_{1\leq i<j\leq 5}{\partial\over \partial p^k}A^{ij}(p)x^1_ix^2_j=0,
\quad
k=1,\ldots,7.
\label{pNxx}
\eeqa
In (\ref{D1}), $r_{\mathfrak{u}(1)}$ is the FI parameter for the
$U(1)$ factor. It is related to the correctly normalized FI parameter $r$ by
$r_{\mathfrak{u}(1)}=r$ or $r_{\mathfrak{u}(1)}=2r$
depending on whether the gauge group is $U(1)\times SU(2)$ or $U(2)$.
See (\ref{Wtree}).
We used $||x_j||=\sum_{a=1}^2|x^a_{\,j}|^2$ and
$||x||^2=\sum_{j=1}^5||x_j||^2$.
 ``$x$'' in (\ref{D2}) stands for the $2\times 5$ matrix $(x^a_{\,\,i})$.

We assume that $A(p)$ is generic so that $A(p)=0$
has no solution other than $p=0$.
Then the Pfaffian $Y_A$ is the locus of ${\rm rank}\,A(p)=2$.
In that case, $Y_A$ is smooth unless it intersects the
orbifold singularity of the ambient space.
In fact, in all cases,
the intersection is impossible under further genericity,
since $Y_A$ is of codimension three in the six-dimensional ambient 
space $\PP$ while the codimensions of the orbifold locus is at least four.
As discussed in the previous section, in the $r\ll 0$ phase,
the theory reduces to the non-linear sigma model
whose target space is this smooth Pfaffian $Y_A$.

The nature of the $r\gg 0$ phase depends on the model, but
two things can be said on all four models:
\begin{center}
\begin{minipage}{14.5cm}
\begin{itemize}
\item[(i)] {\it It is a weakly coupled phase;
the gauge group is broken to a finite subgroup
at any solution to the D-term equations (\ref{D1})-(\ref{D2})},
\item[(ii)] {\it All $p^i$'s must vanish at any solution to the D-term
and F-term equations (\ref{D1})-(\ref{pNxx}), given the genericity
assumption that $Y_A$ is smooth.}
\end{itemize}
\end{minipage}
\end{center}
For $r\gg 0$, i.e. $r_{\mathfrak{u}(1)}\gg 0$,
the $U(1)$ D-term equation (\ref{D1})
forces some $x_j$ to have non-zero values.
Then, the $SU(2)$ D-term equation (\ref{D2}) requires
that the matrix $x$ must have rank two.
This means that two of $x_i$'s, say $x_i$ and $x_j$,
are linearly independent. Then, a stabilizer $(u,U)\in U(1)\times SU(2)$
must satisfy the equations, $u^{q_{x_i}}Ux_i=x_i$
and $u^{q_{x_j}}Ux_j=x_j$. The infinitesimal version can be written as
\beq
\delta U\cdot V+V\cdot\left(
\begin{array}{cc}
q_{x_i}\delta u&0\\
0&q_{x_j}\delta u
\end{array}\right)=0,
\eeq
where $V$ is the invertible $2\times 2$ matrix made of $x_i$ and $x_j$.
Multiplying $V^{-1}$ from left and taking the trace,
we find $(q_{x_i}+q_{x_j})\delta u=0$. Since $q_{x_i}+q_{x_j}$ is positive,
we find $\delta u=0$ and hence $\delta U=0$.
Thus, no continuous subgroup of the gauge group remains unbroken.
This proves (i). Next let us show (ii).
The first set of F-term equations (\ref{Nx}) means that
$x^1=(x^1_i)$ and $x^2=(x^2_i)$ are both in the kernel of $A(p)$.
If $A(p)$ has rank four, the kernel of $A(p)$ is one-dimensional
and hence the rank of $(x^a_i)$ is at most one, failing to satisfy the
requirement from the D-term equation.
If $A(p)$ has rank two, in which case
$p$ represents a point of $Y_A$, $x^1$ and $x^2$ are linearly independent
vectors in the three-dimensional kernel of $A(p)$.
Then, the second set of F-term equations
(\ref{pNxx}) contradicts the smoothness of the Pfaffian $Y_A$ at $p$.
This can be shown as follows. (In this proof, we fix $p=p_*$
with ${\rm rank}\,A(p_*)=2$.)
We take a basis $(x^1,x^2,x^3,x^4,x^5)$ of $\C^5$,
adding three vectors to $x^1$ and $x^2$,
so that the first three elements span the kernel of $A(p_*)$.
In this basis, the matrix $A(p_*)$ is of the form
\begin{equation}
A(p_*)=\left(\begin{array}{ccccc}
0&&&&\\
&0&&\\
&&0&&\\
&&&0&\lambda\\
&&&-\lambda&0\\
\end{array}\right),
\label{Apstar}
\end{equation}
with $\lambda\ne 0$.
In a neighborhood of $p_*$,
the Pfaffian $Y_A$ can be defined as the set of solutions to the
three equations (See Appendix~\ref{app:linearalgebra}),
\beq
{\rm Pf}_{1}A(p)={\rm Pf}_{2}A(p)={\rm Pf}_{3}A(p)=0.
\eeq
Here ${\rm Pf}_{i}A$ is defined to be the Pfaffian of the
$4\times 4$ minor of $A$ obtained by deleting the $i$-th row and $i$-th
column.
Smoothness of $Y_A$ at $p_*$ requires that the $7\times 3$ matrix
of differentials
\beq
\left({\partial\over\partial p^i}{\rm Pf}_{1}A,   
{\partial\over\partial p^i}{\rm Pf}_{2}A,                                   
{\partial\over\partial p^i}{\rm Pf}_{3}A                                    
\right)^{}_{i=1,\ldots,7}
\label{matrix}
\eeq
must be of rank three at $p_*$.
Since $A^{ij}(p_*)$ vanishes unless $i,j=4,5$, we have
\beq
{\partial\over\partial p^i}{\rm Pf}_{3}A(p_*)
={\partial\over \partial p^i}A^{12}(p_*)A^{45}(p_*).
\eeq
These all vanish if we use F-term equations
(\ref{pNxx}) which reads ${\partial\over \partial p^i}A^{12}(p)=0$
in this basis.
Therefore, the matrix (\ref{matrix})
has rank two or less, violating the smoothness condition at $p_*$.
Thus, the only option left for us is that $A(p)$ has rank zero, which means
$p=0$ by the assumption of genericity.
This proves (ii).

Now we move on to the research in the individual cases.
A note on notation:
We shall write $A_{ij...}^{lm}$ for the coefficient of $p^ip^j\cdots$
in $A^{lm}(p)$.

\subsection{(A$^2_{(-1)^4,(-2)^3,1^5}$)}\label{subsec:X5}

The model has gauge group $G=U(1)\times SU(2)$ and the matter consists of
$p^1,\ldots, p^4$ of charge $-1$, $p^5,p^6,p^7$ of charge $-2$ and
doublets $x_1,\ldots, x_5$ of charge $1$.
The matrix $A(p)$ must have charge $-2$ and hence is quadratic in
$p^1,\ldots, p^4$ and linear in $p^5,p^6,p^7$.

First, let us identify the location of the singular points.
Parametrizing the maximal torus of $SU(2)$ in the standard way,
$\sigma_{\mathfrak{su}(2)}={\rm diag}(\sigma_1,-\sigma_1)$,
the effective twisted superpotential on the Coulomb branch is
\begin{eqnarray}
\wt{W}_{\it eff}&=&-t\sigmaU -4(-\sigmaU)(\log(-\sigmaU)-1)
-3(-2\sigmaU)(\log(-2\sigmaU)-1)\nn\\
&&-5(\sigmaU+\sigma_1)(\log(\sigmaU+\sigma_1)-1)
-5(\sigmaU-\sigma_1)(\log(\sigmaU-\sigma_1)-1).
\end{eqnarray}
The singularities are at the locus where the true Coulomb branch exists,
i.e. where there is a non-compact space of solutions to
$\partial_{\sigmaU}\wt{W}_{\it eff}\equiv 0$,
$\partial_{\sigma_1}\wt{W}_{\it eff}\equiv 0$ (mod $2\pi i\Z$),
which read
\beq
\e^{-t}={(\sigmaU+\sigma_1)^5(\sigmaU-\sigma_1)^5\over
(-\sigmaU)^4(-2\sigmaU)^6},
\quad
1={(\sigmaU+\sigma_1)^5\over (\sigmaU-\sigma_1)^5}.
\eeq
Note that $\sigmaU=0,\sigmaU\pm \sigma_1=0$ and $\sigma_1=0$ 
should be excluded since they correspond to the
loci where some of the matter or the W-boson is massless, in which the
analysis is invalid.
Then, the admissible solutions are $(\sigma_1/\sigmaU)^2=-5\pm\sqrt{20}$.
There are four solutions in total but we need to mod out by the
Weyl group action $\sigma_1\mapsto -\sigma_1$.
We therefore have two singular points at
\begin{equation}
\e^{-t}=984\pm440\sqrt{5}.
\label{X5sing}
\end{equation}

The Pfaffian $Y_A$ in the $r\ll 0$ phase is nothing but $X_5$ in
Kanazawa's work \cite{kanazawa10}.  It is a Calabi-Yau threefold with
the following topological data \beq h^{1,1}(X_5)=1,\quad
h^{2,1}(X_5)=51,\quad \int_{X_5}H^3=5,\quad \int_{X_5}c_2(X_5)H=38,
\eeq where $H$ is a generator of $H^2(X_5,\Z)$.  In that paper, the
mirror family was proposed and the Picard-Fuchs equation was
identified with the Calabi-Yau equation No. 238 in
\cite{vanstraten}\footnote{In older versions of the database it had
  the number 302.}.  
The singular points of that equation agrees with (\ref{X5sing}) if we
identify the parameter $\phi$ in \cite{kanazawa10} with our $-\e^t$.

Now let us look into the $r\gg 0$ phase.
We first determine the vacuum manifold.
We already know that $p=0$ is required. Then, the remaining equations are
\beqa
xx^{\dag}&=&{1\over 2}\,r\,{\bf 1}_2,\\
\sum_{i,j=1}^5A^{ij}_k[x_i x_j]&=&0,\qquad k=5,6,7.
\label{m5eqs}
\eeqa
If the gauge group were $U(2)$, the vacuum manifold would be
a complete intersection of three hypersurfaces in the
Grassmannian $G(2,5)$. This is a Fano threefold which we denote by $M_{A'}$,
where $A'=(A^{ij}_k)$.
However, the actual gauge group is $U(1)\times SU(2)$
and hence the vacuum manifold is something like $M_{A'}/\Z_2$
where $\Z_2$ is generated by $(-1,-{\bf 1}_2)$ and acts trivially
on $M_{A'}$.
The fields $p^5,p^6,p^7$ are used up in imposing
the equations (\ref{m5eqs}) and
the remaining fields are $p_{(4)}=(p^1,\ldots, p^4)$.
They are not always massive and hence
we have a hybrid model.
The target space is the quotient
\beq
\mathfrak{X}_{A'}=\left\{\,
(p_{(4)},x)\in \C^4\oplus {\rm Hom}(\C^5,\C^2)\,\,
\Bigr|\,\,{\rm rank}(x)=2,\,\,\,(\ref{m5eqs})\,\right\}
\mbox{\Large $/$}\C^{\times}\!\times SL(\C^2),
\label{fX5}
\eeq
and the superpotential is 
\beq
W_{A''}=\sum_{i,j=1}^5A^{ij}(p_{(4)})[x_ix_j]
=\sum_{l,m=1}^4A_{lm}(x)p^lp^m,
\label{WX5}
\eeq
where $A_{lm}(x):=\sum_{i,j=1}^5A_{lm}^{ij}[x_ix_j]$ and $A''=(A^{ij}_{lm})$.
The space $\mathfrak{X}_{A'}$ 
is the total space of an orbifold vector
bundle of rank $4$ over $M_{A'}/\Z_2$.
It is an orbifold with a $\Z_2$ orbifold singularity
at the zero section $p_{(4)}=0$.
The model $(\mathfrak{X}_{A'},W_{A''})$
may be regarded as a fibration over $M_{A'}$ of 
a $\Z_2$ Landau-Ginzburg orbifold of four variables with
a quadratic superpotential.
Since $M_{A'}$ is three dimensional,
the corank of the mass matrix $A''(x)=(A_{lm}(x))$ can be $0, 1$ or $2$.
The situation is similar to the $r\ll 0$ phase of the model
(S$^{1,+}_{(-1)^4,1^8}$) studied in \cite{Caldararu:2007tc}.
There are two possibilities \cite{Hori:2011pd}.
If the $\Z_2$ is of the type $O_+(1)$, then, the situation is really like
\cite{Caldararu:2007tc}, where we have a double cover of $M_{A'}$,
ramified along the ${\rm corank}\,A''(x)=1$ locus, and have an unresolvable
conifold singularity at the points with 
${\rm corank}\,A''(x)=2$.
If the $\Z_2$ is of the type $O_-(1)$, then, we do not have the double cover.

As an alternative to the above classical
analysis, we may employ a recently developed technique. That is, we
compute the sphere partition function after
\cite{Benini:2012ui,Doroud:2012xw}
and study the behavior of the K\"ahler metric on the moduli space in the
$r\gg 0$ regime, assuming the conjecture of
\cite{Jockers:2012dk}.
(Such a computation in the hybrid phase has been done recently in 
 \cite{Sharpe:2012ji,Halverson:2013eua}.)
The calculation of the partition function for the $r\gg0$
phase of this model is straightforward. Details about the calculation
can be found in appendix \ref{app-loc}. The result is    
\begin{eqnarray}
Z_{S^2}\!\!&=&\!\!
-\frac{(z\bar{z})^q}{2}
\oint\frac{\mathrm{d}\varepsilon_1\mathrm{d}\varepsilon_2}{(2\pi i)^2}
(z\bar{z})^{-\frac{\varepsilon_1}{2}-\varepsilon_2}\pi^3
\frac{\left[\cos\pi\left(\frac{\varepsilon_1}{2}+\varepsilon_2\right)\right]^4
\left[\sin\pi\left(\varepsilon_1+2\varepsilon_2\right)\right]^3}
{\left[\sin\pi\varepsilon_1\right]^5\left[\sin2\pi\varepsilon_2\right]^5}
\times\nonumber\\
&&\!\!\!\!\!\!\!\!\!\!\!\!\!\!\!
\times\left.\vline \sum_{\stackrel{k,l=0}{k+l=even}}^{\infty}\!\!\!
(k-l-\varepsilon_1+2\varepsilon_2)z^{\frac{k+l}{2}}
\frac{\left[\Gamma\left(\frac{1}{2}+\frac{k}{2}+\frac{l}{2}
-\frac{\varepsilon_1}{2}-\varepsilon_2\right)\right]^4
\left[\Gamma\left(1+k+l-\varepsilon_1-2\varepsilon_2 \right)\right]^3}
{\left[\Gamma\left(1+k-\varepsilon_1\right)\right]^5
\left[\Gamma\left(1+l-2\varepsilon_1\right)\right]^5} \vline\right.^2,
\nonumber\\
\end{eqnarray}
where $z:=\e^{-t}$. We define $|f(z,\varepsilon_1,\varepsilon_2)|^2
:=f(z,\varepsilon_1,\varepsilon_2)f(\bar z,\varepsilon_1,\varepsilon_2)$.
The same will be used in what follows.
 Evaluating the residue integrals to lowest order, the correspondence
$Z_{S^2}=\e^{-K}$ gives the following leading behavior
\begin{equation}
\e^{-K}=-\frac{5}{96}(z\bar{z})^q\left(\log^3\frac{z\bar{z}}{2^{16}}
-240\zeta(3)\right)+\ldots
\end{equation}                                                              
The leading behavior of the K\"ahler metric near $z=0$ is thus
\begin{equation}
g_{z\bar{z}}=\frac{3}{z\bar{z}(\log{z\bar{z}\over 2^{16}})^2}+\ldots.
\end{equation}
This shows that the limit $r\to +\infty$ is at infinite distance in the
moduli space. Thus, we are in the true hybrid phase in the sense of
\cite{Aspinwall:2009qy}.

This suggests a possible geometric interpretation of the model.
In \cite{Jockers:2012dk} it has been
observed in examples that the sphere partition function and the
correctly normalized K\"ahler potential are related by a K\"ahler
transformation
\begin{equation}
\e^{-K}\sim\frac{Z_{S^2}}{X^0(z)\overline{X^{0}(z)}},
\label{obsJ}
\end{equation}
where $X^0(z)$ was found to be the fundamental period up to some
rescalings. Therefore we can attempt to extract the fundamental period
using this property. Given the general form of the A-model K\"ahler
potential, the normalization factor is the coefficient of the $\log^3           
z\bar{z}$-term of the partition function. After multiplication with a
constant and rescaling\footnote{Such a scaling factor is also observed
  in \cite{kanazawa10}.} $z\rightarrow -256 z$ we obtain
\begin{equation}
X^{0}(z)=1 - 76 z + 45036 z^2 - 41983600 z^3 + 47990065900 z^4
-61620234426576 z^5+\ldots
\end{equation}
Consulting the database of Calabi-Yau equations \cite{vanstraten},
this is the fundamental period associated to the Picard-Fuchs operator
\begin{eqnarray}
\mathcal{L}&=&\theta^4 +z(2000\theta^4+3904\theta^3+2708\theta^2+756\theta+76)
 \nonumber\\
&&+z^2(63488\theta^4+63488\theta^3-21376\theta^2-18624\theta-2832)
\nonumber\\
&&+z^3(512000\theta^4+24576\theta^3-37888\theta^2+6144\theta+3072)
+z^4(4096(2\theta+1)^4),
\end{eqnarray}
where $\theta=z\frac{d}{dz}$. This is precisely the differential
operator No. 238 
derived in \cite{kanazawa10} via analytic continuation of the
Picard-Fuchs operator of the Pfaffian Calabi-Yau.

We can further check this result using mirror symmetry. Since the
Picard-Fuchs operator in the $r\ll0$-phase is known, we can determine
the Picard-Fuchs operator at $r\gg0$ from it and compute the
periods. The K\"ahler potential can be determined using a procedure
described in \cite{Masuda:1998eh}. The result matches the localization
computation. Further details are discussed in appendix
\ref{app-mirmet}.

It would also be interesting to employ more recent techniques, such as the
one in \cite{Sugishita:2013jca,Honda:2013uca,Hori:2013ika},
to the original and the dual linear
sigma models, to obtain further information about the system.


\subsection{(A$^2_{(-1)^6,(-2),1^4,0}$)}\label{subsec:X10}

The model has gauge group $G=U(1)\times SU(2)$ and the matter consists of
$p^1,\ldots, p^6$ of charge $-1$, $p^7$ of charge $-2$, 
 doublets $x_1,\ldots, x_4$ of charge $1$ and a doublet $x_5$ of
charge $0$. The matrix $A(p)$ must have charge $-2$ in the first $4\times 4$
block (quadratic in $p^{1,...,6}$ and linear in $p^7$)
and charge $-1$ in the remaining off-diagonals (linear
in $p^{1,...,6}$).

The singular points are found in the same way as in the previous subsection.
The vacuum equations read
\beq
\e^{-t}={(\sigmaU+\sigma_1)^4(\sigmaU-\sigma_1)^4\over
(-\sigmaU)^6(-2\sigmaU)^2},
\quad
1={(\sigmaU+\sigma_1)^4\sigma_1\over (\sigmaU-\sigma_1)^4(-\sigma_1)}.
\label{vaceqX10}
\eeq
Taking the admissible solutions and modding out by the Weyl group,
we find that there are two singular points at
\begin{equation}
\e^{-t}=272\pm192\sqrt{2}.
\label{X10sing}
\end{equation}

The Pfaffian $Y_A$ in the $r\ll 0$ phase
is nothing but $X_{10}$ in Kanazawa's work \cite{kanazawa10}.
It is a Calabi-Yau threefold with the following topological data
\beq
h^{1,1}(X_{10})=1,\quad
h^{2,1}(X_{10})=59,\quad
\int_{X_{10}}H^3=10,\quad
\int_{X_{10}}c_2(X_{10})H=52.
\eeq
In that paper, the mirror family was proposed 
and the Picard-Fuchs equation was identified with
the Calabi-Yau equation No. 210 (formerly 263) in \cite{vanstraten}. 
The singular points of that equation
agrees with (\ref{X10sing}) if we identify the 
parameter $\phi$ in  \cite{kanazawa10} with our $-\e^t$.

Let us now look at the $r\gg 0$ phase.
First, we determine the vacuum manifold.
The F-term equations that remain after $p=0$ are
\beqa
&&\sum_{l,m=1}^4A_7^{lm}[x_lx_m]=0,
\label{FeqM10}\\
&&\sum_{l=1}^5A_j^{l5}[x_lx_5]=0,\qquad j=1,\ldots, 6.
\eeqa
By genericity of $A$, the second set of equations yields $[x_lx_5]=0$
for  $l=1,\ldots, 4$.
If $x_5\ne 0$, this means that $x_1,\ldots, x_4$ are all proportional to $x_5$,
violating the D-term constraint ${\rm rank}(x)=2$. Thus, $x_5$ must vanish.
If the gauge group were $U(2)$, then the vacuum manifold would be
the hypersurface  (\ref{FeqM10})
of the Grassmannian $G(2,4)$ spanned by $x_{(4)}=(x_1,\ldots, x_4)$. 
This is a Fano threefold which we denote
by $M_{A_7}$ where $A_7=(A^{lm}_7)$.
Since the gauge group is actually $U(1)\times SU(2)$, the
vacuum manifold is something like $M_{A_7}/\Z_2$ where
$\Z_2=\{(\pm 1,\pm {\bf 1}_2)\}$. Since $p^7$ is used up, the superpotential
at this stage is
\beqa
W&=&\sum_{i,j=1}^6\sum_{l,m=1}^4A_{ij}^{lm}p^ip^j[x_lx_m]
+\sum_{j=1}^6\sum_{l=1}^4A_j^{l5}p^j[x_lx_5]
\nn\\
&=&\sum_{i,j=1}^6A_{ij}(x_{(4)})p^ip^j+\sum_{j=1}^6\sum_{a=1}^2
A_{j,a}^5(x_{(4)})p^jx^a_5,
\label{pWX10}
\eeqa
where $A_{ij}(x_{(4)})=\sum_{l,m}A^{lm}_{ij}[x_lx_m]$
and $A^5_{j,a}(x_{(4)})=\sum_{l,b}A^{l5}_jx_l^b\epsilon_{ba}$
in which $\epsilon_{ab}$ is the symplectic form defining $SU(2)$.
Under the genericity assumption, the $6\times 2$ matrix
$(A^5_{j,a}(x_{(4)}))$ is always of rank $2$ since $(x^b_l)$ is always
of rank $2$. Therefore, the second term on the right hand side of
(\ref{pWX10}) gives mass to $(x^1_5, x^2_5)$ and two of
$p_{(6)}=(p^1,\ldots, p^6)$. Integrating them out, we have the constraint
\beq
\sum_{j=1}^6A^5_{j,a}(x_{(4)})p^j=0,\qquad a=1,2.
\label{anotherX10}
\eeq
The remaining fields are not always massive. Thus, we have the hybrid
model with the target space
\beq
\mathfrak{X}_{A'}={\displaystyle \left\{\,(p_{(6)},x_{(4)})\in
\C^6\oplus {\rm Hom}(\C^4,\C^2)\,\,\Bigr|\,\,
{\rm rank}\,x_{(4)}=2,\,\,
(\ref{FeqM10}),\,\,(\ref{anotherX10})\,\,\right\}
\over\C^{\times}\times SL(\C^2)},
\eeq
and the superpotential
\beq
W_{A''}=\sum_{i,j=1}^6A_{ij}(x_{(4)})p^ip^j.
\eeq
In the above expressions, $A'=(A^{lm}_7,A^{l5}_j)$ and $A''=(A^{lm}_{ij})$.
The space $\mathfrak{X}_{A'}$ 
is the total space of an orbifold vector
bundle of rank $4$ over $M_{A_7}/\Z_2$.
It is an orbifold with a $\Z_2$ orbifold singularity
at the zero section.
The model $(\mathfrak{X}_{A'},W_{A''})$
may be regarded as a fibration over $M_{A_7}$ of a $\Z_2$ Landau-Ginzburg
orbifold of four variables with a quadratic superpotential,
as in the $r\gg 0$ phase of the model (A$^2_{(-1)^4,(-2)^3,1^5}$)
studied in Section~\ref{subsec:X5}
and as in the $r\ll 0$ phase of the model
(S$^{1,+}_{(-1)^4,1^8}$) studied in \cite{Caldararu:2007tc}.
It would be interesting to find the type of the $\Z_2$ orbifold,
$O_+(1)$ or $O_-(1)$, in order to see whether there is a double cover or not.

Let us compute the sphere partition function in order to find
 the behavior of the K\"ahler metric on the moduli space
in the limit $r\to +\infty$. The
calculation is very similar to $X_5$. The result is
\begin{eqnarray}
Z_{S^2}\!\!&=&\!\!-\frac{(z\bar{z})^q}{2}\oint
\frac{\mathrm{d}\varepsilon_1\mathrm{d}\varepsilon_2}{(2\pi i)^2}
(z\bar{z})^{-\frac{\varepsilon_1}{2}-\varepsilon_2}\pi^3
\frac{\left[\cos\pi\left(\frac{\varepsilon_1}{2}+\varepsilon_2\right)
\right]^6
\left[\sin\pi\left(\varepsilon_1+2\varepsilon_2\right)\right]}
{\left[\sin\pi\varepsilon_1\right]^4\left[\sin2\pi\varepsilon_2\right]^4}
\times\nonumber\\
&&\!\!\!\!\!\!\!\!\!\!\!\!\!\!\!\!\!\!
\times\left.\vline \sum_{\stackrel{k,l=0}{k+l=even}}^{\infty}\!\!\!\!
(k-l-\varepsilon_1+2\varepsilon_2)(-z)^{\frac{k+l}{2}}
\frac{\left[\Gamma\left(\frac{1}{2}+\frac{k}{2}+\frac{l}{2}
-\frac{\varepsilon_1}{2}-\varepsilon_2\right)\right]^6
\left[\Gamma\left(1+k+l-\varepsilon_1-2\varepsilon_2 \right)\right]}
{\left[\Gamma\left(1+k-\varepsilon_1\right)\right]^4
\left[\Gamma\left(1+l-2\varepsilon_1\right)\right]^4}
\vline\right.^2,\nonumber\\
\end{eqnarray}
where $z:=\e^{-t}$. The K\"ahler potential to lowest order is
\begin{equation}
e^{-K}=-\frac{\pi^2}{48}(z\bar{z})^q
\left[\log^3\frac{z\bar{z}}{2^{24}}-264\zeta(3)\right]+\ldots
\label{KPX10}
\end{equation}
The leading behavior of the metric is thus
\begin{equation}
g_{z\bar{z}}=\frac{3}{z\bar{z}\log\left(\frac{z\bar{z}}{2^{24}}\right)^2}
+\cdots.
\end{equation}  
Thus, the limit $r\to+\infty$ is at infinite distance in the moduli space.
We are in the true hybrid phase.
                                                             
This result can be confirmed independently by a mirror symmetry
calculation. The fundamental period can be extracted from the sphere
partition function. After rescaling $z\rightarrow -2^{12}z$ one
obtains
\begin{equation}
X^0(z)=1 - 208 z + 531216 z^2 - 2168300800 z^3 + 10900554288400 z^4
-61672477170302208 z^5 +\ldots
\label{XX10exp}
\end{equation}
In agreement with \cite{kanazawa10}, this is annihilated by the
Picard-Fuchs operator No. 211 (formerly 271) of
\cite{vanstraten}:
\begin{eqnarray}
\label{op211}
\mathcal{L}&=&\theta^4 +z(208+2368\theta+9792\theta^2+14848\theta^3
+11264\theta^4) \nonumber\\
&&-z^2(495616+3981312\theta+6684672\theta^2-19267584\theta^3
-23986176\theta^4)
\nonumber\\
&&+z^3(36700160+125829120\theta-314572800\theta^2+503316480\theta^3
+14428405760\theta^4) \nonumber\\
&&+z^4(6710886400(2\theta+1)^4).
\end{eqnarray}


\subsection{(A$^2_{(-2)^7,3,1^4}$)}\label{subsec:X13}

The model has gauge group $G\cong U(2)$ and the matter given
in the following table
\begin{equation}
\begin{tabular}{c|ccc}
&$p^{1,\ldots,7}$&$x_1$&$x_{2,\ldots,5}$\\
\hline
&$\det^{-1}$&$\det\otimes {\bf 2}$&${\bf 2}$
\end{tabular}
\end{equation}
The matrix $A(p)$
must transform as $\det^{-1}$ in the last $4\times 4$ block
(linear in $p$) and
$\det^{-2}$ in the remaining off diagonal block (quadratic in $p$).

Let us identify the singular loci.
Parametrizing the maximal torus as $\sigma={\rm diag}(\sigma_1,\sigma_2)$,
the effective twisted superpotential is
\beq
\wt{W}_{\it eff}=-t(\sigma_1+\sigma_2)+\pi i (\sigma_1-\sigma_2)
-\sum_{\chi}\chi(\sigma)(\log\chi(\sigma)-1),
\eeq
where $t$ is the correctly normalized FI-theta parameter,
$\pi i (\sigma_1-\sigma_2)$ is from the W-boson integral
and the last sum, coming from the
matter one loop integral,
is over the weights,
\begin{equation}
\label{x13characters}
\begin{tabular}{c|ccc}
&$p^{1,\ldots, 7}$&$x_{1}$&$x_{2,\ldots,5}$\\
\hline
$\chi(\sigma)$&$-\sigma_1-\sigma_2$&
$\left(\begin{array}{c}2\sigma_1+\sigma_2\\\sigma_1+2\sigma_2\end{array}        
\right)$
&$\left(\begin{array}{c}\sigma_1\\\sigma_2\end{array}\right)$
\end{tabular}.
\end{equation}
The vacuum equation reads
\beq
-\e^{-t}={(2\sigma_1+\sigma_2)^2(\sigma_1+2\sigma_2)\sigma_1^4\over
(-\sigma_1-\sigma_2)^7}
={(2\sigma_1+\sigma_2)(\sigma_1+2\sigma_2)^2\sigma_2^4\over
(-\sigma_1-\sigma_2)^7}.
\eeq
We need to exclude the solutions such as $\sigma_1+\sigma_2=0$ and
$\sigma_1=\sigma_2$ where some of the matter or the W-bosons are massless.
Then, there are four admissible solutions
but we need to mod out by the
Weyl group action $\sigma_1\leftrightarrow \sigma_2$.
Thus, we find two singular points at
\begin{equation}
\e^{-t}=-\frac{1}{2}(349\pm 85\sqrt{17}).
\label{X13sing}
\end{equation}

The Pfaffian $Y_A$ in the $r\ll 0$ phase
is nothing but the Pfaffian constructed by Tonoli in \cite{Tonoli}
(it is denoted by $X_{13}$ in \cite{kanazawa10}).
It is a Calabi-Yau threefold with the following topological data
\beq
h^{1,1}(X_{13})=1,\quad
h^{2,1}(X_{13})=61,\quad
\int_{X_{13}}H^3=13,\quad
\int_{X_{13}}c_2(X_{13})H=58.
\eeq
In \cite{kanazawa10}, the mirror family was proposed 
and the Picard-Fuchs equation was identified with
the Calabi-Yau equation No. 99 in \cite{vanstraten}. 
The singular points of that equation
agrees with (\ref{X13sing}) if we identify the 
parameter $\phi$ in  \cite{kanazawa10} with our $-\e^t$.

Let us look into the $r\gg 0$ phase.
After $p=0$ the remaining F-term equations are
\begin{equation}
\sum_{\alpha,\beta=2}^5A^{\alpha\beta}_i[x_{\alpha} x_{\beta}]=0
\qquad i=1,\ldots,7.
\end{equation}
There are seven equations for six $[x_{\alpha} x_{\beta}]$'s. 
By genericity, this
implies that they are all zero, which leads to the conclusion that
the matrix $x_{(4)}=(x_2,\ldots,x_5)$ has rank one or less.
Since $x$ must have rank two,
$x_1$ is non-zero, $x_{(4)}$ has rank one, and the
two are linearly independent. Using the left action of
$GL(2,\C)$ we can fix $x_1=(1,0)^T$. The stabilizer of this is
determined as follows.
\begin{equation}
\left(\begin{array}{cc}a&b\\c&d\end{array}\right):
\left(\begin{array}{c}1\\0\end{array}\right)
\longmapsto (ad-bc)
\left(\begin{array}{c}a\\c\end{array}\right)
\stackrel{!}{=}
\left(\begin{array}{c}1\\0\end{array}\right)
\end{equation}
This fixes $c=0,d=a^{-2}$. Using what is left of the symmetry we can
furthermore simplify the shape of $x_{(4)}$. Since this
matrix has rank $1$ and linearly independent of $x_1=(1,0)^T$,
$x_{(4)}$ can be brought to the form
$(\lambda w,w)^T$ with $w=(w_2,\ldots, w_5)$ non-zero. 
Now we apply the residual gauge symmetry
\begin{equation}
\left(\begin{array}{cc}a&b\\0&a^{-2}\end{array}\right)
\left(\begin{array}{c}\lambda w\\w\end{array}\right)=
\left(\begin{array}{c}a\lambda w+bw\\a^{-2}w
\end{array}\right).
\end{equation}
We see that we can eliminate the upper components of $x_{(4)}$.
Thus, the vacuum manifold is the space
$\C^4\setminus \{0\}$ of $w$ modulo $w\mapsto a^{-2}w$
for $a\in \C^{\times}$.
This is nothing but $\CP^3/\Z_2$ or $\PP^3_{2222}$.
Let us write down the superpotential in the gauge $x_1=(1,0)^T$
with the residual gauge symmetry parametrized by
$(a,b)\in \C^{\times}\times\C$: Denoting $x_{(4)}=(z,w)^T$,
it reads
\beqa
W&=&\sum_{i,j=1}^7\sum_{\alpha=2}^5A^{1\alpha}_{ij}p^ip^j[x_1x_{\alpha}]
+\sum_{i=1}^7\sum_{\alpha,\beta=2}^5A^{\alpha\beta}_ip^i
[x_{\alpha}x_{\beta}]\nn\\
&=&\sum_{i,j=1}^7A_{ij}(w)p^ip^j
+\sum_{i=1}^7\sum_{\alpha=2}^5A^{\alpha}_{\,\,\,i}(w)z_{\alpha}p^i
\eeqa
where $A_{ij}(w)=\sum_{\alpha}A^{1\alpha}_{ij}w_{\alpha}$ and
$A^{\alpha}_{\,\,\,i}(w)=\sum_{\beta}A^{\alpha\beta}_iw_{\beta}$.
Note that $\sum_{\alpha}A^{\alpha}_{\,\,\,i}(w)w_{\alpha}=0$
(as it should be since $z\to z+bw$ is a gauge symmetry), and hence
the $4\times 7$ matrix $A'(w)=(A_{\,\,\, i}^{\alpha}(w))$ has rank three or less.
Under the genericity assumption, it is of rank $3$ for any $w\ne 0$.
\footnote{The codimension of the space of rank $\leq k$ matrices in
the space of $m\times n$ matrices is $(m-k)(n-k)$. In the present case, 
if we apply $m=3$, $n=7$, $k=2$ the codimension is
$(3-2)(7-2)=5$ which is too big compared to the dimension
$3$ of the space of $w$'s.}
Integrating out $z$ (modulo shift by $w$), we have the constraint
\beq
A'(w)p=0.
\label{constrX13}
\eeq 
This leaves us $7-3=4$ dimensional space of $p$'s, and none of them
are massive everywhere. Thus, we have the hybrid model with
the target space
\beq
\mathfrak{X}_{A'}={\displaystyle \left\{\,(p,w)\in \C^7\oplus \C^4\,\,
\Bigr|\,\,w\ne 0,\,\,A'(w)p=0.\,\,\right\}\over
\C^{\times}\ni a:(p,w)\mapsto (ap, a^{-2}w)}
\eeq
and the superpotential
\beq
W_{A''}=\sum_{i,j=1}^7A_{ij}(w)p^ip^j,
\eeq
where $A''=(A^{1\alpha}_{ij})$.
The space $\mathfrak{X}_{A'}$ 
is the total space of an orbifold vector
bundle of rank $4$ over $\CP^3/\Z_2$,
with a $\Z_2$ orbifold singularity at the zero section.
The model $(\mathfrak{X}_{A'},W_{A''})$
may be regarded as a fibration over $\CP^3$ of a $\Z_2$ Landau-Ginzburg
orbifold of four variables with a quadratic superpotential,
as in the $r\gg 0$ phase of the models 
studied in Section~\ref{subsec:X5} and \ref{subsec:X10}
and as in the $r\ll 0$ phase of the model
(S$^{1,+}_{(-1)^4,1^8}$) studied in \cite{Caldararu:2007tc}.
It would be interesting to find the type of the $\Z_2$ orbifold,
$O_+(1)$ or $O_-(1)$, in order to see whether there is a double cover or not.

Let us compute the sphere partition function in order to find
 the behavior of the K\"ahler metric on the moduli space
in the limit $r\to +\infty$. 
The detail is outlined in the appendix, and here we only show the result:
\begin{eqnarray}
Z_{S^2}&=&-\frac{(z\bar{z})^q}{32}\oint
\frac{\mathrm{d}\varepsilon_1\mathrm{d}\varepsilon_2}{(2\pi i)^2}
\pi^3(z\bar{z})^{\frac{1}{2}-\frac{\varepsilon_1}{4}-\varepsilon_2}\times
\nonumber\\
&&\times\frac{
\left[\cos\pi\left(\frac{\varepsilon_1}{4}+\varepsilon_2\right)\right]^5
\left[\sin\pi\left(\frac{\varepsilon_1}{2}+2\varepsilon_2\right)\right]^2}
{\left[\sin\pi\frac{\varepsilon_1}{2}\right]^2
\left[\cos\pi\left(\frac{\varepsilon_1}{4}+3\varepsilon_2\right)\right]^2
\left[\cos\pi\left(\frac{\varepsilon_1}{4}-\varepsilon_2\right)\right]^3
\left[\sin2\pi\varepsilon_2\right]^3}\times\nonumber\\
&&\times \left.\vline\sum_{\stackrel{k,l=0}{k+l=even}}^{\infty}
(2-2k+6l+\varepsilon_1-12\varepsilon_2)(-z)^{\frac{k+l}{2}}
\right.\times\nonumber\\
&&\times \left.\frac{\left[\Gamma\left(\frac{1}{2}(1+k+l)
-\frac{\varepsilon_1}{4}-\varepsilon_2\right)\right]^5}
{\left[\Gamma\left(1+k-\frac{\varepsilon_1}{2}\right)\right]^2
\left[\Gamma\left(\frac{3}{2}+\frac{k}{2}+\frac{3l}{2}
-\frac{\varepsilon_1}{4}-3\varepsilon_2\right)\right]^2}\right.
\times\nonumber\\
&&\times\left.\frac{\left[\Gamma\left(1+k+l-\frac{\varepsilon_1}{2}
+2\varepsilon_2\right)\right]^2}
{\left[\Gamma\left(\frac{1}{2}+\frac{k}{2}-\frac{l}{2}
-\frac{\varepsilon_1}{4}+\varepsilon_2\right)\right]^3
\left[\Gamma(1+l-2\varepsilon_2)\right]^3} \vline\right.^2,\nonumber\\
\end{eqnarray}
where $z:=-\e^{-t}$.
The leading term is
\begin{equation}
e^{-K}=-\frac{1}{96}(z\bar{z})^{2q+\frac{1}{2}}
\left[\left(\log\frac{z\bar{z}}{2^{32}}\right)^3+576\zeta(3)\right]+\ldots
\end{equation}
and the K\"ahler metric behaves as
\begin{equation}
g_{z\bar{z}}=
\frac{3}{z\bar{z}\log\left(\frac{z\bar{z}}{2^{32}}\right)^2}+\cdots.
\end{equation}  
Thus, the limit $r\to+\infty$ is at infinite distance in the moduli space.
We are in the true hybrid phase.

Rescaling $z\rightarrow -2^{16}z$, the fundamental period is
\begin{equation}
X^0(z)=1 - 2320 z + 57601296 z^2 - 2373661139200 z^3 +
 121665506430000400 z^4 +\ldots
\end{equation}
This is annihilated by the Picard-Fuchs operator No. $207$ (formerly
$225$) of \cite{vanstraten} in agreement with \cite{kanazawa10}:
\begin{eqnarray}
\mathcal{L}&=&\theta^4 -z(17152\theta^4-285184\theta^3-174208\theta^2
-31616\theta-2320)\nonumber\\
&& -z^2(6696206336\theta^4+15252586496\theta^3-5932843008\theta^2
-1864892416\theta-183107584) \nonumber\\
&&+z^3(255108172480512\theta^4+5360119185408\theta^3
-1702954532864\theta^2\nonumber\\
&&+1340029796352\theta+338497110016) -z^4(2973079441506304(2\theta+1)^4)
\end{eqnarray}


\subsection{(A$^2_{(-2)^5,(-4)^2,3^2,1^3}$)}\label{subsec:X7}

The model has gauge group $G\cong U(2)$ and the matter given in the
following table
\begin{equation}
\begin{tabular}{c|cccc}
&$p^{1,\ldots,5}$&$p^{6,7}$&$x_{1,2}$&$x_{3,\ldots,5}$\\
\hline
&$\det^{-1}$&$\det^{-2}$&$\det\otimes {\bf 2}$&${\bf 2}$
\end{tabular}.
\end{equation}
where $\det$ is the determinant representation and
${\bf 2}$ is the fundamental doublet. The matrix $A(p)$
must transform as $\det^{-3}$ in the first $2\times 2$ block
(cubic in $p^{1,...,5}$ and bilinear in $(p^{1,...,5},p^{6,7})$),
$\det^{-1}$ in the last $3\times 3$ block (linear in $p^{1,...,5}$), 
and $\det^{-2}$ in the off diagonal block (quadratic in $p^{1,...,5}$
and linear in $p^{6,7}$).

Singular points are found in the same way as in the previous subsection.
The vacuum equations read
\beq
-\e^{-t}={(2\sigma_1+\sigma_2)^4(\sigma_1+2\sigma_2)^2\sigma_1^3\over
(-\sigma_1-\sigma_2)^5(-2\sigma_1-2\sigma_2)^4}
={(2\sigma_1+\sigma_2)^2(\sigma_1+2\sigma_2)^4\sigma_2^3\over
(-\sigma_1-\sigma_2)^5(-2\sigma_1-2\sigma_2)^4}.
\eeq
Taking the admissible solutions and modding out by the Weyl group,
we find that there are two singular points at
\begin{equation}
\e^{-t}=540\pm312\sqrt{3}.
\label{X7sing}
\end{equation}

The Pfaffian $Y_A$ in the $r\ll 0$ phase
is nothing but $X_7$ in Kanazawa's work \cite{kanazawa10}.
It is a Calabi-Yau threefold with the following topological data
\beq
h^{1,1}(X_7)=1,\quad
h^{2,1}(X_7)=61,\quad
\int_{X_7}H^3=7,\quad
\int_{X_7}c_2(X_7)H=46.
\eeq
In that paper, the mirror family was proposed 
and the Picard-Fuchs equation was identified with
the Calabi-Yau equation No. 109 in \cite{vanstraten}. 
The singular points of that equation
agrees with (\ref{X7sing}) if we identify the 
parameter $\phi$ in  \cite{kanazawa10} with our $-\e^t$.

Now let us look at the $r\gg 0$ phase.
First what is the vacuum manifold?
Regarding $p=0$, the remaining F-term equations are
\begin{eqnarray}
&&\sum_{\alpha,\beta=3}^5A_{i}^{\alpha\beta}[x_{\alpha} x_{\beta}]=0
\qquad i=1,\ldots,5\nonumber\\
&&\sum_{l=1}^2\sum_{\alpha=3}^5A_{\mu}^{l\alpha}[x_l x_{\alpha}]=0\qquad \mu=6,7.
\end{eqnarray}
From genericity of $A$, the first set of equations imply
$[x_{\alpha}x_{\beta}]=0$
for $\alpha,\beta=3,4,5$. This is equivalent to the requirement
that $x_3,x_4,x_5$ are proportional to each other.
Thus the dimension is cut only by two.
The other two equations are independent, and 
the vacuum manifold has dimension $10-2-2-4=2$.
We do not know the details about this variety other than
the fact that there is one point with $x_{3,4,5}=0$ with the orbifold group
$\Z_3$ and a line $\CP^1$ with the orbifold group $\Z_2$.
We are unable to identify any massive direction, and
we only have a poor description of the model:
It is a hybrid model with the target space given by
the solution space to D-term equation modulo gauge group 
and the original superpotential.
It is similar to the
pseudo-hybrid model discussed in \cite{Aspinwall:2009qy}
in that the vacuum manifold is two-dimensional (and even in that there is
one point with $\Z_3$ orbifold group and a curve with $\Z_2$ orbifold group).

Failing to obtain any idea on the nature of the low energy physics
from the classical analysis, we now
look at the behavior of the K\"ahler metric
of the moduli space via the two sphere partition function.
Details can be found in appendix \ref{app-loc}. 
The leading terms of the partition function are
\begin{equation}
{}\e^{-K}
=\frac{2}{\sqrt{3}\pi}
\frac{\Gamma\left(\frac{1}{3}\right)^{10}}
{\Gamma\left(\frac{2}{3}\right)^8}
(z\bar{z})^{2q+\frac{1}{3}}
-4(z\bar{z})^{2q+\frac{1}{2}}(36+8\log4-3\log{z\bar{z}})+\ldots,
\end{equation}
and hence the K\"ahler metric behaves as
\begin{equation}
g_{z\bar{z}}=-\frac{\pi\Gamma\left(\frac{2}{3}\right)^8
\log^3\left(
\frac{z\bar{z}}{2^{\frac{16}{3}}}\right)}
{6\sqrt{3}\Gamma\left(\frac{1}{3}\right)^{10}
(z\bar{z})^{\frac{5}{6}}}.
\end{equation}
where $z:=-\e^{-t}$.
This shows that the limiting point actually lies at a finite distance
in the moduli space. The very same leading behavior of the K\"ahler
metric has been observed in the pseudo-hybrid model in
\cite{Aspinwall:2009qy} mentioned above.

\section{A new Calabi-Yau threefold from (S) series}\label{sec:(S)}

We now study the model (S$^{2,+}_{(-1)^2,(-2)^3,1^4}$) in detail.
The gauge group is $G=U(1)\times O_+(2)$ and
the matter consists of five $O(2)$ singlets ---
$p^1,p^2$ and $p^3,p^4,p^5$ of $U(1)$ charges 
$-1$ and $-2$ respectively --- and four
$O(2)$ doublets, $x_1,\ldots,x_4$, of $U(1)$ charge $1$.
The matrix $S(p)$ must have charge $-2$. It must be linear in 
$p^{3,4,5}$ and quadratic in $p^{1,2}$.

Let us identify the singular loci.
The effective twisted superpotential is
\beq
\wt{W}_{\it eff}=-t\sigmaU+\pi i \sigma_1
-\sum_{\chi}\chi(\sigma)(\log\chi(\sigma)-1),
\eeq
where $t$ is the correctly normalized FI-theta parameter,
$\pi i \sigma_1$ is the discrete theta angle needed for the
regularity,
and the last sum
is over the weights of the matter representation,
\begin{equation}
\label{Scharacters}
\begin{tabular}{c|ccc}
&$p^{1,2}$&$p^{3,4,5}$&$x_{1,\ldots,4}$\\
\hline
$\chi(\sigma)$&$-\sigmaU$&
$-2\sigmaU$
&$\left(\begin{array}{c}\sigmaU+\sigma_1
\\\sigmaU-\sigma_1\end{array}\right)$
\end{tabular}.
\end{equation}
The vacuum equations read
\begin{equation}
\e^{-t}={(\sigmaU+\sigma_1)^4(\sigmaU-\sigma_1)^4\over
(-\sigmaU)^2(-2\sigmaU)^6},\quad\,\,\,
-1={(\sigmaU+\sigma_1)^4\over (\sigmaU-\sigma_1)^4}.
\label{vaceqS}
\end{equation}
The second equation is solved by $(\sigma_1/\sigmaU)^2=-3\pm 2\sqrt{2}$.
The four solutions are all admissible,
but we need to mod out by the Weyl group action $\sigma_1\mapsto -\sigma_1$.
Thus, we find two singular points at
\begin{equation}
\e^{-t}=17\mp 12\sqrt{2}.
\label{Ssing}
\end{equation}

Let us count the number of parameters that enter into
the matrix $S(p)$. It is a $4\times 4$ symmetric matrix and hence
has $10$ independent entries, each of which is a linear combination of
$(p^1)^2,p^1p^2,(p^2)^2,p^3,p^4, p^5$. Thus, it has $10\cdot 6=60$
parameters. On the other hand, reparametrizations of $x_{1,...,4}$,
$p^{1,2}$ and $p^{3,4,5}$ have $4^2=16$, $2^2=4$ and $3\cdot 6=18$
 parameters respectively.
Since one of these is the gauge symmetry, the net
number of reparametrization is $16+4+18-1=37$.
Therefore, we found that
\beq
\mbox{the number of parameters in $W$}=
60-37=23.
\label{numbercountS}
\eeq

The elliptic genus of the model behaves as
\beq
Z_{T^2}(\tau,z)\,=\,-22\,(\e^{\pi i z}+\e^{-\pi i z})+O(\e^{2\pi i\tau})\qquad
\mbox{as $\tau\to +i\infty$}.
\label{EllGenS}
\eeq
See Appendix~\ref{app:ellipticgenus} for the computation.
In particular, the Witten index is
\beq
{\rm Tr}\,(-1)^F~=~-44.
\label{WindexS}
\eeq

\subsection{$r\ll 0$: the double cover of symmetric determinantal variety}

Let us start with the $r\ll 0$ phase.
As discussed in Section~\ref{subsec:strong}, we have a non-linear sigma
model whose target space is the double cover
$\wt{Y}_S$ of the symmetric determinantal variety
$Y_S=\{\,p\in \PP\,|{\rm rank}\,S(p)\leq 3\}$
which is ramified along 
the curve
\beq
C_S=\left\{\,\,p\in \PP\,\,\Bigr|\,\,{\rm rank}\,S(p)=2\,\,\right\}.
\eeq
Note that ${\rm rank}\,S(p)=1$ is impossible for dimensional reasons 
as long as $S(p)$ is generic.
$Y_S$ has an A$_1$ singularity (i.e. $\C^2/\{\pm 1 \}$ singularity)
along $C_S$ which is unfolded by the double cover $\wt{Y}_S$.
Therefore, $\wt{Y}_S$ is guaranteed to be smooth if $Y_S$ is away from
the orbifold locus of the ambient space $\PP$.
Unfortunately, the orbifold locus of $\PP={\bf WCP}^4_{11222}$,
which is $\mathfrak{S}=\{p^1=p^2=0\}$,
is of codimension $2$ and hence cannot be avoided by $Y_S$.
The intersection is along a curve 
\beq
\Sigma_S=\mathfrak{S}\cap Y_S.
\eeq
This is again a curve of A$_1$ singularity in $Y_S$ (it is away from 
the other singular curve $C_S$).
A closer inspection is needed to see if $\wt{Y}_S$ is smooth or not.

The double cover $\wt{Y}_S$ is defined to be
 the classical vacuum manifold of the dual theory,
\beq
\wt{Y}_S={\displaystyle \left\{\,\,(p,\wt{x})\in \C^5\oplus
{\rm Hom}(\C^4,\C^3)\,\,\Bigr|\,\,\mbox{stability},\,\,
S(p)+(\wt{x}\wt{x})=0\,\,\right\}\over
\C^{\times}\times SO(\C^3)}.
\eeq
The covering map $\wt{Y}_S\to Y_S$ is simply the forgetful map
$(p,\wt{x})\mapsto p$.
We know that the only possible singularity is above $\Sigma_S$. 
Let $(p^{1,2},p^{3,4,5})\in Y_S$ be in a neighborhood of $\Sigma_S$.
In $Y_S$ there is an identification
$(p^{1,2},p^{3,4,5})\equiv (-p^{1,2},p^{3,4,5})$, and this causes the
A$_1$ singularity at $p^{1,2}=0$.
In $\wt{Y}_S$, there are two preimages of this point:
$(p^{1,2},p^{3,4,5},\wt{x})\equiv (-p^{1,2},p^{3,4,5},-\wt{x})$
and $(-p^{1,2},p^{3,4,5},\wt{x})\equiv (p^{1,2},p^{3,4,5},-\wt{x})$.
They are different points if $p^{1,2}\ne 0$
but coalesce as $p^{1,2}\to 0$.
A choice of $\wt{x}$ provides one sheet over $Y_S$ and the deck transform of
the covering map $\wt{Y}_S\to Y_S$ can be described as
\beq
(p^{1,2},p^{3,4,5},\wt{x})\longmapsto (-p^{1,2},p^{3,4,5},\wt{x}).
\eeq
We see that the
$\C^2/\{\pm 1\}$ singularity along $\Sigma_S$ is 
unfolded by the cover $\wt{Y}_S\to Y_S$.

To summarize, $\wt{Y}_S$ is smooth and the covering map
$\wt{Y}_S\to Y_S$ provides a simultaneous unfolding of
the A$_1$ singularity of $Y_S$
along the two disjoint curves, $C_S$ and $\Sigma_S$.

$\wt{Y}_S$ is a Calabi-Yau manifold with the following topological
data\footnote{We thank Hiromichi Takagi for this information.
See Appendix~\ref{app:wtYStop} for the outline.}
\beq
h^{1,1}(\wt{Y}_S)=1,\quad
h^{2,1}(\wt{Y}_S)=23,\quad
\int_{\wt{Y}_S}H^3=2,\quad
\int_{\wt{Y}_S}c_2(\wt{Y}_S)H=20,
\label{wtYStop}
\eeq
where $H$ is some element of $H^2(\wt{Y}_S,\Z)$.
The Hodge numbers confirm that
the one FI-theta parameter is the only exactly marginal
twisted chiral parameter and that, in view of
(\ref{numbercountS}), the parameters entering in $S(p)$
are all the exactly marginal chiral parameters.
The Euler number of $\wt{Y}_S$ is $-44$. This is consistent with
the Witten index (\ref{WindexS}).

Let us compute the two sphere partition function in order to see 
if the metric of 
the K\"ahler moduli space is of the expected form in the geometric phase.
The details of the calculation are again given in the appendix. The result is 
\begin{eqnarray}
Z_{S^2}^{r\ll0}&=&\frac{1}{2}\lim_{\delta\rightarrow0}
\oint\frac{\mathrm{d}\varepsilon_1\mathrm{d}\varepsilon_2}{(2\pi i)^2}
(z\bar{z})^{q-\frac{1}{2}-\varepsilon_2}\pi^5
\frac{\left[\sin\pi(\varepsilon_1+2\varepsilon_2)\right]^4}
{\left[\sin\pi\left(\varepsilon_2\right)\right]^2
\left[\sin 2\pi\varepsilon_2\right]^3
\left[\sin\pi\varepsilon_1\right]^4}
\times\nonumber\\
&&\times \left.\vline\sum_{k,l=0}^{\infty}(-e^{-\delta})^k(-z)^l
\frac{\left[\Gamma(1+k+2l+\varepsilon_1-2\varepsilon_2)\right]^4}
{\left[\Gamma(1+l-\varepsilon_2)\right]^2
\left[\Gamma(1+2l-2\varepsilon_2)\right]^2
\left[\Gamma(1+k+\varepsilon_1)\right]^4} \vline\right.^2
\end{eqnarray}
Here, $\e^{-\delta}$ is a convergence factor which is introduced following
\cite{Jockers:2012dk}. It is hoped by the presence of $(-1)^k$
that the result does not depend on how to take the $\delta\to 0$ limit.
Evaluating this partition function is harder than the previous
examples of $r\gg0$ phases because not all the summation variables
appear as exponents of $z,\bar{z}$. The consequence is that the
coefficients at a given order in $z,\bar{z}$ are infinite sums which
have to be evaluated order by order. 

At this moment, we are unable to find even the leading behavior of
the metric.
A similar difficulty was encountered in \cite{Jockers:2012dk}
(and we indeed followed their procedure) in the strongly coupled phase
of the R\o dland model (A$^2_{(-2)^7,1^7}$).
This difficulty seems to be correlated to the presence of continuous
unbroken gauge group.
We shall find a resolution to this problem momentarily.

Even though we are unable to find the metric, we can extract the
fundamental period by looking at the coefficient of $\log^3z\bar z$.
After an overall rescaling and after setting $z\to -\frac{2^8}{z}$ the 
fundamental period is
\begin{equation}
X^0=1 - 208 z + 531216 z^2
- 2168300800 z^3 + 10900554288400 z^4
- 61672477170302208 z^5 +\ldots
\label{XSexp}
\end{equation}
This is nothing but the fundamental period of the weakly coupled phase
($r\gg 0$) of the model (A$^2_{(-1)^6,(-2),1^4,0}$)!
It is also worth noting that
$\wt{Y}_S$ has the right property (\ref{wtYStop}) for the
Calabi-Yau threefold predicted in \cite{vanStraten2004} for
the Picard-Fuchs operator 211 (formerly 271) that annihilates (\ref{XSexp}).
We will say more on this in a moment.

\subsection{$r\gg 0$: true hybrid phase}

Let us now move on to the study of the opposite phase, $r\gg 0$.
We parametrize the doublets as
$x_j={1\over \sqrt{2}}(u_j+v_j,-iu_j+iv_j)^T$, $j=1,\ldots, 4$.
We also write the gauge group element as
$(g,h,\varepsilon)$ where $g$ is the $U(1)$ element,
$h$ is the element of $SO(2)\cong U(1)$ and 
$\varepsilon=1,\tau$ labels the disconnected part $O(2)/SO(2)\cong \Z_2$.
Note that $\tau$ acts as the exchange of $u_j$ and $v_j$.

The D-term equations read
\beq
-\sum_{\mu=1}^2|p^{\mu}|^2-2\sum_{m=3}^5|p^m|^2+||u||^2+||v||^2=r,\qquad
||u||^2=||v||^2,
\eeq
and the F-term equations read
\beqa
&&\sum_{j=1}^4S^{ij}(p)u_j=\sum_{j=1}^4S^{ij}(p)v_j=0,\quad\,\, 
i=1,\ldots, 4,\label{F1S}\\
&&\sum_{i,j=1}^4\sum_{\nu=4}^5S^{ij}_{\mu\nu}p^{\nu}u_iv_j,\quad
\mu=1,2,\label{F3S}\\
&&\sum_{i,j=1}^4S_m^{ij}u_iv_j=0,\qquad m=3,4,5.\label{F2S}
\eeqa
The D-term equations with $r\gg 0$ require that $u$ and $v$ are both
non-zero. Therefore, the gauge group is broken at most to
$\Z_2\times \langle\tau\rangle$ where the first $\Z_2$ is $g=h=\pm 1$
and the second $\langle\tau\rangle$ is $O(2)/SO(2)\cong\Z_2$. 
Thus, $r\gg 0$ is
 a weakly coupled phase.

Next, we show that $p=0$ follows from the D-term and the F-term equations.
Suppose
 $u$ and $v$ are proportional to each other. The three equations
 (\ref{F2S}) fix $u\propto v$
to a number of points up to constant
multiplication. (There are eight such points.)
If $S(p)$ is chosen generic, the $2\times 2$ matrix
$(S_{\mu\nu}(u))=(\sum_iS^{ij}_{\mu\nu}u_iu_j)$ is invertible at those points
and thus $p^1=p^2=0$ is enforced by (\ref{F3S}).
Similarly the $4\times 3$ matrix $(S^i_m(u))=(\sum_jS^{ij}_mu_j)$
has rank $3$ at those points and thus $p^3=p^4=p^5=0$ is also enforced
by (\ref{F1S}).
Next, suppose $u$ and $v$ are not proportional to each other.
Then, (\ref{F1S}) are solvable only when $S(p)$ has rank two or less.
As discussed above, for a generic $S(p)$, rank one is impossible
and rank zero means $p=0$. Thus, we only have to exclude the rank two case.
This can be done in the same way as in the (A) series in the previous
section: The equations (\ref{F2S})-(\ref{F3S}) are in conflict with
the smoothness of the rank two curve $C_S$.
Choose a point $p_*\in C_S$ and choose a basis of $\C^4$ so that
$S(p_*)$ is represented by a matrix of the form
\beq
S(p_*)=\left(\begin{array}{cccc}
0&&&\\
&0&&\\
&&a&b\\
&&b&d
\end{array}\right).
\label{Spstar}
\eeq
with the last $2\times 2$ block invertible.
In a neighborhood of $p_*$, the rank two curve $C_S$ can be defined
as the set of solutions to the three equations 
(See Appendix~\ref{app:linearalgebra}),
\beq
\Delta_{11}(p)=\Delta_{22}(p)=\Delta_{12}(p)=0,
\eeq
where $\Delta_{ij}(p)$ is the determinant of the minor of $S(p)$
obtained by deleting the $i$-th row and $j$-th column.
Smoothness of $C_S$ at $p_*$ requires that the $5\times 3$ matrix of
differentials
\beq
\left(\,{\partial\over\partial p^k}\Delta_{11},\,
{\partial\over\partial p^k}\Delta_{22},\,
{\partial\over\partial p^k}\Delta_{12}\,\right)_{k=1,\ldots, 5}
\label{themat}
\eeq
must be of rank three at $p_*$. Since $S^{ij}(p_*)$ vanishes unless
$i,j=3,4$. we have
\beq
{\partial\over\partial p^k}\Delta_{11}
={\partial\over\partial p^k}S^{22}(p_*)
\det\left(
\begin{array}{cc}
a&b \\
b&d\end{array}\right),\quad\mbox{etc}.
\eeq
A linear combination of these vanishes for all $k$ by the equations
(\ref{F2S})-(\ref{F3S}). Therefore the matrix (\ref{themat})
has rank two or less, violating the smoothness of $C_S$ at $p_*$.
This completes the proof that all $p$ must vanish.

Now let us determine the vacuum manifold.
It is the space of $(u,v)$, both non-vanishing and satisfying
(\ref{F2S}), modulo the action of the complexified gauge group.
If the gauge group were
$(U(1)\times SO(2))/\{(\pm 1,\pm {\bf 1}_2)\}$, it would be
the complete intersection $M_{S'}$ of the three hypersurfaces
(\ref{F2S}) in $\CP^3\times \CP^3=\{([u],[v])\}$,
where $S'=(S^{ij}_m)$.
But the actual gauge group is $G=U(1)\times O(2)$ and hence the vacuum manifold
is like $M_{S'}/\Z_2\times \langle\tau\rangle$ 
where the first $\Z_2$ acts trivially
and $\tau$ is the exchange of the two $\CP^3$'s.

The fields $p^3, p^4, p^5$ are used up
in imposing the equations (\ref{F2S}) and the remaining fields
$p_{(2)}=(p^1,p^2)$ are not always massive.
Therefore, we have a hybrid model with the target space
\beq
\mathfrak{X}_{S'}={\displaystyle 
\left\{\,\,(p_{(2)},u,v)\in \C^2\oplus \C^4\oplus\C^4\,\,
\Bigr|\,\, u\ne 0, \,\,v\ne 0,\,\,(\ref{F2S})\,\right\}
\over \C^{\times}\times (\C^{\times}\rtimes \langle\tau\rangle)},
\eeq
and the superpotential
\beq
W_{S''}=\sum_{\mu,\nu=1}^2S_{\mu\nu}(u,v)p^{\mu}p^{\nu},
\eeq
where $S_{\mu\nu}(u,v)=\sum_{i,j}S^{ij}_{\mu\nu}u_iv_j$
and $S''=(S^{ij}_{\mu\nu})$.
The space $\mathfrak{X}_{S'}$ 
is the total space of an orbifold vector
bundle of rank two over $M_{S'}/\Z_2\times\langle\tau\rangle$.
It is an orbifold with a $\Z_2$ orbifold singularity
at the zero section and $\Z_2\times \Z_2$ singularity
at the eight fixed points of $\tau:M_{S'}\to M_{S'}$.
The model $(\mathfrak{X}_{S'},W_{S''})$
may be regarded as a fibration over $M_{S'}/\langle\tau\rangle$ 
of a $\Z_2$ Landau-Ginzburg
orbifold of two variables with a quadratic superpotential,
as in the $r\gg 0$ phase of the first three models
studied in Section~\ref{sec:(A)}
and as in the $r\ll 0$ phase of the model
(S$^{1,+}_{(-2)^4,1^8}$) studied in \cite{Caldararu:2007tc}.
It would be interesting to find the type of the $\Z_2$ orbifold,
$O_+(1)$ or $O_-(1)$, in order to see whether there is a double cover or not.

Let us compute the sphere partition function in order to find
 the behavior of the metric on the moduli space
in the limit $r\to +\infty$. 
The result is    
\begin{eqnarray}
Z^{r\gg0}_{S^2}&=&\frac{(z\bar{z})^q}{2}
\oint\frac{\mathrm{d}\varepsilon_1\mathrm{d}\varepsilon_2}{(2\pi i)^2}
(z\bar{z})^{-\frac{\varepsilon_1}{2}-\varepsilon_2}\pi^3
\frac{\left[\cos\pi\left(\frac{\varepsilon_1}{2}+\varepsilon_2\right)\right]^2
\left[\sin\pi\left(\varepsilon_1+2\varepsilon_2\right)\right]^3}
{\left[\sin\pi\varepsilon_1\right]^4\left[\sin2\pi\varepsilon_2\right]^4}
\times\nonumber\\
&&\times\left.\vline \sum_{\stackrel{k,l=0}{k+l=even}}^{\infty}(-z)^{\frac{k+l}{2}}
\frac{\left[\Gamma\left(\frac{1}{2}+\frac{k}{2}+\frac{l}{2}
-\frac{\varepsilon_1}{2}-\varepsilon_2\right)\right]^2
\left[\Gamma\left(1+k+l-\varepsilon_1-2\varepsilon_2 \right)\right]^3}
{\left[\Gamma\left(1+k-\varepsilon_1\right)\right]^4
\left[\Gamma\left(1+l-2\varepsilon_1\right)\right]^4} \vline\right.^2
\end{eqnarray}
The leading term of the K\"ahler potential associated to this is
\begin{equation}
e^{-K}=-\frac{5}{24}\log^3\frac{z\bar{z}}{2^8}+29\zeta(3)+\ldots
\end{equation}
Therefore the metric behaves as
\begin{equation}
g_{z\bar{z}}=
\frac{3}{z\bar{z}\log\left(\frac{z\bar{z}}{2^{8}}\right)^2}+\cdots.
\end{equation}
Thus, the limit $r\to+\infty$ is at infinite distance in the moduli space.
We are in the true hybrid phase.

We may also extract the fundamental period as in the earlier examples.
After rescaling $z\rightarrow                                                   
-16 z$ it is expanded as
\begin{equation}
X^0_{r\gg0}=1 - 28 z + 4716 z^2 - 1226800 z^3 + 389349100 z^4
- 138518544528 z^5 +\ldots
\end{equation}
This is annihilated by the Picard-Fuchs operator No. $210$ (formerly
$263$) of \cite{vanstraten}:
\begin{eqnarray}
\mathcal{L}^{r\gg0}&=&25\theta^4 +z(700+6100\theta+19620\theta^2+27040\theta^3
+13760\theta^4)\nonumber\\
&& -z^2(4240+20160\theta-1536\theta^2-112128\theta^3-93696\theta^4) \nonumber\\
&&+z^3(5120+30720\theta+70656\theta^2+122880\theta^3+180224\theta^4)
\nonumber\\
&&+z^4(4096(2\theta+1)^4).
\label{PFOS}
\end{eqnarray}
Comparing with \cite{kanazawa10}, this is the same Picard-Fuchs
operator as in the strongly coupled Pfaffian phase ($r\ll 0$)
of the model (A$^2_{(-1)^6,(-2),1^4,0}$)!

\subsection{A surprise}

We have encountered a relation between
the present model (S$^{2,+}_{(-1)^2,(-2)^3,1^4}$) and
the model (A$^2_{(-1)^6,(-2),1^4,0}$) (For simplicity, we 
call them (S) and (A) respectively only here):
\begin{itemize}
\item The fundamental period in the weakly coupled $r\gg 0$ phase of (A) is
  the same as the fundamental period for the strongly coupled $r\ll 0$ 
  phase of (S). Compare (\ref{XX10exp}) and (\ref{XSexp}).
\item The Picard-Fuchs operator in the strongly coupled $r\ll 0$ phase of
  (A) matches the Picard-Fuchs operator in the weakly coupled $r\gg 0$
  phase of (S). Compare  \cite{kanazawa10} and (\ref{PFOS}).
\end{itemize}
More fundamentally, we may also add
\begin{itemize}
\item The singular points match. Compare (\ref{X10sing}) and (\ref{Ssing}):
\beq
272\pm 192\sqrt{2}=16(17\pm 12\sqrt{2})
={16\over 17\mp 12\sqrt{2}}.
\eeq
\end{itemize}
All these agreements hold under the relation
\beq
\e^{-t_{\rm A}}=16\cdot\e^{t_{\rm S}}.
\label{relre}
\eeq

These observations suggest that the K\"ahler moduli spaces of
the two models are the same. 
However, it appears to be difficult to check ---
 we have seen that the computation is difficult in the
strongly coupled phase, and there is no phase where both are weakly
coupled.
At this point, we make use of the duality, which maps 
the strongly coupled phase to the weakly coupled phase and vice versa.
We compare the partition functions of the model (A) 
and the dual model of (S).\footnote{The same comparison can be done
also between the dual model of (A) and the model (S).}
In this comparison, as we will see, the equality holds at the level of
the integrand and we actually do not need to evaluate the integral in
either phase.

The dual of (S) has the gauge group
$U(1)\times SO(3)$ and the matter contents as follows:
\beq
\begin{array}{c|cccc}
&p^{1,2}&p^{3,4,5}&\wt{x}^{1,\ldots,4}&s_{ij}=s_{ji}\\
\hline
U(1)\times SO(3)&(-1,{\bf 1})&(-2,{\bf 1})&(-1,{\bf 3})&(2,{\bf 1})\\
U(1)_V&1-\epsilon&2-2\epsilon&1-\epsilon&2\epsilon
\end{array}
\eeq
where ${\bf 1}$ and ${\bf 3}$ stand for the $SO(3)$ singlet and triplet
respectively.
The two sphere partition function of this model with $\epsilon=0$ is
\beq
Z_{S^2}^{(\wt{\rm S})}(r,\theta)
=\sum_{(m_0,m_1)\in \Z^{\oplus 2}}\!\!
\int\limits_{\,\,\,\,\,\,(\R-i\cdot 0)\times \R\!\!\!\!\!\!\!\!\!\!\!\!}
\!\!\!
\dd\sigma_0\,\dd\sigma_1
\e^{2ir\sigma_0+i\theta m_0+\pi i m_1}
\left((2\sigma_1)^2+(m_1)^2\right)
Z^{(\wt{\rm S})}_{\rm matter}(\vec{\sigma}_0,\vec{\sigma}_1).
\eeq
We have shifted the $\sigma_0$ contour slightly, in order to avoid the
pole that came down to $\sigma_0=0$ in the limit $\epsilon\searrow 0$.
The factor $\e^{\pi i m_1}$ is the effect of the W-boson integral
$\e^{2\pi i\rho(m)}$, see (\ref{ZS2}) and (\ref{rhocomp}). The last factor is
\beqa
Z^{(\wt{\rm S})}_{\rm matter}(\vec{\sigma}_0,\vec{\sigma}_1)
&=&f_{-1,1}(\vec{\sigma}_0)^2f_{-2,2}(\vec{\sigma}_0)^3
f_{1,1}(-\vec{\sigma}_0+\vec{\sigma}_1)^4
f_{1,1}(-\vec{\sigma}_0-\vec{\sigma}_1)^4
f_{1,1}(-\vec{\sigma}_0)^4
f_{2,0}(\vec{\sigma}_0)^{10}\nn\\
&=&f_{-1,1}(\vec{\sigma}_0)^6f_{2,0}(\vec{\sigma}_0)^{7}
f_{1,1}(-\vec{\sigma}_0+\vec{\sigma}_1)^4
f_{1,1}(-\vec{\sigma}_0-\vec{\sigma}_1)^4,
\label{secnds}
\eeqa
where we used the notation $\vec{\sigma}=(\sigma,m)$ and
\beq
f_{Q,R}(\vec{\sigma})={\Gamma\left(iQ\sigma-{Qm\over 2}+{R\over 2}\right)
\over \Gamma\left(1-iQ\sigma-{Qm\over 2}-{R\over 2}\right)}.
\eeq
In (\ref{secnds}), we used
\beq
f_{-2,2}(\vec{\sigma}_0)f_{2,0}(\vec{\sigma}_0)
={\Gamma(-2i\sigma_0+m_0)\over
\Gamma(1+2i\sigma_0+m_0)}\cdot
{\Gamma(2i\sigma_0+m_0+1)\over
\Gamma(-2i\sigma_0+m_0)}=1.
\label{secnds2}
\eeq
Let us revisit the
two sphere partition function of the model (A).
The gauge group is $U(1)\times SU(2)$ and the matter content is
\beq
\begin{array}{c|cccc}
&p^{1,\ldots, 6}&p^{7}&x_{1,\ldots,4}&x_5\\
\hline
U(1)\times SU(2)&(-1,{\bf 1})&(-2,{\bf 1})&(1,{\bf 2})&(0,{\bf 2})\\
U(1)_V&1-\epsilon&2-2\epsilon&\epsilon&\epsilon
\end{array}
\eeq
The partition function with $\epsilon=1$ is
\beq
Z_{S^2}^{({\rm A})}(r,\theta)
=\sum_{(m_0,m_1)\in \Z^{\oplus 2}}
\!\!
\int\limits_{\,\,\,\,\,\,(\R+i\cdot 0)\times \R\!\!\!\!\!\!\!\!\!\!\!\!}
\!\!\!
\dd\sigma_0\,\dd\sigma_1
\e^{2ir\sigma_0+i\theta m_0}
\left((2\sigma_1)^2+(m_1)^2\right)
Z^{({\rm A})}_{\rm matter}(\vec{\sigma}_0,\vec{\sigma}_1).
\eeq
We have shifted the $\sigma_0$ contour slightly, in order to avoid the pole
that came up to $\sigma_0=0$ in the limit $\epsilon\nearrow 1$.
The matter factor of the integrand is
\beq
Z^{({\rm A})}_{\rm matter}(\vec{\sigma}_0,\vec{\sigma}_1)
=f_{-1,0}(\vec{\sigma}_0)^6f_{-2,0}(\vec{\sigma}_0)
f_{1,1}(\vec{\sigma}_0+\vec{\sigma}_1)^4
f_{1,1}(\vec{\sigma}_0-\vec{\sigma}_1)^4
f_{1,1}(\vec{\sigma}_1)
f_{1,1}(-\vec{\sigma}_1).
\eeq
Note that
\beq
f_{1,1}(\vec{\sigma}_1)
f_{1,1}(-\vec{\sigma}_1)
={\Gamma\left(i\sigma_1-{m_1\over 2}+{1\over 2}\right)
\over \Gamma\left(-i\sigma_1-{m_1\over 2}+{1\over 2}\right)}\cdot
{\Gamma\left(-i\sigma_1+{m_1\over 2}+{1\over 2}\right)
\over \Gamma\left(i\sigma_1+{m_1\over 2}+{1\over 2}\right)}
=(-1)^{m_1}.
\eeq
and
\beq
f_{-1,1}(\vec{\sigma}_0)^6f_{2,0}(\vec{\sigma}_0)^6|_{\vec{\sigma}_0\to
-\vec{\sigma}_0}
=2^{-24i\sigma_0-6}f_{-1,0}(\vec{\sigma}_0)^6
\eeq
where we have used the identity
$\Gamma(2z)={1\over 2\sqrt{\pi}}2^{2z}\Gamma(z)\Gamma(z+{1\over 2})$.
It follows that
\beq
Z^{(\wt{\rm S})}_{\rm matter}(-\vec{\sigma}_0,\vec{\sigma}_1)
=2^{-24 i\sigma_0-6}(-1)^{m_1}
Z^{({\rm A})}_{\rm matter}(\vec{\sigma}_0,\vec{\sigma}_1).
\eeq
This means that
\beq
Z_{S^2}^{(\wt{\rm S})}(r,\theta)=2^{-6}
Z_{S^2}^{({\rm A})}(-r-\log 2^{12},-\theta)
\eeq
The relation between the FI-theta parameter of the model (S) and 
the one of its dual ($\wt{\rm S}$) can be read off from the 
singular loci: $\e^{-t_{\wt{\rm S}}}=2^{8}\e^{-t_{\rm S}}$.
Taking this duality for granted,
we have
\beqa
Z^{({\rm S})}_{S^2}(r_{\rm S},\theta_{\rm S})
&\sim& Z_{S^2}^{(\wt{\rm S})}(r_{\rm S}-\log 2^8,\theta_{\rm S})
=2^{-6}Z_{S^2}^{({\rm A})}(-r_{\rm S}-\log 2^{4},-\theta_{\rm S})\nn\\
&=&2^{-6}Z_{S^2}^{({\rm A})}(r_{\rm A},\theta_{\rm A})
\eeqa
where the relation (\ref{relre}) is used.
Assuming the conjecture of \cite{Jockers:2012dk},
this proves that the two moduli spaces are exactly the same as 
K\"ahler manifolds.

This is a surprise. 
The Hodge numbers of the corresponding Calabi-Yau threefolds
are different, $h^{2,1}=59$ for (A) and $h^{2,1}=23$ for (S). 
Therefore, the two families of superconformal field fixed points
cannot be the same,
as the number of exactly marginal chiral operators are different.
And yet they have exactly the same moduli space of twisted chiral
parameters. 
What we have observed is perhaps stronger than just the 
equivalence as K\"ahler manifold: assuming that the observation (\ref{obsJ})
of \cite{Jockers:2012dk} extends here, we have also read off
the Picard-Fuchs operators of the two models and they
also match. This would mean the equivalence of the 
$U(1)_V$ R-charge zero sector of the vacuum vector bundle over
the moduli space. 
It would be tempting to conjecture that
the category of B-branes are also the same.
But that cannot be the case
since $H^{0,*}(X,\wedge^*T_X)$ is an invariant of the derived category 
of a variety $X$ and
two Calabi-Yau threefolds of different $h^{2,1}$'s cannot have the same
$H^{0,*}(X,\wedge^*T_X)$.
At this moment, we do not know how to interpret this observation. 
Is this just a special case? and is there some particular reason
behind the equivalence?
Or is this a generic feature of (2,2) superconformal field theories?
We believe that it is worth doing further study in order to clarify the
situation.

\subsection{A practical use of duality}

Recall that we have encountered a difficulty in evaluating the integral
for the sphere partition function in the strongly coupled phase $r\ll 0$.
But the same phase is weakly coupled in the dual theory, and we expect
that the difficulty disappears. Indeed, in the present example, we have
just seen that the partition function of the dual theory
is identical (up to a parameter change and irrelevant overall multiplication)
to the one for the model (A$^2_{(-1)^6,(-2),1^4,0}$)
and the computation in its weakly coupled phase
$r_{\rm A}\gg 0$ (corresponding to $r=r_{\rm S}\ll 0$)
 has been done already in Section~\ref{subsec:X10}.
Inserting the parameter relation (\ref{relre}) into the result
(\ref{KPX10}), we find
\begin{equation}
e^{-K}\sim
\left[\log^3\frac{w\bar w}{2^{16}}-264\zeta(3)\right]+\ldots
\end{equation}
for $w=e^t=z^{-1}$, which yields
\beq
g_{w\bar w}={3\over w\bar w(\log(w\bar w/2^{16}))^2}+\cdots.
\eeq
The dual theory must also be useful in evaluating the integral
in the strongly coupled phase of other models,
such as the Pfaffian phase in R\o dland model (A$^2_{(-2)^7,1^7}$).

\section{Systems with linear $A(p)$ and $S(p)$}\label{sec:linear}

In this section, we study the systems
in which $A(p)$ and $S(p)$ are linear in $p$.
The charges of the $p^i$'s are all $-2$ and the charges
of the $x_j$'s are all $+1$. We shall call the systems
(A$^k_{M,N}$) and (S$^{k,\bullet}_{M,N}$) rather than
(A$^k_{(-2)^M,1^N}$) and (S$^{k,\bullet}_{(-2)^M,1^N}$). 
It turns out that the dual theory can be simplified
to take the same form as the original, 
but with different number of $p$ fields and of course with the dual
gauge group.
The dual pairs are:

(A$^k_{M,N}$) and (A$^{k^{\vee}}_{M^{\vee},N}$),~~ $k+k^{\vee}=N-1$,~
$M+M^{\vee}={N(N-1)\over 2}$,~~~[$N$ odd];

(S$^{k,+}_{M,N}$) and (S$^{k^{\vee},0}_{M^{\vee},N}$),~~  $k+k^{\vee}=N+1$,~
 $M+M^{\vee}={N(N+1)\over 2}$,~~~[$(N-k)$ odd];

(S$^{k,-}_{M,N}$) and (S$^{k^{\vee},-}_{M^{\vee},N}$),~~  $k+k^{\vee}=N+1$,~
 $M+M^{\vee}={N(N+1)\over 2}$,~~~[$N$ odd].

\subsection{Simplifying the dual theory}

\subsubsection{(A) series}

The basic data to specify the model (A$^k_{M,N}$)
is a generic $N\times N$ antisymmetric matrix $A(p)$ which is linear in $p$.
This can be regarded as an embedding
\beq
A:\C^{M}\,\,\hookrightarrow\,\, \wedge^2 V^*,
\eeq
where $V$ is a complex vector space of dimension $N$.
The dual theory can be simplified into
the model (A$^{k^{\vee}}_{M^{\vee},N}$) associated with the orthogonal embedding
\beq
A^{\vee}:\C^{M^{\vee}}\,\,\hookrightarrow\,\, \wedge^2 V.
\eeq
``Orthogonal'' means that the image of $A$ and the image of $A^{\vee}$ are
orthogonal complement of each other under the perfect pairing
$\langle\,\,,\,\rangle: \wedge^2V^*\times \wedge^2V\to \C$.

Before explaining how it works, let us write down
the dual pair of the original system with the data $A$ and
the dual pair of the system with the data $A^{\vee}$.
\beq
\begin{array}{ccc}
\mbox{gauge group}&\mbox{matter}&\mbox{superpotential}\\
\hline
\begin{array}{c}
\\[-0.5cm]
\displaystyle {U(1)\times USp(k)\over \{(\pm 1,\pm {\bf 1}_k)\}}
\\[-0.5cm]
\\
\end{array}&
p\in \C^M(-2),~x\in \C^k\otimes V&
\langle A(p),[xx]\rangle\\
\hline
\begin{array}{c}
\\[-0.5cm]
\displaystyle
{U(1)\times USp(k^{\vee})\over \{(\pm 1,\pm {\bf 1}_{k^{\vee}})\}}
\\[-0.5cm]
\\
\end{array}
&
p\in \C^M(-2),~ \wt{x}\in\C^{k^{\vee}}\otimes V^*,~a\in \wedge^2V&
\langle [\wt{x}\wt{x}]+A(p),a\rangle\\
\hline
\begin{array}{c}
\\[-0.5cm]
\displaystyle
{U(1)\times USp(k^{\vee})\over \{(\pm 1,\pm {\bf 1}_{k^{\vee}})\}}
\\[-0.5cm]
\\
\end{array}
&
p^{\vee}\in \C^{M^{\vee}}\!(2),
~x^{\vee}\in \C^{k^{\vee}}\!\!\otimes V^*&
\langle [x^{\vee}x^{\vee}],A^{\vee}(p^{\vee})\rangle\\
\hline
\begin{array}{c}
\\[-0.5cm]
\displaystyle
{U(1)\times USp(k)\over \{(\pm 1,\pm {\bf 1}_{k})\}}
\\[-0.5cm]
\\
\end{array}&
p^{\vee}\in \C^{M^{\vee}}\!(2),
~ \wt{x}^{\vee}\in\C^{k}\!\!\otimes V,
~a^{\vee}\in \wedge^2V^*&
\langle a^{\vee},[\wt{x}^{\vee}\wt{x}^{\vee}]+A^{\vee}(p^{\vee})\rangle\\
\hline
\end{array}
\label{quadA}
\eeq

\noindent
The number in the parenthesis shows the $U(1)$ charges, and we assigned
$U(1)$ charges $1$ to $V$. Note that we take the non-standard convention
in the dual pair for $A^{\vee}$ where we have assigned positive $U(1)$
charges to $p^{\vee}$ and negative $U(1)$ charges to $x^{\vee}$.

Let us now show how the theory with the data $A^{\vee}$ (the third theory)
arises from the dual theory (the second theory). Integrating out all the
$p$ fields in the second theory, we obtain a constraint that $a$ is
orthogonal to the image of
$A:\C^M\hookrightarrow \wedge^2 V^*$. Then, we can write it as
$a=A^{\vee}(p^{\vee})$ for some $p^{\vee}\in \C^{M^{\vee}}(2)$. This leaves
us with the third theory. Similarly,
the first theory is obtained from the fourth by integrating out 
all the $p^{\vee}$ fields.
The correspondence of the chiral variables is therefore
\beq
\begin{array}{ccccccc}
[xx]&\longleftrightarrow &a&\longleftrightarrow&
A^{\vee}(p^{\vee})&\longleftrightarrow &
-[\wt{x}^{\vee}\wt{x}^{\vee}],\\
-A(p)&\longleftrightarrow & [\wt{x}\wt{x}]
&\longleftrightarrow & [x^{\vee}x^{\vee}]
&\longleftrightarrow & a^{\vee}.
\end{array}
\eeq
Note that, if we keep the relation $[\wt{x}\wt{x}]=[x^{\vee}x^{\vee}]$
then we must have a sign in the relation $[xx]=-[\wt{x}^{\vee}\wt{x}^{\vee}]$.
The top-left and the bottom-right arrows are the relations in the 
quantum duality, whereas the top-right and the bottom left arrows
are the F-term equations.
These quantum relations and the classical relations
 are mapped to each other under the exchange of $A$ and $A^{\vee}$.

Recall that the $r\ll 0$ phase is, apart from the possible presence of
singularity, the non-linear sigma model on the Pfaffian
\beq
Y_A=\left\{\,p\in\CP^{M-1}\,\Bigr|\,\,{\rm rank}\,A(p)\leq
k^{\vee}\,\,\right\}.
\eeq
If we use the simplified dual description, 
we find that the $r\gg 0$ phase can also be regarded in the same way,
with the target space being the linearly dual Pfaffian
\beq
Y_{A^{\vee}}=\left\{\,p^{\vee}\in\CP^{M^{\vee}-1}\,
\Bigr|\,\,{\rm rank}\,A^{\vee}(p^{\vee})\leq k\,\,\right\}.
\eeq

This simplification can also be applied even if $A(p)$ is not
totally linear. In the dual theory,
we may integrate out the $p$'s which enter linearly in
$A(p)$. Then, $a$ can be written as $A^{\prime\vee}(p^{\prime\vee})$
for the orthogonal $A^{\prime\vee}$ to the linear part $A'$ of $A$,
and the dual can be written as the theory with the superpotential
$W=\langle A''(p'')+[\wt{x}\wt{x}],A^{\prime\vee}(p^{\prime\vee})\rangle$
where $A''(p'')$ is the non-linear part of $A(p)$.

\subsubsection{(S) series}

Let us next consider the (S) series.
The basic data to specify the model (S$^{k,+,0,-}_{M,N}$),
a generic $N\times N$ symmetric matrix $S(p)$ which is linear in $p$,
can be regarded as an embedding
\beq
S:\C^{M}\,\,\hookrightarrow\,\, \Sym^2 V^*,
\eeq
where $V$ is a complex vector space of dimension $N$.
The dual theory can be simplified into
the model (S$^{k^{\vee},0,+,-}_{M^{\vee},N}$)
associated with the orthogonal embedding
\beq
S^{\vee}:\C^{M^{\vee}}\,\,\hookrightarrow\,\, \Sym^2 V.
\eeq
``Orthogonal'' means that the image of $S$ and the image of $S^{\vee}$ are
orthogonal complement of each other under the perfect pairing
$\langle\,\,,\,\rangle: \Sym^2V^*\times \Sym^2V\to \C$.

Let us write down
the dual pair of the original system with data $S$
the the dual pair of the system with the data $S^{\vee}$
($N-k$ must be odd here).
\beq
\begin{array}{ccc}
\mbox{gauge group}&\mbox{matter}&\mbox{superpotential}\\
\hline
\begin{array}{c}
\\[-0.5cm]
\displaystyle {U(1)\times O_+(k)\over \{(\pm 1,\pm {\bf 1}_k)\}}
\\[-0.5cm]
\\
\end{array}&
p\in \C^M(-2),~x\in \C^k\otimes V&
\langle S(p),(xx)\rangle\\
\hline
\begin{array}{c}
\\[-0.5cm]
\displaystyle
{U(1)\times SO(k^{\vee})\over \{(\pm 1,\pm {\bf 1}_{k^{\vee}})\}}
\\[-0.5cm]
\\
\end{array}
&
p\in \C^M(-2),~ \wt{x}\in\C^{k^{\vee}}\otimes V^*,~s\in \Sym^2V&
\langle (\wt{x}\wt{x})+S(p),s\rangle\\
\hline
\begin{array}{c}
\\[-0.5cm]
\displaystyle
{U(1)\times SO(k^{\vee})\over \{(\pm 1,\pm {\bf 1}_{k^{\vee}})\}}
\\[-0.5cm]
\\
\end{array}
&
p^{\vee}\in \C^{M^{\vee}}\!(2),
~x^{\vee}\in \C^{k^{\vee}}\!\!\otimes V^*&
\langle (x^{\vee}x^{\vee}),S^{\vee}(p^{\vee})\rangle\\
\hline
\begin{array}{c}
\\[-0.5cm]
\displaystyle
{U(1)\times O_+(k)\over \{(\pm 1,\pm {\bf 1}_{k})\}}
\\[-0.5cm]
\\
\end{array}&
p^{\vee}\in \C^{M^{\vee}}\!(2),
~ \wt{x}^{\vee}\in\C^{k}\!\!\otimes V,
~s^{\vee}\in \Sym^2V^*&
\langle a^{\vee},(\wt{x}^{\vee}\wt{x}^{\vee})+S^{\vee}(p^{\vee})\rangle\\
\hline
\end{array}
\label{quadS}
\eeq

\noindent
There is also a version where all the orthogonal groups are
$O_-$ ($N$ must be odd for that).
We have assigned $U(1)$ charges $1$ to $V$.

The third theory is obtained from the second by simply integrating out
the $p$ fields, while the first is obtained from the fourth by integrating
out the $p^{\vee}$ fields.
The correspondence of the chiral variables is
\beq
\begin{array}{ccccccc}
(xx)&\longleftrightarrow &s&\longleftrightarrow&
S^{\vee}(p^{\vee})&\longleftrightarrow &
-(\wt{x}^{\vee}\wt{x}^{\vee}),\\
-S(p)&\longleftrightarrow & (\wt{x}\wt{x})
&\longleftrightarrow & (x^{\vee}x^{\vee})
&\longleftrightarrow & s^{\vee}.
\end{array}
\eeq
The top-left and the bottom-right arrows are the relations in the 
quantum duality, whereas the top-right and the bottom left arrows
are the F-term equations.
These quantum relations and the classical relations
 are mapped to each other under the exchange of $S$ and $S^{\vee}$.

Recall that the $r\ll 0$ phase is, apart from the possible presence of a
singularity, the non-linear sigma model on 
a double cover $\wt{Y}_S$ of
the symmetric determinantal variety
\beq
Y_S=\left\{\,p\in\CP^{M-1}\,\Bigr|\,\,{\rm rank}\,S(p)\leq
k^{\vee}\,\,\right\}.
\eeq
If we use the simplified dual description, 
we find that the $r\gg 0$ phase can also be regarded in the similar way,
with the target space being the linearly dual 
symmetric determinantal variety
\beq
Y_{S^{\vee}}=\left\{\,p^{\vee}\in\CP^{M^{\vee}-1}\,
\Bigr|\,\,{\rm rank}\,S^{\vee}(p^{\vee})\leq k\,\,\right\}.
\eeq
In the $O_-$ version, we have $Y_S$ itself in the $r\ll 0$ phase.

This simplification can also be applied even if $S(p)$ is not
totally linear. In the dual theory,
we may integrate out the $p$'s which enter linearly in
$S(p)$. Then, $s$ can be written as $S^{\prime\vee}(p^{\prime\vee})$
for the orthogonal $S^{\prime\vee}$ to the linear part $S'$ of $S$,
and the dual can be written as the theory with the superpotential
$W=\langle S''(p'')+(\wt{x}\wt{x}),S^{\prime\vee}(p^{\prime\vee})\rangle$
where $S''(p'')$ is the non-linear part of $S(p)$.
In fact, we have already encountered such a simplification
in the computation of the two sphere partition function.
See (\ref{secnds}) and (\ref{secnds2}).

\subsection{Coulomb branch analysis}

Let us study the effective theory on the Coulomb branch, in order to identify
the location of the singularity (Calabi-Yau case)
and the massive vacua (non-Calabi-Yau case).
We shall see that the dual pair yield the same results,
under a certain relationship among the FI-theta parameters.
In what follows, by ``the dual theory'' we mean the simplified dual theory.
We denote the FI-theta parameters of the original theory and the (simplified)
dual theory by $t$ and $t^{\vee}$.
For completeness,  we shall also briefly mention the relation to
the FI-theta parameters of the dual theories before the simplification,
denoted by $\wt{t}$ and $\wt{t}^{\vee}$.

\subsubsection{(A) series}

The effective twisted superpotential (\ref{Weff})
of the model (A$^k_{M,N}$) is
\beqa
\wt{W}_{\it eff}&=&-2t\sigmaU
-{l(l+1)\over 2}\pi i\sigma_0'
-M(-2\sigmaU)(\log(-2\sigmaU)-1)
\nn\\
&&
-N\sum_{a=1}^l(\sigmaU+\sigma_a)
(\log(\sigmaU+\sigma_a)-1)
-N\sum_{a=1}^l(\sigmaU-\sigma_a)
(\log(\sigmaU-\sigma_a)-1)\nn\\
&=&-t\sigma'_0
-{l(l+1)\over 2}\pi i\sigma_0'
-M\sigma'_0(\log(-\sigma'_0)-1)
\nn\\
&&
-N\sum_{a=1}^l\sigma'_a(\log\sigma'_a-1)
-N\sum_{a=1}^l(\sigma'_0-\sigma'_a)
(\log(\sigma'_0-\sigma'_a)-1).
\eeqa
The vacuum equation yields
\beq
(-1)^{{l(l+1)\over 2}+M}\e^{-t}=
{\prod_{a=1}^l(\sigma_0'-\sigma_a')\over(\sigma_0')^M},\qquad
(\sigma_0'-\sigma_a')^N=(\sigma_a')^N\,\,\,(a=1,\ldots, l).
\eeq
The latter equation is solved by
$\sigma'_0/\sigma_a'=1+\omega_a$ where $\omega_a^N=1$
and we find
\beq
(-1)^{{l(l+1)\over 2}+M}\e^{-t}=(\sigma_0')^{Nl-M}
\prod_{a=1}^l(1+\omega_a)^{-N}.
\label{singA}
\eeq
Note that the region with $\sigma_a=-\sigma_b$ (any $a,b$) and
$\sigma_a=\sigma_b$ ($a\ne b$) is excluded and that
the Weyl group acts as permutations and sign flips of the $\sigma_a$'s.
This corresponds to excluding solutions with $\omega_a=\omega_b^{-1}$
  (any $a,b$) and $\omega_a=\omega_b$ ($a\ne b$), and modding out the solutions
by permutations and inversions of of $\omega_a$'s.
Then, there are
\beq
n(k,N)={{N-1\over 2}\choose {k\over 2}}
\label{nkNA}
\eeq
inequivalent choices of $\{\omega_a\}$. In the Calabi-Yau case,
$Nl=M$, this is the number of singular points in the FI-theta
parameter space and (\ref{singA}) shows the location.
In the non-Calabi-Yau case, for each such $\{\omega_a\}$,
the equation (\ref{singA}) has $|Nl-M|$ solutions for $\sigma_0'$.
Thus, the number of massive vacua is 
\beq
{1\over 2}\,|Nk-2M|\, n(k,N).
\eeq
The value of the twisted superpotential at such a vacuum is
\beq
\wt{W}_{\it eff}={1\over 2}(Nk-2M)\sigma_0'.
\eeq

Computation in the dual theory is similar.
The vacuum equation yields
\beq
(-1)^{{l^{\vee}(l^{\vee}+1)\over 2}+M^{\vee}}\e^{t^{\vee}}
=(-\sigma_0^{\vee\prime})^{Nl^{\vee}-M^{\vee}}
\prod_{a=1}^{l^{\vee}}(1+\omega_a)^{-N},
\label{singAdual}
\eeq
and the value of the twisted superpotential at each solution is
\beq
\wt{W}_{\it eff}=-{1\over 2}(Nk^{\vee}-2M^{\vee})\sigma_0^{\vee\prime}.
\eeq
We need to choose $l^{\vee}={k^{\vee}\over 2}$ distinct roots among the
${N-1\over 2}$ possibilities.
Recalling that
$l+l^{\vee}={N-1\over 2}$, we find that a one to one correspondence
between the solutions is given by complementary choices of the roots.
Noting that $Nk-2M=-(Nk^{\vee}-2M^{\vee})$,
we find a one to one correspondence between the Coulomb branch
vacua, with the same values of the twisted superpotential (non Calabi-Yau case)
and the same location of the singular points (Calabi-Yau case),
up to an overall normalization.
The matching is perfect if the FI-theta parameters
of the dual pair are related by:
\beq
\e^{-t}
=\e^{-t^{\vee}}
(-1)^{{l(l+1)\over 2}+{l^{\vee}(l^{\vee}+1)\over 2}+M^{\vee}+Nl}
\prod_{a=1}^{N-1\over 2}(1+\e^{2\pi i a\over N})^{-N}.
\eeq
This is the duality map between the parameters.
In the non-Calabi-Yau case, this is to be regarded as the relation
at the same scale. The FI-theta parameters after and before the simplification
are related by $\e^{-t}=(-1)^{M^{\vee}}\e^{\wt{t}^{\vee}}$
and $\e^{-t^{\vee}}=(-1)^M\e^{-\wt{t}}$.

Note that the number $n(k,N)$ in (\ref{nkNA})
agrees with the Witten index for the $USp(k)$ gauge theory with
$N$ fundamentals with generic twisted masses \cite{Hori:2011pd}.
This is not a coincidence: the problem of 
solving the above equations for $\sigma_a$'s
with fixed $\sigmaU$ is exactly the same as the
problem of finding the vacua in that theory with twisted masses
$\wt{m}_i=-\sigmaU$.
This observation will be useful when we discuss the more complicated (S)
series.

\subsubsection{(S) series}

We only show the results and sketch the outline,
since it is lengthy to describe all the detail of the
analysis that depends on the cases.
We denote by $n(k,N)$ the Witten index of the $H$ theory with $N$ fundamentals
with a generic twisted masses, where $H$ is the orthogonal group factor
in the gauge group $G=(U(1)\times H)/\{(\pm 1, \pm {\bf 1}_k)\}$
of the theory (S$^{k,\pm}_{M,N}$).
Concretely, it is given in the following table: (It is
an extract from \cite{Hori:2011pd}. Not all is needed here,
because of the constraint discussed in Section~\ref{sec:AandS}.)

{\footnotesize
\beq
\begin{array}{c|c|c|c}
\mbox{group}&k&N&n(k,N)\\
\hline
\Ost(k)&\mbox{even}&\mbox{even}&
{\displaystyle {{N\over 2}\choose {k\over 2}}}\\[0.4cm]
\Ons(k)&\mbox{even}&\mbox{even}&
{\displaystyle {{N\over 2}\choose {k\over 2}}}\\[0.4cm]
SO(k)&\mbox{even}&\mbox{even}&
{\displaystyle 2{{N\over 2}\choose {k\over 2}}}
\\[0.4cm]
\hline
\Ost(k)&\mbox{even}&\mbox{odd}&
{\displaystyle {{N-1\over 2}\choose {k\over 2}}
+2{{N-1\over 2}\choose {k\over 2}-1}}\\[0.4cm]
\Ons(k)&\mbox{even}&\mbox{odd}&
{\displaystyle {{N+1\over 2}\choose {k\over 2}}}\\[0.4cm]
SO(k)&\mbox{even}&\mbox{odd}&
{\displaystyle 2{{N-1\over 2}\choose {k\over 2}}
+{{N-1\over 2}\choose {k\over 2}-1}}
\end{array}
\quad
\begin{array}{c|c|c|c}
\mbox{group}&k&N&n(k,N)\\
\hline
\Ost(k)&\mbox{odd}&\mbox{even}&
{\displaystyle 2{{N\over 2}\choose {k-1\over 2}}}\\[0.4cm]
\Ons(k)&\mbox{odd}&\mbox{even}&
{\displaystyle {{N\over 2}\choose {k-1\over 2}}}\\[0.4cm]
SO(k)&\mbox{odd}&\mbox{even}&
{\displaystyle {{N\over 2}\choose {k-1\over 2}}}\\[0.4cm]
\hline
\Ost(k)&\mbox{odd}&\mbox{odd}&
{\displaystyle {{N-1\over 2}\choose {k-1\over 2}}}\\[0.4cm]
\Ons(k)&\mbox{odd}&\mbox{odd}&
{\displaystyle 2{{N-1\over 2}\choose {k-1\over 2}}}\\[0.4cm]
SO(k)&\mbox{odd}&\mbox{odd}&
{\displaystyle {{N-1\over 2}\choose {k-1\over 2}}}
\end{array}
\label{WItable}
\eeq
}

\noindent
First, we describe an overview of the results. 
If the theory is not Calabi-Yau, $kN\ne 2M$, the
number of massive vacua is 
\beq
{1\over 2}\,|kN-2M|\,n(k,N).
\label{mvS}
\eeq
If the theory is Calabi-Yau, $kN=2M$, the number of singular points
in the FI-theta parameter space, including the multiplicity, is
\beq
n(k,N),
\label{NsingS}
\eeq
except in the series (S$^{k,+}_{M,N}$) with $k$ odd
where the number is one half of this.

If the group $H$ is not $O_{\pm}(k)$
with $k$ odd, the reason we get these numbers is the same
as in the (A) series. We find $n(k,N)$ vacua from the $H$ sector.
And we find ${1\over 2}|kN-2M|$ vacua from the $U(1)$ sector in
non-Calabi-Yau case. If $H=O_{\pm}(k)$ with $k$ odd, the gauge group
$G$ is isomorphic to $U(1)\times SO(k)$ and we need to use the $SO(k)$ results.
The Witten index of the $SO(k)$ theory is one half of $n(k,N)$
in these cases (i.e. $O_+(k)$ with $k$ odd and $N$ even as well as
$O_-(k)$ with $k$ and $N$ both odd).
But the number of vacua from the $U(1)$ sector is
twice as much, $|kN-2M|$, getting (\ref{mvS}).
In the Calabi-Yau case with $H=O_+(k)$ with $k$ odd and $N$ even, the
number of singular points is one half of (\ref{NsingS}), 
while it is (\ref{NsingS}) itself in the dual, which has
$H=SO(k^{\vee})$ with $k^{\vee}$ even.
Let us describe the results in more detail.

\noindent
\underline{(S$^{k,\pm}_{M,N}$) with $k$ even ($N$ odd)}:
The vacuum equation yields
\beq
(-1)^{{l(l-1)\over 2}+M}\e^{-t}=(\sigma_0')^{Nl-M}
\prod_{a=1}^l(1+\omega_a)^{-N},
\label{singS1}
\eeq
where $\omega_a^N=1$, $\omega_a\ne\omega_b^{\pm 1}$ for $a\ne b$;
$\omega_a$'s related by permutations and inversions are regarded as the same.
Those involving the fixed point $\omega_a=1$ of the inversion
supports two vacua ({\it resp}. one vacuum) in the theory
(S$^{k,+}_{M,N}$) ({\it resp}.  (S$^{k,-}_{M,N}$)).
The count is
${{N-1\over 2}\choose l}+2{{N-1\over 2}\choose l-1}=n(k,N)$
({\it resp}. ${{N-1\over 2}\choose l}+
{{N-1\over 2}\choose l-1}=n(k,N)$).
The value of the twisted superpotential at each vacuum is
\beq
\wt{W}_{\it eff}={1\over 2}(Nk-2M)\sigma_0'.
\eeq
Computations in the dual theory yield exactly the same
results, provided the parameters are related by the duality map
\beq
\e^{-t}
=\e^{-t^{\vee}}
(-1)^{{l(l-1)\over 2}+{l^{\vee}(l^{\vee}-1)\over 2}+M^{\vee}+Nl}
\prod_{a=0}^{N-1\over 2}(1+\e^{2\pi i a\over N})^{-N}.
\eeq
Note also the relations $\e^{-t}=(-1)^{M^{\vee}}\e^{\wt{t}^{\vee}}$
and $\e^{-t^{\vee}}=(-1)^M\e^{-\wt{t}}$.

\noindent
\underline{(S$^{k,+}_{M,N}$) with $k$ odd ($N$ even)}:
The theory has gauge group $G\cong U(1)\times SO(k)$ and the vacuum
equation is
\beqa
\e^{-t}&=&{\sigmaU^N
\prod_{a=1}^l(\sigmaU+\sigma_a)^N
(\sigmaU-\sigma_a)^N
\over (-2\sigmaU)^{2M}},\nn\\
&&\!\!\!\!\!\!\!\!\!\!\!\!\!\!\!\!\!\!
(\sigmaU+\sigma_a)^N
=-(\sigmaU-\sigma_a)^N~~(a=1,\ldots, l).
\eeqa
We require $\sigma_a\ne\pm\sigma_b$ for $a\ne b$ and identify solutions related
by permutations and arbitrary sign flips of the $\sigma_a$'s. The count is
${{N\over 2}\choose l}={1\over 2}n(k,N)$.
The value of the twisted superpotential at each vacuum is
\beq
\wt{W}_{\it eff}=(Nk-2M)\sigmaU.
\eeq
On the other hand, the gauge group of the dual theory is
$G^{\vee}=(U(1)\times SO(k^{\vee}))/\{(\pm 1,\pm{\bf 1}_{k^{\vee}})\}$
and the vacuum equation is
\beqa
(-1)^{l^{\vee}(l^{\vee}+1)\over 2}\e^{-t^{\vee}}&=&\left[
{\prod_{a=1}^{l^{\vee}}(\sigma_0^{\vee}-\sigma^{\vee}_a)^N
\over (2\sigma_0^{\vee})^{M^{\vee}}}\right]^{-1},\nn\\
&&\!\!\!\!\!\!\!\!\!\!\!\!\!\!\!\!\!\!\!\!\!\!\!\!\!\!\!\!\!\!\!\!\!\!\!\!
(\sigma_0^{\vee}+\sigma^{\vee}_a)^N
=-(\sigma_0^{\vee}-\sigma^{\vee}_a)^N~~(a=1,\ldots, l^{\vee}).
\eeqa
We require $\sigma^{\vee}_a\ne\pm\sigma^{\vee}_b$ for
$a\ne b$ and identify solutions related
by permutations and an even number of sign flips of the $\sigma^{\vee}_a$'s.
The count is
$2{{N\over 2}\choose l^{\vee}}=n(k,N)$.
The value of the twisted superpotential at each vacuum is
\beq
\wt{W}_{\it eff}=-(Nk^{\vee}-2M^{\vee})\sigmaU^{\vee}.
\eeq
We see that the two theories yield the same results
concerning the massive vacua in non-Calabi-Yau cases but different results
concerning the singular points in the Calabi-Yau case.
In fact, there is a good reason for the mismatch.
An odd number of sign flips of $\sigma^{\vee}_a$'s corresponds to
the $\Z_2$ symmetry associated with $O(k^{\vee})/SO(k^{\vee})\cong\Z_2$,
and, as discussed in Section~\ref{subsec:theta}, that symmetry
induces a shift of the theta angle by $\pi$. 
In particular, the theories with $\e^{-t^{\vee}}$ and $-\e^{-t^{\vee}}$
are physically equivalent.
Indeed, there is a $2:1$ map from $\e^{-t^{\vee}}$ to $\e^{-t}$, 
\beq
\e^{-t}
=\e^{-2t^{\vee}}(-1)^{l^{\vee}}2^{-N(N+1)}\prod_{a=1}^N
\left(1+\omega_a\right)^N,
\eeq
where $\omega_a$ are the $N$ solutions to $(1+\omega)^N=-(1-\omega)^N$.
The FI-theta parameters after and before the simplification
are related by $\e^{-t}=\e^{\wt{t}^{\vee}}$
and $\e^{-t^{\vee}}=(-1)^M\e^{-\wt{t}}$.

\noindent
\underline{(S$^{k,-}_{M,N}$) with $k$ odd ($N$ odd)}:
The gauge group is
$G\cong U(1)\times SO(k)$ and the vacuum equation is
\beqa
\e^{-t}&=&{\sigmaU^N
\prod_{a=1}^l(\sigmaU+\sigma_a)^N
(\sigmaU-\sigma_a)^N
\over (-2\sigmaU)^{2M}},\nn\\
&&\!\!\!\!\!\!\!\!\!\!\!\!\!\!\!\!\!\!
(\sigmaU+\sigma_a)^N
=(\sigmaU-\sigma_a)^N~~(a=1,\ldots, l).
\eeqa
We require $\sigma_a\ne\pm\sigma_b$ for $a\ne b$ and $\sigma_a\ne 0$
for any $a$,
 and identify solutions related
by permutations and arbitrary sign flips of $\sigma_a$'s. The count is
${{N-1\over 2}\choose l}={1\over 2}n(k,N)$.
The count for the $U(1)$ part is $|kN-2M|$ yielding
(\ref{mvS}).
The value of the twisted superpotential at each vacuum is
\beq
\wt{W}_{\it eff}=(Nk-2M)\sigmaU.
\eeq
The dual theory yield exactly the same results, provided
the parameters at the same scale are related by
\beq
\e^{-t}=\e^{-t^{\vee}}2^{-N(N+1)}
\prod_{a=1}^N
\left(1+\omega_a\right)^{2N}.
\eeq
where $\omega_a$ are the solutions to $(1+\omega)^N=(1-\omega)^N$.
We also have $\e^{-t}=\e^{\wt{t}^{\vee}}$
and $\e^{-t^{\vee}}=\e^{-\wt{t}}$.
There is no Calabi-Yau case, since $kN-2M$ can never vanish if
$k$ and $N$ are both odd.

\subsection{Category of B-branes}

The category of B-type D-branes (B-branes for short)
is believed to be invariant under renormalization group flow as well as
under deformation of the twisted chiral parameters of
the theory. 
If the resulting target variety $Y$ is smooth, the category of B-branes in the
series (A) or (S$^+$) is 
the same as the derived category $D^b(Y)$ 
of coherent sheaves on $Y$.
Even if they are singular, it is possible that the quantum gauge theory
itself is perfectly fine. Although we do not have a proof,
the absence of a Coulomb branch may be regarded as one indication for that.
If the theory is indeed fine, there must be a good category of B-branes.
Such a good category has been constructed in some cases
under the name of ``non-commutative resolution''.
Here we assume that the theory is fine and
denote the good category of B-branes simply by $D(Y)$.
We also denote the category for the (S$^-$) series by  $D_{(-1)^{F_s}}(Y)$.

The above results then yield the following
prediction concerning equivalences of the categories.
\beq
\begin{array}{rcll}
D_-&\cong&D_+,&\mbox{if $kN=2M$,}\\
D_-+\cdots
&\cong& D_+,&\mbox{if $kN>2M$},\\
D_-&\cong&D_++\cdots,
&\mbox{if $kN<2M$},
\end{array}
\eeq
with
\beq
\begin{array}{c||cc}
&D_-&D_+\\
\hline
\begin{array}{c}
\\[-0.5cm]
\mbox{(A)}
\\[-0.5cm]
\\
\end{array}&D(Y_A)&D(Y_{A^{\vee}})\\
\begin{array}{c}
\\[-0.57cm]
\mbox{(S$^+$)}
\\[-0.5cm]
\\
\end{array}
&D(\wt{Y}_S)&D(Y_{S^{\vee}})\\
\begin{array}{c}
\\[-0.57cm]
\mbox{(S$^-$)}
\\[-0.5cm]
\\
\end{array}
&D_{(-1)^{F_s}}(Y_S)&D_{(-1)^{F_s}}(Y_{S^{\vee}})
\end{array}
\eeq
$+\cdots$ is the
collection of ${1\over 2}\,|kN-2M|\,n(k,N)$ objects
associated to the massive vacua on the Coulomb branch.

Such equivalences of categories seem to fit with the framework
of ``homological projective duality'' by A. Kuznetsov \cite{KuznetsovHPD}.
Interestingly, the latter seems to have something to say about the (A)
series with even $N$.
It would be interesting to understand the reason of the similarity
between the quantum duality in gauge theory
and the homological projective duality in the study of derived categories.

We should also note that we expect equivalences of categories
$D_-\cong D_+$ in the first three systems of Section~\ref{sec:(A)}
and in the system of Section \ref{sec:(S)}. $D_-$ is
the derived category of the smooth compact Calabi-Yau threefolds
(the Pfaffian or the double cover of the symmetric determinantal variety).
$D_+$ is the category of B-branes in the true hybrid model.
It would be interesting to provide a useful description of $D_+$,
such as modules over sheaves of algebras on the base Fano threefold,
as in \cite{Caldararu:2007tc}.

Obviously, it is an interesting problem to determine and apply
the grade restriction rule \cite{Herbst:2008jq}
in the above situations, in order to physically
find the equivalences of categories as well as monodromy along
the closed loops in the FI-theta parameter space.
We would like to mention that there is a recent mathematical
progress in the classical grade restriction rule
\cite{Segal,DHL,BFK,DoSe}.
More recently a way to determine the quantum grade restriction rule
in non-Abelian gauged linear sigma model
was suggested in \cite{Hori:2013ika}.

\subsection{Smooth Calabi-Yau examples}

Let us see when we have a smooth Calabi-Yau manifold
as either or both of the Pfaffians $Y_A$ and $Y_{A^{\vee}}$,
or of the (double cover of) symmetric determinants,
$\wt{Y}_S$ and $Y_{S^{\vee}}$.
Calabi-Yau condition reads
\beq
kN-2M=0,
\eeq
which can also be stated as $k^{\vee}N-2M^{\vee}=0$.
The dimension of the variety $Y_A$ ({\it resp}. $Y_S$) is
\beq
d=M-1-{k(k+1)\over 2}~\left({\it resp}.~M-1-{k(k-1)\over 2}\right)
={kk^{\vee}\over 2}-1.
\eeq

$Y_A$ is smooth when the corank $k+3$ degeneration of $A(p)$
is absent, which is the case when the dimension $M-1$ of the ambient space
is smaller than ${(k+3)(k+2)\over 2}$. That is, $N\leq k+5+{6\over k}$.
Recalling also $N> k+1$ is needed for $d\geq 0$, we find that
$Y_A$ is smooth when
\beqa
k=2:&&N=5,\,7,\, 9\,\,~(d=1,\, 3,\, 5),\nn\\
k\geq 4:&&N=k+3,~k+5~(d=k-1,~2k-1).\nn
\eeqa
The smoothness condition for $Y_{A^{\vee}}$ is obtained by the replacement
$k\to k^{\vee}$. Both are smooth in the cases below:
\beq
\begin{array}{c|c|c|c}
&k^{\vee}=2&k^{\vee}=4&k^{\vee}=6\\
\hline
k=2&
\begin{array}{c}
\mbox{\footnotesize $N=5,\,\,M=M^{\vee}=5$}\\
d=1
\end{array}
&
\begin{array}{c}
\mbox{\footnotesize $N=7,\,\,M=7,\,M^{\vee}=14$}\\
d=3
\end{array}
&
\begin{array}{c}
\mbox{\footnotesize $N=9,\,\,M=9,\,M^{\vee}=27$}\\
d=5
\end{array}\\
\hline
k=4&-&
\begin{array}{c}
\mbox{\footnotesize $N=9,\,\,M=M^{\vee}=18$}\\
d=7
\end{array}&\mbox{$Y_A$ not smooth}\\
\hline
k=6&-&\mbox{$Y_{A^{\vee}}$ not smooth}&\mbox{both not smooth}
\end{array}
\eeq
We do not write the lower triangular part by the redundancy from
$Y_A\leftrightarrow Y_{A^{\vee}}$.
The threefold pair is the pair in R\o dland's work.
The pair of elliptic curves is in the same family. In fact, from
the equivalence $D^b(Y_A)\cong D^b(Y_{A^{\vee}})$ and from
the fact that they are elliptic curves, they
must be biholomorphic, $Y_A\cong Y_{A^{\vee}}$.
In the fivefold case, $Y_{A^{\vee}}$ can also be regarded as
the complete intersection of nine hypersurfaces in the Grassmannian $G(2,9)$. 
($k=2$ case is always like that.)
The ninefold pair are in the same family. We do not know if they are
biholomorphic to each other. At this moment,
the derived equivalences $D^b(Y_A)\cong D^b(Y_{A^{\vee}})$ 
for these fivefold and ninefold pairs
seem to be an interesting prediction.

$\wt{Y}_S$ is smooth when the corank $(k+1)$ degeneration is absent, 
which is the case when $M-1$ is smaller than ${(k+1)(k+2)\over 2}$.
That is, $N\leq k+3+{2\over k}$. Recalling also $N\geq k$ for $d\geq 0$
and that $N-k$ is odd, we find that $\wt{Y}_S$ is smooth when
\beqa
k=1:&&N=2,\,4,\,6\,\,~(d=0,\,1,\,2),\nn\\
k\geq 2:&&N=k+1,~k+3~(d=k-1,~2k-1).\nn
\eeqa
$Y_{S^{\vee}}$ is smooth when the corank $k^{\vee}$ degeneration is absent,
i.e., $M^{\vee}-1\leq {k^{\vee}(k^{\vee}+1)\over 2}-1$.
Recalling also that $N\geq k^{\vee}$, we find that $Y_{S^{\vee}}$
when
\beq
N=k^{\vee},~k^{\vee}+1,\quad \mbox{i.e.}\,\,\,k=1,\,\,2.
\eeq
We should also bear in mind that $k^{\vee}$ must be even for regularity.
Both are smooth and regularity is satisfied in the cases below:
\beq
\begin{array}{c|c|c|c}
&k^{\vee}=2&k^{\vee}=4&k^{\vee}=6\\
\hline
k=1&
\begin{array}{c}
\mbox{\footnotesize $N=2,\,\,M=1,\,M^{\vee}=2$}\\
d=0
\end{array}&
\begin{array}{c}
\mbox{\footnotesize $N=4,\,\,M=2,\,M^{\vee}=8$}\\
d=1
\end{array}&
\begin{array}{c}
\mbox{\footnotesize $N=6,\,\,M=3,\,M^{\vee}=18$}\\
d=2
\end{array}\\
\hline
k=2&
\begin{array}{c}
\mbox{\footnotesize $N=3,\,\,M=3,\,M^{\vee}=3$}\\
d=1
\end{array}&
\begin{array}{c}
\mbox{\footnotesize $N=5,\,\,M=5,\,M^{\vee}=10$}\\
d=3
\end{array}&
\mbox{$\wt{Y}_S$ not smooth}
\end{array}
\eeq
The two cases with $d=1$ (elliptic curve) must be such that
$\wt{Y}_S$ and $Y_{S^{\vee}}$ are biholomorphic to each other.
The threefold pair are the pair in Hosono-Takagi's work.

Finally, among the model involving $O_-(k)$ gauge group, there is only one
example where both $Y_S$ and $Y_{S^{\vee}}$ are smooth. This is the case
where $k=k^{\vee}=2$ and the varieties the the elliptic curves.

\subsection{The partition function and the fundamental period in
$d=1$ examples}

Let us comment on the results of the computation of the
two sphere partition function and the fundamental periods 
in two of the examples in which the elliptic curve appears
in both phases.

\subsubsection{(A$^2_{5,5}$)}

This model is the one-dimensional version of the R{\o}dland model 
(A$^2_{7,7}$).
As discussed above, both phases correspond to elliptic curves and
they must be biholomorphic to each other.

The calculation of the sphere partition function is the same as for
the R{\o}dland example. Therefore we can use the results of
\cite{Jockers:2012dk}. The partition function in the $r\gg0$-phase is
\begin{eqnarray}
\label{su2torus}
Z^{r\gg0}_{S^2}&=&\frac{(z\bar{z})^{2q}}{2}\oint
\frac{d^2\varepsilon}{(2\pi i)^2}
\frac{\pi^5\sin^5\pi(\varepsilon_1+\varepsilon_2)}
{\sin^5\pi\varepsilon_1\sin^5\pi\varepsilon_2}
(z\bar{z})^{\varepsilon_1+\varepsilon_2}\times\nonumber\\
&&\times \left.\vline\sum_{K=0}^{\infty}(-z)^K\sum_{k=0}^K
(2k-K+\varepsilon_1-\varepsilon_2)
\frac{\Gamma(1+K+\varepsilon_1+\varepsilon_2)^5}
{\Gamma(1+k+\varepsilon_1)^5\Gamma(1+K-k+\varepsilon_2)^5} \vline\right.^2.
\end{eqnarray}
The residue integrals can be evaluated and the the fundamental
period can be extracted from the $\log z\bar{z}$-term of the result. We get
\begin{eqnarray}
X^0_{r\gg0}&=&1+3 z+19 z^2+147 z^3+1251 z^4+11253 z^5+\ldots.\nonumber\\
&=&\sum_{n=0}^{\infty}\sum_{\alpha=0}^n
\left(\begin{array}{c}n\\\alpha\end{array}\right)^2
\left(\begin{array}{c}n+\alpha\\\alpha\end{array}\right) z^n.
\end{eqnarray}
Note that the expansion coefficients are the  Ap\'ery numbers.
The fundamental period is annihilated by the Picard-Fuchs operator
\begin{equation}
\mathcal{L}=\theta^2-z(11\theta^2+11\theta+3)-z^2(\theta+1)^2
\end{equation}
which also appears in the database \cite{vanstraten}. The second
solution of the Picard-Fuchs equation is
\begin{equation}
X^1_{r\gg0}=\log z+(3 \log z +5) z +\left(19 \log
z+\frac{75}{2}\right)z ^2 +\left(147 \log z+\frac{1855}{6}\right) z
^3+\ldots
\end{equation}
We can compute the complexified K\"ahler modulus of the elliptic curve,
\beq
\rho={B\over 2\pi}+i{\rm Area}
\eeq
via the ``mirror map'' $2\pi i \rho=\frac{X^1}{X^0}$. 
With the definition $q=e^{2\pi i \rho}$ we obtain
\begin{equation}
q( z)= z +5  z ^2+35  z ^3+280  z ^4+2410  z ^5+\dots
\end{equation}
The inverse of this series is
\begin{equation}
q-5 q^2+15 q^3-30 q^4+40 q^5+\ldots
\end{equation}
Actually this can be given in a closed form (see for instance \cite{avz}):
\begin{equation}
q\prod_{n=1}^{\infty}(1-q^n)^{5\left(\frac{5}{n}\right)},
\end{equation}
where $\left(\frac{m}{n}\right)$ denotes the Jacobi symbol. This is a
modular function with respect to the congruence subgroup $\Gamma_1(5)$
of $SL(2,\mathbb{Z})$.

Let us compare the result with the computation of the two sphere
partition function for the non-linear sigma model whose target
space is the elliptic curve.
Note that there is no instanton effect as there is no topologically
non-trivial map from the two sphere to the elliptic curve.
Note also that the contribution from the non-zero modes is insensitive to
the moduli. Therefore, the non-trivial part is just the zero mode integral.
If we take the
usual path-integral measure induced from the (K\"ahler) metric of
the elliptic curve, the result is the
 area\footnote{We thank Hirosi Ooguri and Yuji Tachikawa for
discussions including this result.}
\beq
Z_{S^2}^{\rm nlsm}=\mbox{Area}={\rm Im}\,\rho.
\label{Znlsm}
\eeq
Note that this is indeed $\e^{-K}$ for the the K\"ahler moduli space which
is known to have the metric $\dd s^2=|\dd \rho|^2/({\rm Im}\,\rho)^2$.
We would like to see if the partition function (\ref{su2torus})
of the gauged linear sigma model reproduces this result.
Let us perform the following K\"ahler transform,
\begin{equation}
\e^{-K'}=\frac{1}{40\pi}
\frac{Z_{S^2}^{r\gg0}}{X^0_{r\gg0}(z)\bar{X}^0_{r\gg0}(\bar{z})}.
\end{equation}
 After expanding in $z$ and $\bar{z}$ we obtain
\begin{equation}
\e^{-K'}=\frac{1}{2i}\frac{1}{2\pi i}
\left[\left(\log z+5z+\frac{45}{2}z^2+\frac{440}{3}z^3+\ldots\right)
+c.c.\right].
\end{equation}
where we used $\log z\bar{z}=\log z+\log\bar{z}$. We compare this to
the solutions of the Picard-Fuchs equation, where we computed
\begin{equation}
2\pi i \rho=
\frac{X^1(z)}{X^0(z)}=\log z+5z+\frac{45}{2}z^2+\frac{440}{3}z^3+\ldots
\end{equation}
Using this, we find
\begin{equation}
\e^{-K'}=\frac{1}{2i}(\rho-\bar{\rho})=\mathrm{Im}\,\rho,
\end{equation}
which is indeed the same as the result (\ref{Znlsm})
 of the non-linear sigma model.

The sphere partition function in the $r\ll0$ phase is
\begin{eqnarray}
Z_{S^2}^{r\ll0}&=&\frac{(z\bar{z})^{2q-1}}{2}\lim_{\delta\rightarrow 0}
\oint\frac{d^2\varepsilon}{(2\pi i)^2}
\frac{\pi^5\sin^5\pi(\varepsilon_1+\varepsilon_2)}
{\sin^5\pi\varepsilon_1\sin^5\pi\varepsilon_2}
(z\bar{z})^{-\varepsilon_1}\times\nonumber\\
&&\times \left.\vline \sum_{K,k\geq 0}(-e^{-\delta})^k(-z)^{-K}
(1+K+k+\varepsilon_1+2\varepsilon_2)
\frac{\Gamma(1+K+k+\varepsilon_1+2\varepsilon_2)^5}
{\Gamma(1+K+\varepsilon_1)^5\Gamma(1+k+\varepsilon_2)^5}
\vline\right.^2\nonumber\\
\end{eqnarray}
Since the coefficients to a given order in $z$ are infinite sums, the
partition function is hard to evaluate. Alternatively, one can also
use the fact that the mirror on the $r\ll0$ phase is known
\cite{boehmthesis} to calculate the period directly or one can compute
the Picard-Fuchs operator at $r\ll0$ by transforming the Picard-Fuchs
operator of the $r\gg0$ phase. Details about this can be found in
appendix \ref{app-mirsmall}. In any case, one finds that the period in
the $r\ll 0$ phase is the same as in the $r\gg0$ phase.

\subsubsection{(S$^{2,+}_{3,3}$)}

This is the one-dimensional analogue on the Hosono-Takagi model
(S$^{2,+}_{5,5}$). Both phases correspond to elliptic curves and
they must be biholomorphic to each other.

The sphere partition function in the $r\gg0$ phase is
\begin{eqnarray}
\label{o2torus}
Z_{S^2}^{r\gg0}&=&(z\bar{z})^{2q}\oint\frac{d^2\varepsilon}{(2\pi i)^2}
\frac{\pi^3\sin^3\pi(\varepsilon_1+\varepsilon_2)}
{\sin^3\pi\varepsilon_1\sin^3\pi\varepsilon_2}
(z\bar{z})^{\varepsilon_1+\varepsilon_2}\times\nonumber\\
&&\times\left.\vline\sum_{K=0}^{\infty}(-z)^K\sum_{k=0}^K
\frac{\Gamma(1+K-\varepsilon_1-\varepsilon_2)^3}
{\Gamma(1+k-\varepsilon_1)^3\Gamma(1+K-k-\varepsilon_2)^3}\vline\right.^2
\end{eqnarray}
If we replaced the $3$'s by $5$'s everywhere, we would get the partition
function for the Hosono-Takagi model. Again, we can extract the
fundamental period from the $\log z\bar{z}$ term. The result is (after
a shift $z\rightarrow -z$):
\begin{eqnarray}
X^0_{r\gg0}&=&1 + 2 z + 10 z^2 + 56 z^3 + 346 z^4+\ldots\nonumber\\
&=&\sum_{n=0}^{\infty}\sum_{k=0}^n
\left(\begin{array}{c}n\\k\end{array}\right)^3z^n
\end{eqnarray}
Since the mirror is a free $\mathbb{Z}_2$-quotient of complete
intersection of codimension $3$ in $\mathbb{P}^2\times \mathbb{P}^2$
we can use toric methods to confirm this results. The procedure is
completely analogous to the mirror construction of the Hosono-Takagi
model given in \cite{hosonotakagi11}. Some more details can be found
in appendix \ref{app-mirsmall}. The Picard-Fuchs operator which
annihilates the fundamental period is
\begin{equation}
\mathcal{L}=\theta^2-z(7\theta^2+7\theta+2)-8z^2(\theta+1)^2.
\end{equation} 
Solving the Picard-Fuchs equation, we can determine the
second period and the mirror map. Its inverse has following power
series expansion in terms of $q=e^{2\pi i \rho}$:
\begin{equation}
q-3 q^2+3 q^3+5 q^4-18 q^5+15 q^6+\ldots.
\end{equation}
This series can be obtained as the expansion around $q=0$ of \cite{oeis}:
\begin{equation}
\frac{q}{\left(\frac{\chi(-q^3)}{\chi(-q)}\right)^3},
\end{equation}
where $\chi(q)$ is the Ramanujan theta function given by
\begin{equation}
\chi(q)=\prod_{k\geq0}(1+q^{2k+1}).
\end{equation}
It would be interesting to find out more about the modular properties
of this function.

Let us also check if the K\"ahler transform of (\ref{o2torus})
computes the area of the elliptic curve. For this purpose we define
\begin{equation}
e^{-K'}=\frac{1}{24\pi}
\frac{Z_{S^2}^{r\gg0}}{X^0_{r\gg0}(z)\bar{X}^0_{r\gg0}(\bar{z})}
\end{equation}
Expanding in $z$ and $\bar{z}$ we get 
\begin{equation}
e^{-K'}=\frac{1}{2i}\frac{1}{2\pi i}\left[\left(\log z-3z+\frac{21}{2}z^2-49z^3
+\ldots\right)+c.c.\right].
\end{equation}
Comparing with the solutions of the Picard-Fuchs equation and setting
$z\rightarrow -z$, one finds that the expression in the parentheses is
indeed the quotient of the periods $\frac{X^1(z)}{X^0(z)}=2\pi i\rho$,
which gives the expected result (\ref{Znlsm}).

In the $r\ll0$ regime the sphere partition function is
\begin{eqnarray}
Z_{S^2}^{r\ll 0}&=&(z\bar{z})^{2q-1}\lim_{\delta\rightarrow 0}\oint
\frac{d^2\varepsilon}{(2\pi i)^2}
\frac{\sin^2\pi(\varepsilon_1+\varepsilon_2)}
{\sin^3\pi\varepsilon_1\sin^3\pi\varepsilon_2}
(z\bar{z})^{-\varepsilon_1}\times\nonumber\\
&&\times \left.\vline\sum_{K,k\geq 0}(-e^{-\delta})^k(-z)^{-K}
\frac{\Gamma(1+K+k+\varepsilon_1+\varepsilon_2)^3}
{\Gamma(1+K+\varepsilon_1)^3\Gamma(1+k+\varepsilon_2)^3} \vline\right.^2
\end{eqnarray}
Extracting the period is awkward due to the involvement of infinite
sums but we can nevertheless read off the period from the coefficient
of the $\log z\bar{z}$ term. The result is:
\begin{equation}
X^0_{r\ll0}\sim-\frac{3}{2^{14}}\left(1+\frac{z}{4}+\frac{5z^2}{32}
+\frac{7z^3}{64}+\ldots\right)
\end{equation}
Redefining $z\rightarrow 8z$ the expression in the parentheses is the
same is the period in the $r\gg0$ phase. Alternatively, one can
transform the Picard-Fuchs operator to the $r\ll0$ phase which
immediately shows that $\mathcal{L}$ transforms into itself.

\appendix{Some linear algebra}\label{app:linearalgebra}

In this appendix, we show (i) in the $r\ll 0$ phase of the linear
sigma model (A) or (S), the D-term and F-term equations require
$x_j=0$ for all $j=1,\ldots,N$.
(ii) the Pfaffian $Y_A$ for $k=2$, $M=7$, $N=5$
is locally defined by ${\rm Pf}_iA(p)=0$, $i=1,2,3$
in a neighborhood of a point $p_*$ with (\ref{Apstar}),
and (iii) the rank two curve $C_S$ that appears in the model
(S$^{2,+}_{(-1)^2,(-2)^3,1^4}$) is defined by $\Delta_{ij}(p)=0$
for $(i,j)=(1,1), (2,2), (1,2)$ in a neighborhood of a point $p_*$ with
(\ref{Spstar}).

\noindent
{\bf (i)} We first consider a model in the (A) series.
Let us write down the F-term equations:
\beqa
&&\sum_{j=1}^NA(p)^{ij}x^a_j=0,\quad i=1,\ldots, N,
\label{1stF}\\
&&\sum_{i,j=1}^N{\partial\over\partial p^k}A^{ij}(p)[x_ix_j],\quad
k=1,\ldots,M.\label{2ndF}
\eeqa
We know from the D-term equation for $U(1)$ that $p$ cannot vanish and span
a weighted projective space $\PP$.
For an odd number $l$, let $Y_A^{(l)}$ be the locus 
of $p\in \PP$ such that $A(p)$ has rank $N-l$ or less.
It has codimenion ${l(l-1)\over 2}$ in $\PP$
and dimension $M-1-{l(l-1)\over 2}$.
If the latter number is negative, $Y_A^{(l)}$ is empty.
The dimension of the kernel of $A(p)$ is $l$, and we consider
the matrix of first derivatives
${\partial\over \partial p^k}A^{ij}(p)$ in which $(i,j)$ is restricted
to the kernel direction. This may be regarded as an $M\times {l\choose 2}$
matrix. It has generically rank ${l\choose 2}$
because we consider $l$ such that $M-1-{l(l-1)\over 2}\geq 0$.
Its rank drops from maximal if we tune $M-({l\choose 2}-1)$ parameters
but that is too big compared to the dimension of $Y^{(l)}_A$.
Therefore, the rank of ${\partial\over \partial p^k}A^{ij}(p)$
stays maximal, i.e. ${l\choose 2}$.
By (\ref{1stF}), $x^a=(x^a_j)_{j=1}^N$ belongs to the kernel of $A(p)$
for each $a=1,\ldots, k$ and by (\ref{2ndF}) and by the observation on
the rank of the matrix of first derivatives, we find $[x_ix_j]=0$
for all $(i,j)$. By the D-term equation for $USp(k)$,
this means $x^a_i=0$ for all $(a,i)$. 
The argument is essentially the same for a model in the (S) series.

\noindent
{\bf (ii)} In the situation described above, we would like to
show that ${\rm Pf}_iA(p)=0$ for $i=1,2,3$ imply
${\rm Pf}_jA(p)=0$ for any $j$. Suppose ${\rm Pf}_5A(p)\ne 0$.
This means that $A^1(p),\ldots, A^4(p)$ are linearly independent vectors in
$\C^N$, where $A^i(p)=(A^{ij}(p))_{j=1}^N$.
On the other hand, $A^4(p)$ and $A^5(p)$ are also linearly independent
by the assumption. By the antisymmetry of $A^{ij}(p)$,
$A^1(p),\ldots, A^5(p)$ cannot be linearly independent.
Thus, there is $(c_1,\ldots, c_5)\ne 0$ such that
$\sum_{i=1}^5c_iA^i(p)=0$.
By the linear independence of $A^1(p),\ldots, A^4(p)$, we have $c_5\ne 0$.
By the linear independence of $A^4(p)$ and $A^5(p)$, we have
$(c_1,c_2,c_3)\ne 0$, say, $c_1\ne 0$.
$A^1(p)$ can then be written as a linear combination of 
$A^2(p),\ldots, A^5(p)$ which then mean that
$A^2(p),\ldots, A^5(p)$ are linearly independent.
That is, ${\rm Pf}_1A(p)$ is not zero. A contradiction.

\noindent
{\bf (iii)} Under the situation described above,
we have 
\beqa
&&\left(\begin{array}{c}
S^{11}\\
S^{31}\\
S^{41}
\end{array}\right)=
c_{13}\left(\begin{array}{c}
S^{13}\\
S^{33}\\
S^{43}
\end{array}\right)
+c_{14}\left(\begin{array}{c}
S^{14}\\
S^{34}\\
S^{44}
\end{array}\right),
\quad
\left(\begin{array}{c}
S^{22}\\
S^{32}\\
S^{42}
\end{array}\right)=
c_{23}\left(\begin{array}{c}
S^{23}\\
S^{33}\\
S^{43}
\end{array}\right)
+c_{24}\left(\begin{array}{c}
S^{24}\\
S^{34}\\
S^{44}
\end{array}\right),\nn\\
&&\left(\begin{array}{c}
S^{12}\\
S^{32}\\
S^{42}
\end{array}\right)=
d_{23}\left(\begin{array}{c}
S^{13}\\
S^{33}\\
S^{43}
\end{array}\right)
+d_{24}\left(\begin{array}{c}
S^{14}\\
S^{34}\\
S^{44}
\end{array}\right),
\quad
\left(\begin{array}{c}
S^{21}\\
S^{31}\\
S^{41}
\end{array}\right)=
d_{13}\left(\begin{array}{c}
S^{23}\\
S^{33}\\
S^{43}
\end{array}\right)
+d_{14}\left(\begin{array}{c}
S^{24}\\
S^{34}\\
S^{44}
\end{array}\right).\nn
\eeqa
Since the last $2\times 2$ block of $S(p)$ is invertible,
we find $c_{ij}=d_{ij}$.
This means that 
\beq
\left(\begin{array}{c}
S^{11}\\
S^{21}\\
S^{31}\\
S^{41}
\end{array}\right)=
c_{13}\left(\begin{array}{c}
S^{13}\\
S^{23}\\
S^{33}\\
S^{43}
\end{array}\right)
+c_{14}\left(\begin{array}{c}
S^{14}\\
S^{24}\\
S^{34}\\
S^{44}
\end{array}\right),
\quad
\left(\begin{array}{c}
S^{12}\\
S^{22}\\
S^{32}\\
S^{42}
\end{array}\right)=
c_{23}\left(\begin{array}{c}
S^{13}\\
S^{23}\\
S^{33}\\
S^{43}
\end{array}\right)
+c_{24}\left(\begin{array}{c}
S^{14}\\
S^{24}\\
S^{34}\\
S^{44}
\end{array}\right).
\eeq
This clearly shows that $S(p)$ is of rank 2.

\appendix{Details on the two sphere partition function}
\label{app-loc}

In this appendix we give details on the computation of the two sphere
partition function in the hybrid-like phases of the Pfaffian GLSMs, as
well as both phases of the $U(1)\times O(2)$ model.

The two sphere partition function of the $(2,2)$ supersymmetric gauge theory
has been computed in \cite{Benini:2012ui,Doroud:2012xw}. There is a
correction noticed in \cite{Hori:2013ika,Hori:2013ewa}
(and also in \cite{Honda:2013uca} for a special gauge group).
When applied to a linear sigma model of Calabi-Yau type, it reads as follows:
\begin{eqnarray}
Z_{S^2}&=&(\ell\Lambda)^{\wh{c}}\sum_{m\in {\rm Q}^{\vee}}
\int\limits_{\mbox{$i\mathfrak{t}$}}\dd^{l_G}\sigma\,
\exp\Bigl(\,2ir(\sigma)+i(\theta+2\pi\rho)(m)\Bigr)\label{ZS2}
\\[-0.2cm]
&&~~~~~~~~~~~~~~~~~~
\times\prod_{\alpha>0}\left({\alpha(m)^2\over 4}+\alpha(\sigma)^2\right)
\prod_{i}
{\Gamma\left(i \chi_i(\sigma)-{\chi_i(m)\over 2}+{R_i\over 2}\right)\over
\Gamma\left(1-i \chi_i(\sigma)-{\chi_i(m)\over 2}-{R_i\over 2}\right)}.\nn
\end{eqnarray}
$\ell$ is the radius of the two sphere, $\Lambda$ is an energy scale
introduced in the renormalization, 
$\wh{c}$ is one third of the central charge of the infra-red fixed point, 
$\mathfrak{t}$ is the Lie algebra of a maximal torus of the gauge group $G$
($i\mathfrak{t}$ is the real part of its complexification $\mathfrak{t}_{\C}$),
${\rm Q}^{\vee}\subset i\mathfrak{t}$ is the coroot lattice  of $G$
(i.e., the dual to the weight lattice ${\rm P}\subset i\mathfrak{t}^*$ of $G$),
 $l_G$ is the rank of $G$,
$r$ is the FI parameter, $\theta$ is the theta parameter,
$\rho$ is half the sum of the positive roots (for a choice of Weyl
chamber), $\prod_{\alpha>0}$ is the product over the positive roots,
$\prod_i$ is the product over the weights of the matter representation,
and $R_i$ is the vector R-charge of the $i$-th chiral multiplet.
For the above contour, we need to choose $0<R_i<2$.
The factor $\e^{2\pi i\rho(m)}$ is the correction noticed
in \cite{Hori:2013ika,Hori:2013ewa}. Since $2\rho$ is always a weight,
it is a sign factor. It may affect the sum over topological types of
the $G$ bundles over the sphere.
It is therefore trivial if $G$ is simply connected, but can be non-trivial
otherwise.

In this appendix, we take a non-standard sign convention for the theta angle.
What is written below as $\theta$ is actually $-\theta$ in the main text
and also in Eqn (\ref{ZS2}).
This does not matter since
the final result is invariant under $\theta\to -\theta$.
Also, what is written as $2\pi r$ below stands for $r$
in the main text and in Eqn (\ref{ZS2}).
We take simple names of the models as
(A$^2_{(-1)^4,(-2)^3,1^5}$) $= X_5$,
(A$^2_{(-1)^6,(-2),1^4,0}$) $=X_{10}$,
(A$^2_{(-2)^7,3,1^4}$) $=X_{13}$,
(A$^2_{(-2)^5,(-4)^2,3^2,1^3}$) $=X_7$,
after the name of the Pfaffian Calabi-Yau given in \cite{kanazawa10}.

\subsection{$X_5$}
This example has gauge group $G=U(1)\times SU(2)$.
The weights and the R-charges of the fields are
\begin{equation}
\begin{tabular}{c|ccc}
&$p^{1,\ldots, 4}$&$p^{5,6,7}$&$x_{1,\ldots,5}$\\
\hline
$\chi(\sigma)$&$-\sigmaU$&$-2\sigmaU$&$\left(\begin{array}{c}
\sigmaU+\sigma_1\\
\sigmaU-\sigma_1
\end{array}\right)$\\
$U(1)_V$&$1-2q$&$2-4q$&$2q$
\end{tabular}
\end{equation}
We will assume from now on that $0<q<\frac{1}{2}$ such that all the
charges are positive. Inserting into the general formula for the
sphere partition function one gets
\begin{eqnarray}
Z_{gauge}&=&m_1^2+4\sigma_1^2\\
Z_{class}&=&e^{-4\pi ir\sigma_0-i\theta m_0}
\end{eqnarray}
\begin{equation}
Z_{p^{1,\ldots,4}}=\left[
\frac{\Gamma\left(\frac{1}{2}-q+i\sigma_0+\frac{m_0}{2}\right)}
{\Gamma\left(\frac{1}{2}+q-i\sigma_0+\frac{m_0}{2}\right)}\right]^4\qquad
Z_{p^{5,\ldots,7}}=\left[
\frac{\Gamma\left(1-2q+2i\sigma_0+m_0\right)}
{\Gamma\left(2q-2i\sigma_0+m_0 \right)}\right]^3
\end{equation}
\begin{equation}
Z_{x_{1,\ldots,5}}=\left[
\frac{\Gamma\left(q-i\sigma_0-i\sigma_1-\frac{(m_0+m_1)}{2}\right)
\Gamma\left(q-i\sigma_0+i\sigma_1-\frac{(m_0-m_1)}{2}\right)}
{\Gamma\left(1-q+i\sigma_0+i\sigma_1-\frac{(m_0+m_1)}{2}\right)
\Gamma\left(1-q-i\sigma_0-i\sigma_1-\frac{(m_0-m_1)}{2}\right)}\right]^5
\end{equation}
Introducing the coordinates $\tau_0=q-i\sigma_0$, $\tau_1=-i\sigma_1$
the partition function becomes
\begin{eqnarray}
Z_{S^2}&=&\frac{1}{2}\sum_{m_0,m_1=-\infty}^{\infty}\int_{q-i\infty}^{q+i\infty}
\frac{\mathrm{d}\tau_0}{2\pi i}\int_{-i\infty}^{i\infty}
\frac{\mathrm{d}\tau_1}{2\pi i}
\left(m_1^2-4\tau_1^2\right)e^{-4\pi rq-i\theta m_0}e^{4\pi r \tau_0}
\times\nonumber\\
&&\left[\frac{\Gamma\left(\frac{1}{2}-\tau_0+\frac{m_0}{2}\right)}
{\Gamma\left(\frac{1}{2}+\tau_0+\frac{m_0}{2}\right)}\right]^4
\left[\frac{\Gamma\left(1-2\tau_0+m_0\right)}
{\Gamma\left(2\tau_0+m_0\right)}\right]^3\times\nonumber\\
&&\left[\frac{\Gamma\left(\tau_0+\tau_1-\frac{(m_0+m_1)}{2}\right)}
{\Gamma\left(1-\tau_0-\tau_1-\frac{(m_0+m_1)}{2}\right)}
\frac{\Gamma\left(\tau_0-\tau_1-\frac{(m_0-m_1)}{2}\right)}
{\Gamma\left(1-\tau_0+\tau_1-\frac{(m_0-m_1)}{2}\right)}\right]^5.
\end{eqnarray}
We aim to compute the sphere partition function in the $r\gg0$
phase. For the result to be convergent, the $\tau_0$ contour has to be
closed in the left half-plane. For the $\tau_1$-integration we can
close the contour either to the left or to the right. The result will
be independent of the choice since the integrand is symmetric under
the exchange $\tau_1\rightarrow -\tau_1,m_1\rightarrow -m_1$. We start
with the $\tau_1$-integration and close the contour to the left. The
possible poles come from the numerator of $Z_{x_{1,\ldots,5}}$. The
factor $\Gamma\left(\tau_0+\tau_1-\frac{(m_0+m_1)}{2}\right)$ has a
pole if
\begin{equation}
\tau_1=-k-\tau_0+\frac{(m_0+m_1)}{2}\qquad k\in\mathbb{Z}\geq 0.
\end{equation}
The poles lie inside the contour if $m_0+m_1\leq 2k$ ($\tau_0$ is at
this point just an imaginary parameter, so it does not enter the
condition). These poles may be canceled by poles in the
denominator. The factor
$\Gamma\left(1-\tau_0-\tau_1-\frac{(m_0+m_1)}{2}\right)$ has a pole
for $\tau_1=1+k_1-\tau_0-\frac{(m_0+m_1)}{2}$ with $k_1\geq0$. The
poles will cancel if
\begin{equation}
-k+\frac{(m_0+m_1)}{2}=1+k_1-\frac{(m_0+m_1)}{2}
\quad\Rightarrow\quad k_1=-k-1+(m_0+m_1).
\end{equation}
Since $k_1\geq 0$, cancellation is not possible for $m_0+m_1\leq k$.
This happens when the pole in the numerator is within the contour,
and therefore there will a non-zero contribution to the residue
integral if
\begin{equation}
\tau_1=-k-\tau_0+\frac{(m_0+m_1)}{2}\qquad k\geq0\qquad m_0+m_1\leq k.
\end{equation}
The second quotient in $Z_{x_{1,\ldots,5}}$ does not
contribute. $\Gamma\left(\tau_0-\tau_1-\frac{(m_0-m_1)}{2}\right)$ has
a pole inside the contour if $\tau_1=l+\tau_0-\frac{(m_0-m_1)}{2}$
with $l\geq0$ and $m_0-m_1\geq 2l$. This is canceled against a pole of
$\Gamma\left(1-\tau_0+\tau_1-\frac{(m_0-m_1)}{2}\right)$ in the
denominator if $m_0-m_1\geq l+1$. Thus, there are no cancellations if
$m_0-m_1\leq 0$. This is always outside the contour, which shows that
there is indeed no contribution. Therefore the expression for the
partition function becomes
\begin{eqnarray}
Z_{S^2}&=&\frac{1}{2}\sum_{k=0}^{\infty}\sum_{\stackrel{m_0,m_1}{m_0+m_1\leq k}}
\int_{q-i\infty}^{q+i\infty}\frac{\mathrm{d}\tau_0}{2\pi i}\oint
\frac{\mathrm{d}\varepsilon_1}{2\pi i}
\left[m_1^2-4\left(-k-\tau_0+\frac{(m_0+m_1)}{2}+\varepsilon_1\right)^2\right]
\times\nonumber\\
&&\times e^{-4\pi rq-i\theta m_0}e^{4\pi r \tau_0}
\left[\frac{\Gamma\left(\frac{1}{2}-\tau_0+\frac{m_0}{2}\right)}
{\Gamma\left(\frac{1}{2}+\tau_0+\frac{m_0}{2}\right)}\right]^4
\left[\frac{\Gamma\left(1-2\tau_0+m_0\right)}
{\Gamma\left(2\tau_0+m_0\right)}\right]^3\times\nonumber\\
&&\times\left[\frac{\Gamma\left(-k+\varepsilon_1\right)}
{\Gamma\left(1+k-(m_0+m_1)-\varepsilon_1\right)}
\frac{\Gamma\left(2\tau_0+k-m_0-\varepsilon_1\right)}
{\Gamma\left(1-2\tau_0-k+m_1+\varepsilon_1\right)}\right]^5
\end{eqnarray}
No poles of $Z_{p^{1,\ldots,4}}$ and $Z_{p^{5,\ldots,7}}$ contribute
to the $\tau_0$-integral. The only poles which are inside the contour
are
\begin{equation}
\tau_0=\frac{1}{2}(-l-k+m_0+\varepsilon_1)\qquad l\in\mathbb{Z}_{\geq 0}
\end{equation}
These poles can be canceled by poles in the denominator. A
calculation like the one for the $\tau_1$-integration shows that the
poles of the denominator of $Z_{x_{1,\ldots,5}}$ cancel the poles in
the numerator, unless $m_0-m_1\leq l$. There are additional
cancellations from the poles of the denominators of
$Z_{p^{1,\ldots,4}}$ and $Z_{p^{5,\ldots,7}}$. In order to cancel the
pole in the numerator both factors in the denominator have to have a
pole. At first we show that the denominator of $Z_{p^{5,\ldots,7}}$
always reduces the pole order of the numerator. It has a pole if
$\tau_0=-\frac{l_1}{2}-\frac{m_0}{2}$ with integer $l_1\geq 0$. 
Therefore there is a cancellation of poles if
\begin{equation}
l_1=k+l-m_0.
\end{equation}
Inside the contour the right-hand side is always a positive integer
and the equality can always be satisfied. The denominator of
$Z_{p^{1,\ldots,4}}$ has a pole if
$\tau_0=-\frac{1}{2}-l_1-\frac{m_0}{2}$ This cancels the pole in the
numerator if
\begin{equation}
l_1=\frac{1}{2}(-1+k+l)-m_0.
\end{equation}
The condition implies that $l_1$ is only an integer if $k+l$ is
odd. In this case the pole in the numerator is canceled. There is no
cancellation only if $k+l$ is an even integer. Therefore we only have
a contribution to the residue integral if
\begin{equation}
\tau_0=\frac{1}{2}(-l-k+m_0+\varepsilon_1)
\qquad l\geq 0,\quad m_0-m_1\leq l,\quad k+l=even
\end{equation}
Defining
\begin{equation}
k'=k-(m_0+m_1)\qquad l'=l-(m_0-m_1),
\end{equation}
the partition function reduces to
\begin{eqnarray}
Z_{S^2}&=&\frac{1}{2}\sum_{\stackrel{k,k',l,l'=0}{k(k')+l(l')=even}}^{\infty}
\oint\frac{\mathrm{d}\varepsilon_1\mathrm{d}\varepsilon_2}{(2\pi i)^2}
\left[-(k-l-\varepsilon_1+2\varepsilon_2)
(k'-l'-\varepsilon_1+2\varepsilon_2)\right]e^{-4\pi r q}\times\nonumber\\
&&\times e^{2\pi r\frac{1}{2}(-k-l-k'-l'+2\varepsilon_1+4\varepsilon_2)}
e^{-i\theta\frac{1}{2}(k-k'+l-l')}
\left[\frac{\Gamma\left(\frac{1}{2}+\frac{k}{2}+\frac{l}{2}
-\frac{\varepsilon_1}{2}-\varepsilon_2\right)}
{\Gamma\left(\frac{1}{2}-\frac{k'}{2}-\frac{l'}{2}+\frac{\varepsilon_1}{2}
+\varepsilon_2\right)}\right]^4\times\nonumber\\
&&\times\left[\frac{\Gamma\left(1+k+l-\varepsilon_1-2\varepsilon_2 \right)}
{\Gamma\left(-k'-l'+\varepsilon_1+2\varepsilon_2 \right)}\right]^3
\left[\frac{\Gamma\left(-k+\varepsilon_1\right)}
{\Gamma\left(1+k'-\varepsilon_1\right)}
\frac{\Gamma\left(-l+2\varepsilon_2\right)}
{\Gamma\left(1+l'-2\varepsilon_1\right)}\right]^4.
\end{eqnarray}
Finally, we use the identity $\Gamma(x)\Gamma(1-x)=\frac{\pi}{\sin\pi x}$ 
and define $z=e^{-2\pi r+i\theta}$. With the condition that
$k+l$ and $k'+l'$ are even we can remove all the dependence on the
summation variables from the $\sin$-factors and the partition function
factorizes nicely. The final result is
\begin{eqnarray}
Z_{S^2}&=&-\frac{(z\bar{z})^q}{2}\oint
\frac{\mathrm{d}\varepsilon_1\mathrm{d}\varepsilon_2}{(2\pi i)^2}
(z\bar{z})^{-\varepsilon_1-2\varepsilon_2}\pi^3
\frac{\left[\cos\pi\left(\frac{\varepsilon_1}{2}+\varepsilon_2\right)\right]^4
\left[\sin\pi\left(\varepsilon_1+2\varepsilon_2\right)\right]^3}
{\left[\sin\pi\varepsilon_1\right]^5
\left[\sin2\pi\varepsilon_2\right]^5}
\times\nonumber\\
&&\times\left.\vline \sum_{\stackrel{k,l=0}{k+l=even}}^{\infty}
(k-l-\varepsilon_1+2\varepsilon_2)z^{\frac{k+l}{2}}
\frac{\left[\Gamma\left(\frac{1}{2}+\frac{k}{2}+\frac{l}{2}
-\frac{\varepsilon_1}{2}-\varepsilon_2\right)\right]^4
\left[\Gamma\left(1+k+l-\varepsilon_1-2\varepsilon_2 \right)\right]^3}
{\left[\Gamma\left(1+k-\varepsilon_1\right)\right]^5
\left[\Gamma\left(1+l-2\varepsilon_1\right)\right]^5}
\vline\right.^2.\nonumber\\
\end{eqnarray}

\subsection{$X_{10}$}
Up to a tiny subtlety the localization computation for this model is
almost identical to the $X_5$-case. The gauge group is $U(1)\times SU(2)$
and the weights and the R-charges of the fields are
\begin{equation}
\begin{tabular}{c|cccc}
&$p^{1,\ldots, 6}$&$p^{7}$&$x_{1,\ldots,4}$&$x_5$\\ \hline
  $\chi(\sigma)$&$-\sigma_0$&$-2\sigma_0$&
$\left(\begin{array}{c}\sigma_0+\sigma_1\\\sigma_0-\sigma_1 \end{array}
\right)$&
$\left(\begin{array}{c}\sigma_1\\-\sigma_1 \end{array}\right)$\\
$U(1)_V$&$1-2q$&$2-4q$&$2q$&$1$
\end{tabular}
\end{equation}
($0<q<{1\over 2}$ is assumed as usual.)
The partition function is thus
\begin{eqnarray}
Z_{S^2}&=&\frac{1}{2}\sum_{m_0,m_1=-\infty}^{\infty}\int_{q-i\infty}^{q+i\infty}
\frac{\mathrm{d}\tau_0}{2\pi i}\int_{-i\infty}^{i\infty}
\frac{\mathrm{d}\tau_1}{2\pi i}\left(m_1^2-4\tau_1^2\right)
e^{-4\pi rq-i\theta m_0}e^{4\pi r \tau_0}Z_{p^{1,\ldots,6}}Z_{p^7}
Z_{x_{1,\ldots,4}}Z_{x_5}\nonumber\\
&=&\frac{1}{2}\sum_{m_0,m_1=-\infty}^{\infty}\int_{q-i\infty}^{q+i\infty}
\frac{\mathrm{d}\tau_0}{2\pi i}\int_{-i\infty}^{i\infty}
\frac{\mathrm{d}\tau_1}{2\pi i}\left(m_1^2-4\tau_1^2\right)
e^{-4\pi rq-i\theta m_0}e^{4\pi r \tau_0}
\left[\frac{\Gamma\left(\frac{1}{2}-\tau_0+\frac{m_0}{2}\right)}
{\Gamma\left(\frac{1}{2}+\tau_0+\frac{m_0}{2}\right)}\right]^6\times
\nonumber\\
&&\times \left[\frac{\Gamma\left(1-2\tau_0+m_0\right)}
{\Gamma\left(2\tau_0+m_0\right)}\right]
\left[\frac{\Gamma\left(\tau_0+\tau_1-\frac{(m_0+m_1)}{2}\right)}
{\Gamma\left(1-\tau_0-\tau_1-\frac{(m_0+m_1)}{2}\right)}
\frac{\Gamma\left(\tau_0-\tau_1-\frac{(m_0-m_1)}{2}\right)}
{\Gamma\left(1-\tau_0+\tau_1-\frac{(m_0-m_1)}{2}\right)}\right]^4
\times\nonumber\\
&&\times\frac{\Gamma\left(\frac{1}{2}+\tau_1-\frac{m_1}{2}\right)
\Gamma\left(\frac{1}{2}-\tau_1+\frac{m_1}{2}\right)}
{\Gamma\left(\frac{1}{2}-\tau_1-\frac{m_1}{2}\right)
\Gamma\left(\frac{1}{2}+\tau_1+\frac{m_1}{2}\right)}.
\end{eqnarray}
$Z_{x_5}$ does not contribute any poles. To see this, use the identity
$\Gamma\left(\frac{1}{2}+x\right)\Gamma\left(\frac{1}{2}-x\right)
=\frac{\pi}{\cos\pi x}$ to show that
$Z_{x_5}=\frac{\cos\pi(y+m_1)}{\cos\pi y}
=(-1)^{m_1}$ with $y=\tau_1-\frac{m_1}{2}$. $Z_{x_5}$ thus reduces
to the contribution of a theta angle of $\pi$. From this point onward
the calculation carries over line by line from the $X_5$-example. The
result is
\begin{eqnarray}
Z_{S^2}&=&-\frac{(z\bar{z})^q}{2}\oint
\frac{\mathrm{d}\varepsilon_1\mathrm{d}\varepsilon_2}{(2\pi i)^2}
(z\bar{z})^{-\varepsilon_1-2\varepsilon_2}\pi^3
\frac{\left[\cos\pi\left(\frac{\varepsilon_1}{2}+\varepsilon_2\right)
\right]^6
\left[\sin\pi\left(\varepsilon_1+2\varepsilon_2\right)\right]}
{\left[\sin\pi\varepsilon_1\right]^4\left[\sin2\pi\varepsilon_2\right]^4}
\times\nonumber\\
&&\times\left.\vline \sum_{\stackrel{k,l=0}{k+l=even}}^{\infty}
(k-l-\varepsilon_1+2\varepsilon_2)(-z)^{\frac{k+l}{2}}
\frac{\left[\Gamma\left(\frac{1}{2}+\frac{k}{2}+\frac{l}{2}
-\frac{\varepsilon_1}{2}-\varepsilon_2\right)\right]^6
\left[\Gamma\left(1+k+l-\varepsilon_1-2\varepsilon_2 \right)\right]}
{\left[\Gamma\left(1+k-\varepsilon_1\right)\right]^4
\left[\Gamma\left(1+l-2\varepsilon_1\right)\right]^4} \vline\right.^2
\nonumber\\
\end{eqnarray}
The sphere partition function for the $r\ll0$-phase of $X_{10}$ is
\begin{eqnarray}
Z^{r\ll0}_{S^2}&=&\frac{1}{2}\lim_{\delta\rightarrow 0}\oint
\frac{\mathrm{d}\varepsilon_1\mathrm{d}\varepsilon_2}{(2\pi i)^2}
(z\bar{z})^{q-\frac{1}{2}-\varepsilon_2}\pi^7
\frac{\left[\sin\pi(\varepsilon_1+2\varepsilon_2)\right]^4}
{\left[\sin\pi\varepsilon_2\right]^6
\left[\sin 2\pi\varepsilon_2\right]
\left[\sin\pi\varepsilon_1\right]^4}\times\nonumber\\
&&\times \left.\vline\sum_{k,l=0}^{\infty}(-e^{-\delta})^k(-z)^{-l}
(1+2k+2l+2\varepsilon_1+2\varepsilon_2)\times\right.\nonumber\\
&&\times\left.\frac{\left[\Gamma(1+k+2l+\varepsilon_1+2\varepsilon_2)
\right]^4}
{\left[\Gamma(1+l+\varepsilon_2)\right]^6
\left[\Gamma(1+2l+2\varepsilon_2)\right]
\left[\Gamma(1+k+\varepsilon_1)\right]^4} \vline\right.^2,
\end{eqnarray}

\subsection{$X_{13}$}
While the localization procedure for $X_5$ and $X_{10}$ is very
similar to other examples in the literature, $X_{13}$ is more
challenging at the technical level. The gauge group for this model is
$G=U(2)$ and the weights and R-charges of the matter fields are
\begin{equation}
\begin{tabular}{c|ccc}
&$p^{1,\ldots, 7}$&$x_1$&$x_{2,\ldots,5}$\\
\hline
$\chi(\sigma)$&$-(\sigma_1+\sigma_2)$&
$\left(\begin{array}{c}2\sigma_1+\sigma_2\\\sigma_1+2\sigma_2 \end{array}
\right)$ &
$\left(\begin{array}{c}\sigma_1\\\sigma_2\end{array}\right)$\\
$U(1)_V$&$-4q$&$1+6q$&$1+2q$
\end{tabular}
\end{equation}
($-\frac{1}{6}<q<0$ is assumed.) Defining
$\tau_{1,2}=q-i\sigma_{1,2}$ the partition function can be written as
\begin{eqnarray}
Z_{S^2}&=&\frac{1}{2}\sum_{m_1,m_2=-\infty}^{\infty}\int_{q-i\infty}^{q+i\infty}
\frac{\mathrm{d}\tau_1\mathrm{d}\tau_2}{(2\pi i)^2}
\left[\frac{(m_1-m_2)^2}{4}-(\tau_1-\tau_2)^2\right]\times\nonumber\\
&&\times e^{4\pi r(\tau_1+\tau_2)}e^{-8\pi r q}e^{-i\theta(m_1+m_2)}e^{-i\pi (m_1-m_2)}
\times\nonumber\\ 
&&\times Z_{p^{1,\ldots,7}}Z_{x_1}Z_{x_{2,\ldots,5}}
\end{eqnarray}
where
\begin{eqnarray}
Z_{p^{1,\ldots,7}}&=&\left[
\frac{\Gamma\left(-\tau_1-\tau_2+\frac{m_1+m_2}{2}\right)}
{\Gamma\left(1+\tau_1+\tau_2+\frac{m_1+m_2}{2}\right)}\right]^7\\
Z_{x_1}&=&\left[
\frac{\Gamma\left(\frac{1}{2}+2\tau_1+\tau_2-\frac{2m_1+m_2}{2}\right)}
{\Gamma\left(\frac{1}{2}-2\tau_1-\tau_2-\frac{2m_1+m_2}{2}\right)}
\frac{\Gamma\left(\frac{1}{2}+\tau_1+2\tau_2-\frac{m_1+2m_2}{2}\right)}
{\Gamma\left(\frac{1}{2}-\tau_1-2\tau_2-\frac{m_1+2m_2}{2}\right)}\right]\\
Z_{x_{2,\ldots,5}}&=&\left[
\frac{\Gamma\left(\frac{1}{2}+\tau_1-\frac{m_1}{2}\right)}
{\Gamma\left(\frac{1}{2}-\tau_1-\frac{m_1}{2}\right)}
\frac{\Gamma\left(\frac{1}{2}+\tau_2-\frac{m_2}{2}\right)}
{\Gamma\left(\frac{1}{2}-\tau_2-\frac{m_2}{2}\right)}\right]^4.
\end{eqnarray}
The following transformation tremendously simplifies the calculation:
\begin{equation}
a=\tau_1+\tau_2\qquad b=\tau_1-\tau_2.
\end{equation}
The partition function in the new coordinates is
\begin{eqnarray}
Z_{S^2}&=&\frac{1}{4}\sum_{m_1,m_2=-\infty}^{\infty}
\int_{2q-i\infty}^{2q+i\infty}
\frac{\mathrm{d}a}{2\pi i}\int_{-i\infty}^{i\infty}\frac{\mathrm{d}b}{2\pi i}
\left[\frac{(m_1-m_2)^2}{4}-b^2\right]\times\nonumber\\ 
&&\times e^{4\pi r a}e^{-8\pi q r}e^{-i\theta(m_1+m_2)}e^{-i\pi(m_1-m_2)}\times\nonumber\\
&&\times\left[
\frac{\Gamma\left(-a+\frac{m_1+m_2}{2}\right)}
{\Gamma\left(1+a+\frac{m_1+m_2}{2}\right)}\right]^7
\left[\frac{\Gamma\left(\frac{1}{2}+\frac{3a}{2}+\frac{b}{2}
-\frac{2m_1+m_2}{2}\right)}
{\Gamma\left(\frac{1}{2}-\frac{3a}{2}-\frac{b}{2}-\frac{2m_1+m_2}{2}\right)}
\frac{\Gamma\left(\frac{1}{2}+\frac{3a}{2}-\frac{b}{2}
-\frac{m_1+2m_2}{2}\right)}
{\Gamma\left(\frac{1}{2}-\frac{3a}{2}+\frac{b}{2}-\frac{m_1+2m_2}{2}\right)}
\right]\times\nonumber\\
&&\times\left[
\frac{\Gamma\left(\frac{1}{2}+\frac{a}{2}+\frac{b}{2}-\frac{m_1}{2}\right)}
{\Gamma\left(\frac{1}{2}-\frac{a}{2}-\frac{b}{2}-\frac{m_1}{2}\right)}
\frac{\Gamma\left(\frac{1}{2}+\frac{a}{2}-\frac{b}{2}-\frac{m_2}{2}\right)}
{\Gamma\left(\frac{1}{2}-\frac{a}{2}+\frac{b}{2}-\frac{m_2}{2}\right)}
\right]^4
\end{eqnarray}
The structure of the transformed integral is that of a partition
function for a model with gauge group $G=U(1)\times SU(2)$. It is
strongly advised to do the the $b$-integration first. We choose to
choose the contour on the left half-plane. An analysis of the pole
structure shows that there are two types of poles, stemming from the
first $\Gamma$-factors in $Z_{x_1}$ and $Z_{x_{2,\ldots,5}}$,
respectively. They are
\begin{eqnarray}
\mbox{pole 1}:&\qquad&b=-2k-1-3a+(2m_1+m_2)\qquad 2m_1+m_2\leq k\\
\mbox{pole 2}:&\qquad&b=-2k-1-a+m_1\qquad m_1\leq k.
\end{eqnarray}
Since the $m_i$-dependence is different the poles cannot be dealt with
at once and the partition function splits into contributions from the
two types of poles: $Z_{S^2}=Z_{S^2}^{(1)}+Z_{S^2}^{(2)}$, where the
first term is
\begin{eqnarray}
Z_{S^2}^{(1)}&=&\frac{1}{4}\sum_{k=0}^{\infty}
\sum_{\stackrel{m_1,m_2}{2m_1+m_2\leq k}}\int\frac{\mathrm{d}a}{2\pi i}\oint
\frac{\mathrm{d}\varepsilon_1}{2\pi i}
\left[\frac{(m_1-m_2)^2}{4}-(-1-3a-2k+2m_1+m_2+\varepsilon_1)^2\right]
\times\nonumber\\
&&\times e^{4\pi ra}e^{-8\pi rq}e^{-i\theta(m_1+m_2)}e^{-i\pi(m_1-m_2)}
\left[\frac{\Gamma\left(-a+\frac{m_1+m_2}{2}\right)}
{\Gamma\left(1+a+\frac{m_1+m_2}{2}\right)}\right]^7\times\nonumber\\
&&\times\left[\frac{\Gamma\left(-k+\frac{\varepsilon_1}{2}\right)}
{\Gamma\left(1+k-2m_1-m_2-\frac{\varepsilon_1}{2}\right)}
\frac{\Gamma\left(1+3a+k-\frac{3(m_1+m_2)}{2}-\frac{\varepsilon_1}{2}\right)}
{\Gamma\left(-3a-k+\frac{m_1+m_2}{2}+\frac{\varepsilon_1}{2}\right)}
\right]\nonumber\\
&&\times\left[
\frac{\Gamma\left(-a-k+\frac{m_1+m_2}{2}+\frac{\varepsilon_1}{2}\right)}
{\Gamma\left(1+a+k-\frac{3m_1}{2}-\frac{m_2}{2}-\frac{\varepsilon_1}{2}
\right)}
\frac{\Gamma\left(1+2a+k-m_1-m_2-\frac{\varepsilon_1}{2}\right)}
{\Gamma\left(-2a-k+m_1+\frac{\varepsilon_1}{2}\right)}\right]^4
\end{eqnarray}
The second term is
\begin{eqnarray}
Z_{S^2}^{(2)}&=&\frac{1}{4}\sum_{k=0}^{\infty}\sum_{\stackrel{m_1,m_2}{m_1\leq k}}
\int\frac{\mathrm{d}a}{2\pi i}\oint\frac{\mathrm{d}\varepsilon_1}{2\pi i}
\left[\frac{(m_1-m_2)^2}{4}-(-1-a-2k+m_1+\varepsilon_1)^2\right]
\times\nonumber\\
&&\times e^{4\pi ra}e^{-8\pi rq} e^{-i\theta(m_1+m_2)}e^{-i\pi(m_1-m_2)} 
\left[\frac{\Gamma\left(-a+\frac{m_1+m_2}{2}\right)}
{\Gamma\left(1+a+\frac{m_1+m_2}{2}\right)}\right]^7\times\nonumber\\
&&\times\left[\frac{\Gamma\left(a-k-\frac{m_1+m_2}{2}
+\frac{\varepsilon_1}{2}\right)}
{\Gamma\left(1-a+k-\frac{3m_1}{2}-\frac{m_2}{2}-\frac{\varepsilon_1}{2}
\right)}
\frac{\Gamma\left(1+2a+k-m_1-m_2-\frac{\varepsilon_1}{2}\right)}
{\Gamma\left(-2a-k-m_2+\frac{\varepsilon_1}{2}\right)}\right]\times\nonumber\\
&&\times \left[\frac{\Gamma\left(-k+\frac{\varepsilon_1}{2}\right)}
{\Gamma\left(1+k-m_1-\frac{\varepsilon_1}{2}\right)}
\frac{\Gamma\left(1+a+k-\frac{(m_1+m_2)}{2}-\frac{\varepsilon_1}{2}\right)}
{\Gamma\left(-a-k+\frac{m_1-m_2}{2}+\frac{\varepsilon_1}{2}\right)}\right]^4
\end{eqnarray}
Next, we take care of the $a$-integration. All the possible poles
except for those of $Z_{p^{1,\ldots,7}}$ contribute. Therefore each of
the two expressions will split into three pieces. Let us start with
$Z_{S^{2}}^{(1)}=Z_{S^{2}}^{(1a)}+Z_{S^{2}}^{(1b)}+Z_{S^{2}}^{(1c)}$. The
three poles are
\begin{eqnarray}
\mbox{pole 1a}:&&a=\frac{1}{3}\left(-1-k-l+\frac{3(m_1+m_2)}{2}
+\frac{\varepsilon_1}{2}\right)
\quad m_1+2m_2\leq l,\mathrm{mod}(-2+k+l,3)\neq 0\nonumber\\
\mbox{pole 1b}:&&a=l-k+\frac{m_1+m_2}{2}+\frac{\varepsilon_1}{2}
\quad m_1\leq l,l\leq k,\ldots\nonumber\\
\mbox{pole 1c}:&&a=\frac{1}{2}\left(-1-l-k+m_1+m_2
+\frac{\varepsilon_1}{2}\right)
\qquad m_2\leq l,k+l=even
\end{eqnarray}
For poles 1a and 1c the restrictions in $k,l$ are due to extra
cancellations of poles coming from the denominator of
$Z_{p^{1,\ldots,7}}$. For pole 1b the condition that it is enclosed by
the contour results in an extra condition $k\leq l$. There may be
further conditions which we do not check because the contribution will
cancels completely. Inserting we get for the first term
\begin{eqnarray}
Z_{S^2}^{(1a)}&=&\frac{1}{4}
\sum_{\stackrel{k,l,k',l'=0}{\mathrm{mod}(-2+k(k')+l(l'),3)\neq 0}}^{\infty}
\oint
\frac{\mathrm{d}\varepsilon_1\mathrm{d}\varepsilon_2}{(2\pi i)^2}
\left[-\frac{1}{4}\left(2(k-l)-\varepsilon_1+6\varepsilon_2\right)
\left(2(k'-l')-\varepsilon_1+6\varepsilon_2\right)\right]\times\nonumber\\
&&\times e^{-8\pi q r}e^{4\pi r\frac{1}{6}(-2-k-k'-l-l'+\varepsilon_1
+6\varepsilon_2)}
e^{-i\theta\frac{1}{3}(k-k'+l-l')}e^{-i\pi(k-k'-l+l')}\times\nonumber\\ 
&&\times\left[
\frac{\Gamma\left(\frac{1}{3}+\frac{k}{3}+\frac{l}{3}
-\frac{\varepsilon_1}{6}-\varepsilon_2\right)}
{\Gamma\left(\frac{2}{3}-\frac{k'}{3}-\frac{l'}{3}
+\frac{\varepsilon_1}{6}+\varepsilon_2\right)}\right]^7
\left[\frac{\Gamma\left(-k+\frac{\varepsilon_1}{2}\right)}
{\Gamma\left(1+k'-\frac{\varepsilon_1}{2}\right)}
\frac{\Gamma\left(-l+3\varepsilon_2\right)}
{\Gamma\left(1+l'-3\varepsilon_2\right)}\right]\times\nonumber\\
&&\times\left[\frac{\Gamma\left(\frac{1}{3}-\frac{2k}{3}+\frac{l}{3}
+\frac{\varepsilon_1}{3}-\varepsilon_2\right)}
{\Gamma\left(\frac{2}{3}+\frac{2k'}{3}-\frac{l'}{3}
-\frac{\varepsilon_1}{3}+\varepsilon_2\right)}
\frac{\Gamma\left(\frac{1}{3}+\frac{k}{3}-\frac{2l}{3}
-\frac{\varepsilon_1}{6}+2\varepsilon_2\right)}
{\Gamma\left(\frac{2}{3}-\frac{k'}{3}+\frac{2l'}{3}
+\frac{\varepsilon_1}{6}-2\varepsilon_2\right)}\right]^4,
\end{eqnarray}
where we defined $k'=k-2m_1-m_2$ and $l'=l-m_1-2m_2$. For the second
term one gets
\begin{eqnarray}
Z_{S^2}^{(1b)}&=&\frac{1}{4}\sum_{\stackrel{k,l,k',l'=0}{l\leq k}}^{\infty}\oint
\frac{\mathrm{d}\varepsilon_1\mathrm{d}\varepsilon_2}{(2\pi i)^2}
\left[-\frac{1}{4}\left(2-2k+6l+\varepsilon_1+6\varepsilon_2\right)
\left(2-2k'+6l'+\varepsilon_1+6\varepsilon_2\right)\right]\times\nonumber\\
&&\times e^{-8\pi q r}e^{4\pi r\frac{1}{2}(-k-k'+l+l'+\varepsilon_1+2\varepsilon_2)}
e^{-i\theta(k-k'-l+l')}e^{-i\pi(-k+k'+3l-3l')}\times\nonumber\\ 
&&\times\left[
\frac{\Gamma\left(k-l-\frac{\varepsilon_1}{2}-\varepsilon_2\right)}
{\Gamma\left(1-k'+l'+\frac{\varepsilon_1}{2}+\varepsilon_2\right)}\right]^7
\left[\frac{\Gamma\left(-k+\frac{\varepsilon_1}{2}\right)}
{\Gamma\left(1+k'-\frac{\varepsilon_1}{2}\right)}
\frac{\Gamma\left(1-2k+3l+\varepsilon_1+3\varepsilon_2\right)}
{\Gamma\left(2k'-3l'-\varepsilon_1-3\varepsilon_2\right)}\right]
\times\nonumber\\
&&\times\left[\frac{\Gamma\left(-l-\varepsilon_2\right)}
{\Gamma\left(1+l'+\varepsilon_2\right)}
\frac{\Gamma\left(1-k+2l+\frac{\varepsilon_1}{2}+2\varepsilon_2\right)}
{\Gamma\left(k'-2l'-\frac{\varepsilon_1}{2}-2\varepsilon_2\right)}\right]^4
\end{eqnarray}
with $k'=k-2m_1-m_2$, $l'=l-m_1$ and finally
\begin{eqnarray}
Z_{S^2}^{(1c)}&=&\frac{1}{4}
\sum_{\stackrel{k,l,k',l'=0}{k(k')+l(l')=even}}^{\infty}
\oint\frac{\mathrm{d}\varepsilon_1\mathrm{d}\varepsilon_2}{(2\pi i)^2}
\left[-\frac{1}{16}\left(2-2k+6l+\varepsilon_1-12\varepsilon_2\right)
\left(2-2k'+6l'+\varepsilon_1-12\varepsilon_2\right)\right]\times\nonumber\\
&&\times 
e^{-8\pi q r}e^{4\pi r\frac{1}{4}(-2-k-k'-l-l'+\varepsilon_1+4\varepsilon_2)}
e^{-i\theta\frac{1}{2}(k-k'+l-l')}e^{-i\pi\frac{1}{2}(k-k'-l+l')}\times\nonumber\\
&&\times\left[
\frac{\Gamma\left(\frac{1}{2}+\frac{k}{2}+\frac{l}{2}
-\frac{\varepsilon_1}{4}-\varepsilon_2\right)}
{\Gamma\left(\frac{1}{2}-\frac{k'}{2}-\frac{l'}{2}
+\frac{\varepsilon_1}{4}+\varepsilon_2\right)}\right]^7
\left[\frac{\Gamma\left(-k+\frac{\varepsilon_1}{2}\right)}
{\Gamma\left(1+k'-\frac{\varepsilon_1}{2}\right)}
\frac{\Gamma\left(-\frac{1}{2}-\frac{k}{2}-\frac{3l}{2}
+\frac{\varepsilon_1}{4}+3\varepsilon_2\right)}
{\Gamma\left(\frac{3}{2}+\frac{k'}{2}+\frac{3l'}{2}
-\frac{\varepsilon_1}{4}-3\varepsilon_2\right)}\right]\times\nonumber\\
&&\times\left[\frac{\Gamma\left(\frac{1}{2}-\frac{k}{2}+\frac{l}{2}
+\frac{\varepsilon_1}{4}-\varepsilon_2\right)}
{\Gamma\left(\frac{1}{2}+\frac{k'}{2}-\frac{l'}{2}
-\frac{\varepsilon_1}{4}+\varepsilon_2\right)}
\frac{\Gamma\left(-l+2\varepsilon_2\right)}
{\Gamma\left(1+l'-2\varepsilon_2\right)}\right]^4,
\end{eqnarray}
where $k'=k-2m_1-m_2$ and $l'=l-m_2$. The evaluation of
$Z_{S^2}^{(2)}$ proceeds in exactly the same way. The two poles are
\begin{eqnarray}
\mbox{pole 2a}:&&a=-l+k+\frac{m_1+m_2}{2}-\frac{\varepsilon_1}{2}
\qquad 2m_1+m_2\leq l,k\leq l,\ldots\nonumber\\
\mbox{pole 2b}:&&a=
\frac{1}{2}\left(-1-l-k+m_1+m_2+\frac{\varepsilon_1}{2}\right)
\qquad m_1+2m_2\leq l, k+l=even\nonumber\\
\end{eqnarray}
The numerator has a additional pole at
$a=-1-l-k+\frac{m_1+m_2}{2}+\frac{\varepsilon_1}{2}$ coming from
$Z_{x_{2,\ldots,5}}$, which is however always canceled by poles in the
denominator.  From the first pole we get
\begin{eqnarray}
Z_{S^2}^{(2a)}&=&\frac{1}{4}\sum_{\stackrel{k,l,k',l'=0}{k\leq l}}^{\infty}\oint
\frac{\mathrm{d}\varepsilon_1\mathrm{d}\varepsilon_2}{(2\pi i)^2}
\left[-\frac{1}{4}\left(2+6k-2l-3\varepsilon_1+2\varepsilon_2\right)
\left(2+6k'-2l'-3\varepsilon_1+2\varepsilon_2\right)\right]\times\nonumber\\
&&\times e^{-8\pi q r}e^{4\pi r\frac{1}{2}(k+k'-l-l'-\varepsilon_1+2\varepsilon_2)}
e^{-i\theta(-k+k'+l-l')}e^{-i\pi(k-k'-l+l')}\times\nonumber\\ 
&&\times\left[
\frac{\Gamma\left(-k+l+\frac{\varepsilon_1}{2}-\varepsilon_2\right)}
{\Gamma\left(1+k'-l'+\frac{\varepsilon_1}{2}+\varepsilon_2\right)}\right]^7
\left[\frac{\Gamma\left(-l+{\varepsilon_2}\right)}
{\Gamma\left(1+l'-{\varepsilon_2}\right)}
\frac{\Gamma\left(1+3k-2l-\frac{3\varepsilon_1}{2}+2\varepsilon_2\right)}
{\Gamma\left(-3k'+2l'+\frac{3\varepsilon_1}{2}-2\varepsilon_2\right)}\right]
\times\nonumber\\
&&\times\left[\frac{\Gamma\left(-k-\frac{\varepsilon_1}{2}\right)}
{\Gamma\left(1+k'+\frac{\varepsilon_1}{2}\right)}
\frac{\Gamma\left(1+2k-l-\varepsilon_1+\varepsilon_2\right)}
{\Gamma\left(-2k'+l'+\varepsilon_1-\varepsilon_2\right)}\right]^4,
\end{eqnarray}
with $k'=k-m_1$, $l'=l-2m_1-m_2$.  The second pole gives
\begin{eqnarray}
Z_{S^2}^{(2b)}&=&
\frac{1}{4}\sum_{\stackrel{k,l,k',l'=0}{k(k')+l(l')=even}}^{\infty}
\oint
\frac{\mathrm{d}\varepsilon_1\mathrm{d}\varepsilon_2}{(2\pi i)^2}
\left[-\frac{1}{16}\left(2+6k-2l-3\varepsilon_1+4\varepsilon_2\right)
\left(2+6k'-2l'-3\varepsilon_1+4\varepsilon_2\right)\right]\times\nonumber\\
&&\times e^{-8\pi q r}
e^{4\pi r\frac{1}{4}(-2-k-k'-l-l'+\frac{\varepsilon_1}{4}+\varepsilon_2)}
e^{-i\theta\frac{1}{2}(k-k'+l-l')}e^{-i\pi(-k+k'+3l-3l')}\times\nonumber\\
&&\times\left[\frac{\Gamma\left(\frac{1}{2}+\frac{k}{2}+\frac{l}{2}
-\frac{\varepsilon_1}{4}-\varepsilon_2\right)}
{\Gamma\left(\frac{1}{2}-\frac{k'}{2}-\frac{l'}{2}
+\frac{\varepsilon_1}{4}+\varepsilon_2\right)}\right]^7
\left[\frac{\Gamma\left(-l+2\varepsilon_2\right)}
{\Gamma\left(1+l'-2\varepsilon_2\right)}
\frac{\Gamma\left(-\frac{1}{2}-\frac{3k}{2}-\frac{l}{2}
+\frac{3\varepsilon_1}{4}+\varepsilon_2\right)}
{\Gamma\left(\frac{3}{2}+\frac{3k'}{2}+\frac{l'}{2}
-\frac{3\varepsilon_1}{4}-\varepsilon_2\right)}\right]
\times\nonumber\\
&&\times\left[\frac{\Gamma\left(\frac{1}{2}+\frac{k}{2}-\frac{l}{2}
-\frac{\varepsilon_1}{4}+\varepsilon_2\right)}
{\Gamma\left(\frac{1}{2}-\frac{k'}{2}+\frac{l'}{2}
+\frac{\varepsilon_1}{4}-\varepsilon_2\right)}
\frac{\Gamma\left(-k+\frac{\varepsilon_1}{2}\right)}
{\Gamma\left(1+k'-\frac{\varepsilon_1}{4}\right)}\right]^4,
\end{eqnarray}
with $k'=k-m_1$, $l'=l-m_1-2m_2$.
Looking at the five terms, there are transformations such that
\begin{eqnarray}
k\leftrightarrow l,\; \varepsilon_1\rightarrow2\varepsilon_2,\;
\varepsilon_2\rightarrow-\frac{\varepsilon_1}{2}:
&\qquad& Z_{S^2}^{(1b)}\rightarrow -Z_{S^2}^{(2a)}\\
k\leftrightarrow l,\; \varepsilon_1\rightarrow4\varepsilon_2,\;
\varepsilon_2\rightarrow\phantom{-}\frac{\varepsilon_1}{4}:
&\qquad& Z_{S^2}^{(1c)}\rightarrow \phantom{-}Z_{S^2}^{(2b)}.
\end{eqnarray}
Therefore the complete sphere partition function simplifies to
\begin{equation}
Z_{S^2}=Z_{S^2}^{(1a)}+2Z_{S^2}^{(1c)}.
\end{equation}
With the same tools as for the previous examples and the definition
$z=-e^{-2\pi r+i\theta}$ the three terms can be further manipulated to
give the following.  
\begin{eqnarray}
Z_{S^2}^{(1a)}&=&-\frac{(z\bar{z})^{2q}}{16}\oint
\frac{\mathrm{d}\varepsilon_1\mathrm{d}\varepsilon_2}{(2\pi i)^2}
\sum_{\stackrel{k,k',l,l'=0}{\mathrm{mod}(-2+k(k')+l(l'),3)\neq 0}}^{\infty}\pi^3
(z\bar{z})^{\frac{1}{3}-\frac{\varepsilon_1}{6}-\varepsilon_2}
(-1)^{k+l}(-1)^{k-k'-l+l'}\times\nonumber\\ 
&&\times
\frac{\left[\sin\pi\left(\frac{1}{3}(2-k'-l')+\frac{\varepsilon_1}{6}
+\varepsilon_2\right)\right]^7}
{\sin\pi\frac{\varepsilon_1}{2}\sin 3\pi\varepsilon_2
\left[\sin\pi\left(\frac{2}{3}(2-k-l)+\frac{\varepsilon_1}{3}
-\varepsilon_2\right)\right]^4
\left[\sin\pi\left(\frac{2}{3}(2-k-l)+\frac{\varepsilon_1}{6}
-2\varepsilon_2\right)\right]^4} \times\nonumber\\
&&\times\bigg\{z^{\frac{k+l}{3}}(2(k-l)-\varepsilon_1+6\varepsilon_2)
\frac{\left[\Gamma\left(\frac{1}{3}(1+k+l)-\frac{\varepsilon_1}{6}
-\varepsilon_2\right)\right]^7}
{\Gamma\left(1+k-\frac{\varepsilon_1}{2}\right)
\Gamma\left(1+l-3\varepsilon_2\right)} \times\nonumber\\
&&\times \frac{1}{\left[\Gamma\left(\frac{1}{3}(2+2k-l)
-\frac{\varepsilon_1}{3}+\varepsilon_2 \right)\right]^4
\left[\Gamma\left(\frac{1}{3}(2-k+2l)+\frac{\varepsilon_1}{6}
-2\varepsilon_2 \right)\right]^4}   \bigg\}\{k',l',\bar{z}\}
\end{eqnarray}
The expression $\{k',l',\bar{z}\}$ means to multiply with the same
factor and the indicated replacements. Note that $Z_{S^2}^{(1a)}$ does
not factorize, which is unexpected. Computing the residue gives a
non-zero expression which does not factorize either. However, when one
explicitly evaluates the sums one actually finds that
$Z_{S^2}^{(1a)}=0$!
To see that $Z_{S^2}^{(1a)}=0$ we note that after evaluating the
residue the expression has the following form:
\begin{eqnarray}
Z_{S^2}^{(1a)}&=&\frac{2\pi}{3}(z\bar{z})^{\frac{1}{3}+2q}
\sum_{\stackrel{k,k',l,l'=0}{\mathrm{mod}(-2+k(k')+l(l'),3)\neq 0}}^{\infty}
\frac{\left[\sin\pi\left(\frac{1}{3}\left(-2+k'+l'\right)\right)\right]^7}
{\left[\sin\pi\left(\frac{2}{3}\left(-2+k+l\right)\right)\right]^8}
(-1)^{k+l}(-1)^{-k'+l'}\pi\times\nonumber\\ 
&&\times\left\{(k-l)z^{\frac{k+l}{3}}
\frac{\left[\Gamma\left(\frac{1}{3}\left(1+k+l\right)\right)\right]^7}
{\left[\Gamma\left(\frac{1}{3}\left(2-k+2l\right)\right)\right]^4
\left[\Gamma\left(\frac{1}{3}\left(2+2k-l\right)\right)\right]^4
\Gamma(1+k)\Gamma(1+l)} \right\}\{k',l',\bar{z}\}\nonumber\\
\end{eqnarray}
This immediately shows that for fixed $(k,l)$ every summand with a
particular $(k',l')$ cancels against the summand where
$k'\leftrightarrow l'$, which is the same up the an overall
sign. Therefore the sum is $0$, even though the individual summands
are not.

The only thing that is left is $Z_{S^2}^{(1c)}$ which indeed factorizes:
\begin{eqnarray}
Z_{S^2}^{(1c)}&=&\frac{(z\bar{z})^q}{64}\oint
\frac{\mathrm{d}\varepsilon_1\mathrm{d}\varepsilon_2}{(2\pi i)^2}\pi^3
(z\bar{z})^{\frac{1}{2}-\frac{\varepsilon_1}{4}-\varepsilon_2}
\frac{\left[\cos\pi\left(\frac{\varepsilon_1}{4}+\varepsilon_2\right)\right]^7}
{\sin\pi\frac{\varepsilon_1}{2}
\cos\pi\left(\frac{\varepsilon_1}{4}+3\varepsilon_2\right)
\left[\cos\pi\left(\frac{\varepsilon_1}{4}-\varepsilon_2\right)\right]^4
\left[\sin2\pi\varepsilon_2\right]^4}\times\nonumber\\
&&\times \left.\vline\sum_{\stackrel{k,l=0}{k+l=even}}^{\infty}
(2-2k+6l+\varepsilon_1-12\varepsilon_2)(-z)^{\frac{k+l}{2}}
\right.\times\nonumber\\
&&\times\left.
\frac{\left[\Gamma\left(\frac{1}{2}(1+k+l)-\frac{\varepsilon_1}{4}
-\varepsilon_2\right)\right]^7}
{\Gamma\left(1+k-\frac{\varepsilon_1}{2}\right)
\Gamma\left(\frac{3}{2}+\frac{k}{2}+\frac{3l}{2}-\frac{\varepsilon_1}{4}
-3\varepsilon_2\right)
\left[\Gamma\left(\frac{1}{2}+\frac{k}{2}-\frac{l}{2}-\frac{\varepsilon_1}{4}
+\varepsilon_2\right)\right]^4
\left[\Gamma(1+l-2\varepsilon_2)\right]^4} \vline\right.^2\nonumber\\
\end{eqnarray}

\subsection{$X_{7}$}
The field content of this model only differs very slightly from
$X_{13}$, yet it exhibits completely different behavior in the
$r\gg0$-region. The gauge group is $U(2)$ and the weights 
and R-charges of the matter fields are
\begin{equation}
\begin{tabular}{c|cccc}
&$p^{1,\ldots, 5}$&$p^{6,7}$&$x_{1,2}$&$x_{3,\ldots,5}$\\
\hline
$\chi(\sigma)$&$-\sigma_1-\sigma_2$&$-2\sigma_1-2\sigma_2$&
$\left(\begin{array}{c}2\sigma_1+\sigma_2\\\sigma_1+2\sigma_2 \end{array}
\right)$ &
$\left(\begin{array}{c}\sigma_1\\\sigma_2\end{array}\right)$\\
$U(1)_V$&$-4q$&$-8q$&$1+6q$&$1+2q$
\end{tabular}.
\end{equation}
($-\frac{1}{6}<q<0$ is assumed.) In terms of the variables
$\tau_{1,2}=q-\sigma_{1,2}$ the sphere partition function looks like
the one of $X_{13}$ except for $Z_{p^{6,7}}$ and different
multiplicities.
\begin{eqnarray}
Z_{S^2}&=&\frac{1}{2}\sum_{m_1,m_2=-\infty}^{\infty}\int_{q-i\infty}^{q+i\infty}
\frac{\mathrm{d}\tau_1\mathrm{d}\tau_2}{(2\pi i)^2}
\left[\frac{(m_1-m_2)^2}{4}-(\tau_1-\tau_2)^2\right]\times\nonumber\\
&&\times e^{4\pi r(\tau_1+\tau_2)}e^{-8\pi r q}e^{-i\theta(m_1+m_2)}
e^{-i\pi(m_1-m_2)}\times\nonumber\\ 
&&\times Z_{p^{1,\ldots,5}}Z_{p^{6,7}}Z_{x_{1,2}}Z_{x_{3,\ldots,5}}
\end{eqnarray}
where
\begin{eqnarray}
Z_{p^{1,\ldots,5}}&=&\left[
\frac{\Gamma\left(-\tau_1-\tau_2+\frac{m_1+m_2}{2}\right)}
{\Gamma\left(1+\tau_1+\tau_2+\frac{m_1+m_2}{2}\right)}\right]^5\\
Z_{p^{6,7}}&=&\left[\frac{\Gamma\left(-2\tau_1-2\tau_2+{m_1+m_2}\right)}
{\Gamma\left(1+2\tau_1+2\tau_2+{m_1+m_2}\right)}\right]^2\\
Z_{x_{1,2}}&=&\left[
\frac{\Gamma\left(\frac{1}{2}+2\tau_1+\tau_2-\frac{2m_1+m_2}{2}\right)}
{\Gamma\left(\frac{1}{2}-2\tau_1-\tau_2-\frac{2m_1+m_2}{2}\right)}
\frac{\Gamma\left(\frac{1}{2}+\tau_1+2\tau_2-\frac{m_1+2m_2}{2}\right)}
{\Gamma\left(\frac{1}{2}-\tau_1-2\tau_2-\frac{m_1+2m_2}{2}\right)}\right]^2\\
Z_{x_{3,\ldots,5}}&=&\left[
\frac{\Gamma\left(\frac{1}{2}+\tau_1-\frac{m_1}{2}\right)}
{\Gamma\left(\frac{1}{2}-\tau_1-\frac{m_1}{2}\right)}
\frac{\Gamma\left(\frac{1}{2}+\tau_2-\frac{m_2}{2}\right)}
{\Gamma\left(\frac{1}{2}-\tau_2-\frac{m_2}{2}\right)}\right]^3.
\end{eqnarray}
Since we are interested in the $r\gg0$ region the poles of
$Z_{p^{1,\ldots,5}}$ and $Z_{p^{6,7}}$ will not contribute to the
residue integrals. The only difference to $X_{13}$ is that for the
calculation of $Z_{S^2}^{(1a)}$ there is an additional condition
$\mathrm{mod}(-1+2k+2l,3)\neq0$ due to extra pole cancellations coming
from the denominator of $Z_{p^{6,7}}$. The partition function is
$Z_{S^2}=Z_{S^2}^{(1a)}+2Z_{S^2}^{(1c)}$ with
\begin{eqnarray}
Z_{S^2}^{(1a)}&=&-\frac{(z\bar{z})^{2q}}{16}\oint
\frac{\mathrm{d}\varepsilon_1\mathrm{d}\varepsilon_2}{(2\pi i)^2}
\sum_{\stackrel{k,k',l,l'=0}{cond.}}^{\infty}\pi^3
(z\bar{z})^{\frac{1}{3}-\frac{\varepsilon_1}{6}-\varepsilon_2}(-1)^{k+l}
(-1)^{k-k'-l+l'}\times\nonumber\\ 
&&\times \frac{\left[\sin\pi\left(\frac{1}{3}(2-k'-l')+\frac{\varepsilon_1}{6}
+\varepsilon_2\right)\right]^5
\left[\sin\pi\left(\frac{1}{3}(1-2k'-2l')+\frac{\varepsilon_1}{3}
+2\varepsilon_2\right)\right]^2}
{\left[\sin\pi\frac{\varepsilon_1}{2}\right]^2
\left[\sin 3\pi\varepsilon_2\right]^2
\left[\sin\pi\left(\frac{2}{3}(2-k-l)+\frac{\varepsilon_1}{3}
-\varepsilon_2\right)\right]^3
\left[\sin\pi\left(\frac{2}{3}(2-k-l)+\frac{\varepsilon_1}{6}
-2\varepsilon_2\right)\right]^3} \times\nonumber\\
&&\times\bigg\{z^{\frac{k+l}{3}}(2(k-l)-\varepsilon_1+6\varepsilon_2)
\frac{\left[\Gamma\left(\frac{1}{3}(1+k+l)-\frac{\varepsilon_1}{6}
-\varepsilon_2\right)\right]^5
\left[\Gamma\left(\frac{2}{3}(1+k+l)+\frac{\varepsilon_1}{3}
+2\varepsilon_2\right)\right]^2}
{\left[\Gamma\left(1+k-\frac{\varepsilon_1}{2}\right)\right]^2
\left[\Gamma\left(1+l-3\varepsilon_2\right)\right]^2} \times\nonumber\\
&&\times \frac{1}{\left[\Gamma\left(\frac{1}{3}(2+2k-l)
-\frac{\varepsilon_1}{3}+\varepsilon_2 \right)\right]^3
\left[\Gamma\left(\frac{1}{3}(2-k+2l)+\frac{\varepsilon_1}{6}
-2\varepsilon_2 \right)\right]^3}   \bigg\}\{k',l',\bar{z}\},
\end{eqnarray}
where $cond.$ stands for $\mathrm{mod}(-2+k+l,3)\neq0$ and
$\mathrm{mod}(-1+2k+2l,3)\neq0$. Furthermore
\begin{eqnarray}
Z_{S^2}^{(1c)}&=&-\frac{(z\bar{z})^q}{64}\oint
\frac{\mathrm{d}\varepsilon_1\mathrm{d}\varepsilon_2}{(2\pi i)^2}\pi^3
(z\bar{z})^{\frac{1}{2}-\frac{\varepsilon_1}{4}-\varepsilon_2}\times\nonumber\\
&&\times
\frac{\left[\cos\pi\left(\frac{\varepsilon_1}{4}+\varepsilon_2\right)\right]^5
\left[\sin\pi\left(\frac{\varepsilon_1}{2}+2\varepsilon_2\right)\right]^2}
{\left[\sin\pi\frac{\varepsilon_1}{2}\right]^2
\left[\cos\pi\left(\frac{\varepsilon_1}{4}+3\varepsilon_2\right)\right]^2
\left[\cos\pi\left(\frac{\varepsilon_1}{4}-\varepsilon_2\right)\right]^3
\left[\sin2\pi\varepsilon_2\right]^3}\times\nonumber\\
&&\times \left.\vline\sum_{\stackrel{k,l=0}{k+l=even}}^{\infty}
(2-2k+6l+\varepsilon_1-12\varepsilon_2)(-z)^{\frac{k+l}{2}}\right.
\times\nonumber\\
&&\times\left. \frac{\left[\Gamma\left(\frac{1}{2}(1+k+l)
-\frac{\varepsilon_1}{4}-\varepsilon_2\right)\right]^5}
{\left[\Gamma\left(1+k-\frac{\varepsilon_1}{2}\right)\right]^2
\left[\Gamma\left(\frac{3}{2}+\frac{k}{2}+\frac{3l}{2}
-\frac{\varepsilon_1}{4}-3\varepsilon_2\right)\right]^2}
\right.\times \nonumber\\
&&\times\left.\frac{\left[\Gamma\left(1+k+l-\frac{\varepsilon_1}{2}
+2\varepsilon_2\right)\right]^2}
{\left[\Gamma\left(\frac{1}{2}+\frac{k}{2}-\frac{l}{2}
-\frac{\varepsilon_1}{4}+\varepsilon_2\right)\right]^3
\left[\Gamma(1+l-2\varepsilon_2)\right]^3} \vline\right.^2
\end{eqnarray}

The leading term in the $z\to 0$ limit comes from
$Z_{S^{2}}^{(1a)}$. However, since it is
\begin{equation}
Z_{S^{2}}^{(1a)}=\frac{2}{\sqrt{3}\pi}
\frac{\Gamma\left(\frac{1}{3}\right)^{10}}
{\Gamma\left(\frac{2}{3}\right)^8}
(z\bar{z})^{2q+\frac{1}{3}}, \qquad -\frac{1}{6}<q<0
\end{equation}
the contribution to $g_{z\bar{z}}$ is zero. Therefore we have to take
into account the next to leading order contributed by
$Z_{S^{2}}^{(1c)}$. With that we get,
\begin{equation}
e^{-K}=\frac{2}{\sqrt{3}\pi}
\frac{\Gamma\left(\frac{1}{3}\right)^{10}}
{\Gamma\left(\frac{2}{3}\right)^8}
(z\bar{z})^{2q+\frac{1}{3}}
-4(z\bar{z})^{2q+\frac{1}{2}}(36+8\log4-3\log{z\bar{z}})+\ldots
\end{equation}
The leading term of the K\"ahler metric is
\begin{equation}
g_{z\bar{z}}=-\frac{\pi\Gamma\left(\frac{2}{3}\right)^8
\log^3\left(
\frac{z\bar{z}}{2^{\frac{16}{3}}}\right)}
{6\sqrt{3}\Gamma\left(\frac{1}{3}\right)^{10}
(z\bar{z})^{\frac{5}{6}}}.
\end{equation}

\subsection{(S$^{2,+}_{(-1)^2,(-2)^3,1^4}$)}
The gauge group is $U(1)\times O(2)$ and the weights and the R-charges of
the matter fields are
\begin{equation}
\begin{tabular}{c|ccc}
&$p^{1,2}$&$p^{3,\ldots,5}$&$x_{1,\ldots,4}$\\
\hline
$\chi(\sigma)$&$-\sigmaU$&$-2\sigmaU$&
$\left(\begin{array}{c}\sigmaU+\sigma_1\\
\sigmaU-\sigma_1\end{array}\right)$\\
$U(1)_V$&$1-2q$&$2-4q$&$2q$
\end{tabular}.
\end{equation}
Since $O(2)$ has no root, we have
\begin{equation}
Z_{gauge}=1.
\end{equation}
To compute $Z_{class}$ we note that only the $U(1)$-factor contributes
an $FI$-parameter $r$. The one from $O(2)$ is zero. However, since
$N-k=4-2=2$ is even, we need the discrete $\theta$-angle of $\pi$ for
the regularity. Therefore, the one gets
\begin{equation}
Z_{class}=e^{-4\pi r\sigma_0-i\theta m_0-i\pi m_1}.
\end{equation}
Defining $\tau_{0}=q-i\sigma_{0}$ and $\tau_{1}=-i\sigma_{1}$, the
contribution from the chiral fields is
\begin{equation}
Z_{p^{1,2}}=\left[\frac{\Gamma\left(\frac{1}{2}-\tau_0+\frac{m_0}{2}\right)}
{\Gamma\left(\frac{1}{2}+\tau_0+\frac{m_0}{2}\right)}\right]^2\qquad
Z_{p^{3,\ldots,5}}=\left[\frac{\Gamma\left(1-2\tau_0+m_0\right)}
{\Gamma\left(2\tau_0+m_0\right)}\right]^3,
\end{equation}
\begin{equation}
Z_{x_{1,\ldots,4}}=\left[\frac{\Gamma\left(\tau_0+\tau_1-\frac{(m_0+m_1)}{2}
\right)}
{\Gamma\left(1-\tau_0-\tau_1-\frac{(m_0+m_1)}{2}\right)}
\frac{\Gamma\left(\tau_0-\tau_1-\frac{m_0-m_1}{2}\right)}
{\Gamma\left(1-\tau_0+\tau_1-\frac{m_0-m_1}{2}\right)}\right]^4.
\end{equation}
Therefore the sphere partition function is
\begin{eqnarray}
Z_{S^2}&=&\sum_{m_0,m_1=-\infty}^{\infty}
\int_{q-i\infty}^{q+i\infty}\int_{-i\infty}^{i\infty}
\frac{d\tau_0d\tau_1}{(2\pi i)^2}e^{-4\pi r q-i\theta m_0}(-1)^{m_1}e^{4\pi r\tau_0}
Z_{p^{1,2}}Z_{p^{3,\ldots,5}}Z_{x_{1,\ldots,4}}
\end{eqnarray}
The calculation proceeds exactly like in the previous examples. It is
recommended to do the $\tau_1$ integration first. While the closing of
the contour can be chosen freely, we always choose to close it in the
same direction as for $\tau_0$. Since the $r\ll0$-phase is also new in
this case, we start by computing the partition function for this
case. For the $\tau_1$-integration where we close the contour on the
right half, the only contribution comes from the pole at
\begin{equation}
\tau_1=\tau_0+k-\frac{(m_0-m_1)}{2}\qquad m_0-m_1\leq k
\end{equation}
For the $\tau_0$-integration, there are two possible poles coming from
the singularities of $Z_p$:
\begin{eqnarray}
\label{poleo2}
\tau_0&=&\frac{1}{2}+l+\frac{m_0}{2}\nonumber\\
\tau_0&=&\frac{1}{2}+\frac{l}{2}+\frac{m_0}{2}
\end{eqnarray}
Both types of poles get canceled by the pole
$\tau_0=\frac{1}{2}(1+l_1-k-m_1)$ in the denominator of
$Z_{x_{1,\ldots,4}}$. However, this is only a fourth order pole,
wheres the combined pole order of $Z_p$ if five. So there will always
be something left. We only have to take into account the first type of
pole in (\ref{poleo2}) because for the second type terms with odd $l$
have no poles in $Z_{p^{1,2}}$ and $l$ thus has to be even, whereupon
the second type of pole becomes the first. After the usual
manipulations, one gets
\begin{eqnarray}
Z^{r\ll0}_{S^2}&=&\frac{(z\bar{z})^{q-\frac{1}{2}}}{2}\lim_{\delta\rightarrow 0}
\oint\frac{\mathrm{d}\varepsilon_1\mathrm{d}\varepsilon_2}{(2\pi i)^2}
(z\bar{z})^{q-\frac{1}{2}-\varepsilon_2}\pi^5
\frac{\left[\sin\pi(\varepsilon_1-2\varepsilon_2)\right]^4}
{\left[\sin\pi\left(\varepsilon_2\right)\right]^2
\left[\sin 2\pi\varepsilon_2\right]^3
\left[\sin\pi\varepsilon_1\right]^4}\times\nonumber\\
&&\times \left.\vline\sum_{k,l=0}^{\infty}(-e^{-\delta})^k(-z)^{-l}
\frac{\left[\Gamma(1+k+2l+\varepsilon_1-2\varepsilon_2)\right]^4}
{\left[\Gamma(1+l-\varepsilon_2)\right]^2
\left[\Gamma(1+2l-2\varepsilon_2)\right]^3
\left[\Gamma(1+k+\varepsilon_1)\right]^4} \vline\right.^2,\nonumber\\
\end{eqnarray}

For the $r\gg0$-phase we close the contours of both $\tau_0$ and
$\tau_1$ in the negative half plane. The calculation is identical to
the $U(1)\times SU(2)$-examples and the result is
\begin{eqnarray}
Z^{r\gg0}_{S^2}&=&\frac{(z\bar{z})^q}{2}
\oint\frac{\mathrm{d}\varepsilon_1\mathrm{d}\varepsilon_2}{(2\pi i)^2}
(z\bar{z})^{-\varepsilon_1-2\varepsilon_2}\pi^3
\frac{\left[\cos\pi\left(\frac{\varepsilon_1}{2}+\varepsilon_2\right)
\right]^2
\left[\sin\pi\left(\varepsilon_1+2\varepsilon_2\right)\right]^3}
{\left[\sin\pi\varepsilon_1\right]^4\left[\sin2\pi\varepsilon_2\right]^4}
\times\nonumber\\
&&\times\left.\vline \sum_{\stackrel{k,l=0}{k+l=even}}^{\infty}
(-z)^{\frac{k+l}{2}}
\frac{\left[\Gamma\left(\frac{1}{2}+\frac{k}{2}+\frac{l}{2}
-\frac{\varepsilon_1}{2}-\varepsilon_2\right)\right]^2
\left[\Gamma\left(1+k+l-\varepsilon_1-2\varepsilon_2 \right)\right]^3}
{\left[\Gamma\left(1+k-\varepsilon_1\right)\right]^4
\left[\Gamma\left(1+l-2\varepsilon_1\right)\right]^4} \vline\right.^2\nonumber\\
\end{eqnarray}

\appendix{Elliptic genus of (S$^{2,+}_{(-1)^2,(-2)^3,1^4}$)}
\label{app:ellipticgenus}

\newcommand{\vtheta}{\vartheta}

We compute the $q\to 0$ limit of the
elliptic genus of the model (S$^{2,+}_{(-1)^2,(-2)^3,1^4}$)
applying the recently developed technique from 
\cite{Benini:2013nda,Benini:2013xpa}.
Recall that the model has gauge group $U(1)\times O(2)$.
It is initially expressed as an integration over the moduli space of
flat $U(1)\times O(2)$ connections on the torus, which can further
be simplified into a contour integral of a meromorphic form on
the moduli space
along a certain middle dimensional cycle in the complement of the poles.

The moduli space of flat $U(1)\times O(2)$ connections on the torus 
is the direct product $\mathfrak{M}^{U(1)}\times \mathfrak{M}^{O(2)}$
of the moduli spaces for the groups $U(1)$ and $O(2)$.
The first factor is parametrized by the $U(1)$ holonomies along
the two directions. With a natural complex structure, it can be identified
as the torus itself, $\mathfrak{M}^{U(1)}\cong \C/(\Z+\tau\Z)$,
i.e. $u_0\equiv u_0+1\equiv u_0+\tau$ where 
$\tau$ is the complex structure of the torus.
The second factor $\mathfrak{M}^{O(2)}$ consists of
seven components:\footnote{Here we regard $O(2)$
as the semi-direct product $U(1)\rtimes \ZZ_2$, where the generator
$\gamma\in \ZZ_2$ acts on $U(1)$ as the inversion.}
one is the space of $u_1\in \C/(\Z+\tau\Z)$ modulo
$u_1\equiv -u_1$ and the others are six points represented by
the commuting pairs of holonomies
$((1,\gamma),(\pm 1, 1))$, $((\pm 1,1),(1,\gamma))$,
$((\pm 1, \gamma),(1,\gamma))$. 
Accordingly, the moduli space of flat $U(1)\times O(2)$ connections 
consists of seven components ---
one two-dimensional component 
$\mathfrak{M}_{\rm cont}\cong (\C/(\Z+\tau\Z))^2$
and six one-dimensional components
$\mathfrak{M}_a\cong \C/(\Z+\tau\Z)$, where $a$ parametrizes the above six
discrete $O(2)$ holonomy pairs.

Under $O(2)\cong U(1)\rtimes \ZZ_2$, an $O(2)$ doublet corresponds to
a multiplet of
two fields $\phi_1,\phi_2$ of $U(1)$ charge $1,-1$ which are exchanged by
the element $(1,\gamma)\in U(1)\rtimes \ZZ_2$.
For the six discrete $O(2)$ holonomies, it is convenient to
take the combinations $\phi_{\pm}=\phi_1\pm \phi_2$.
Both of them are odd under the element
$(-1,1)$ while $\phi_+$ ({\it resp}. $\phi_-$) is even ({\it resp}. odd)
under $(1,\gamma)$.
The $U(1)$ gauge multiplet is even under $(-1,1)$ and odd under
$(1,\gamma)$.

The elliptic genus for the theory with gauge group 
$U(1)\times O_{\epsilon}(2)$ with the discrete theta angle $\theta_D$
($\epsilon=\pm$, $\theta_D=0,\pi$) is given by
\beqa
Z_{T^2}^{\epsilon,\theta_D}(\tau,z)
&=&{1\over 2}\sum_{u_*\in\mathfrak{M}_{\rm cont}^*}
\oint_{u_*}\dd u_1\,\dd u_0\,Z_{\rm cont}(\tau,u_0,u_1)\nn\\
&&+\,{1\over 4}\,\sum_a\epsilon_a\e^{i\theta_a}
\!\!\!\sum_{u_{0*}\in \mathfrak{M}_{a}^*}\oint_{u_{0*}}
\dd u_0\,Z_{\rm disc}(\tau,u_0,a).
\eeqa
$\mathfrak{M}_{\rm cont}^*$ is a part of the intersection points of
the singular hyperplanes where the matter scalars have zero modes,
selected by a choice of a generic weight vector $\eta=(\eta^0,\eta^1)$.
For $\eta$ near the positive $\eta^0$ axis,
the intersection points of the singular hyperplanes, 
$u_0+u_1\equiv 0$ and $u_0-u_1\equiv 0$, for
the $O(2)$ doublets are selected,
\beq
\mathfrak{M}_{\rm cont}^*=\Bigl\{\,\textstyle{(0,0),\,
({1\over 2},{1\over 2}),\,({\tau\over 2},{\tau\over 2}),\,
({1+\tau\over 2},{1+\tau\over 2})\,}\Bigr\}.
\eeq
These are all non-degenerate and the contour
around each $u_*\in \mathfrak{M}_{\rm cont}^*$ is uniquely 
specified.
$\mathfrak{M}_{a}^*$ is the set of $u_0$'s
where the $O(2)$ doublets have zero modes,
\beq
\mathfrak{M}_{a}^*=\{-a_+,-a_-\}.
\eeq
See below for the definition of $a_{\pm}$. The integrands are given by
the one-loop determinants
\beqa
Z_{\rm cont}(\tau,u_0,u_1)&=&
\left({-i\eta(\tau)^3\over \vtheta(-z)}\right)^2
\left({\vtheta(-{z\over 2}-u_0)\over\vtheta({z\over 2}-u_0)}\right)^2
\left({\vtheta(-2u_0)\over\vtheta(z-2u_0)}\right)^3\times\nn\\
&&\left({\vtheta(-z+u_0+u_1)\over\vtheta(u_0+u_1)}\right)^4
\left({\vtheta(-z+u_0-u_1)\over\vtheta(u_0-u_1)}\right)^4,\\
Z_{\rm disc}(\tau,u_0,a)&=&
{-i\eta(\tau)^3\over \vtheta(-z)}
{\vtheta(a_{\rm v})\over\vtheta(-z+a_{\rm v})}
\left({\vtheta(-{z\over 2}-u_0)\over\vtheta({z\over 2}-u_0)}\right)^2
\left({\vtheta(-2u_0)\over\vtheta(z-2u_0)}\right)^3\times\nn\\
&&\left({\vtheta(-z+u_0+a_+)\over\vtheta(u_0+a_+)}\right)^4
\left({\vtheta(-z+u_0+a_-)\over\vtheta(u_0+a_-)}\right)^4,
\eeqa
where
\beq
\vtheta(z):=\theta_1(\tau|z)
=-iq^{1/8}y^{1/2}\prod_{k=1}^{\infty}(1-q^k)(1-yq^k)(1-y^{-1}q^{k-1})
\eeq
with $q=\e^{2\pi i \tau}$ and $y=\e^{2\pi i z}$. Finally, 
the data $\epsilon_a$, $\theta_a$, $a_{\rm v}$ and $a_{\pm}$ are given by
\beq
{\footnotesize
\begin{array}{c|ccccccc}
a&\!\!((1,\gamma),(1,1))\!\!
&\!\!((1,\gamma),(-1,1))\!\!
&\!\!((1,1),(1,\gamma))\!\!
&\!\!((-1,1),(1,\gamma))\!\!
&\!\!((1,\gamma),(1,\gamma))\!\!
&\!\!((-1,\gamma),(1,\gamma))\!\!
\\
\hline
\!\!(a_{\rm v},a_+,a_-)\!\!&({1\over 2},0,{1\over 2})&
({1\over 2},-{\tau\over 2},{1+\tau\over 2})&
({\tau\over 2},0,{\tau\over 2})&
({\tau\over 2},-{1\over 2},{1+\tau\over 2})&
({1+\tau\over 2},0,{1+\tau\over 2})&
({1+\tau\over 2},-{1\over 2},{\tau\over 2})
\\
(\epsilon_a,\theta_a)&(1,1)&(1,\theta_D)&
(\epsilon,1)&(\epsilon,\theta_D)&
(1,1)&(1,\theta_D)
\end{array}}
\eeq
In the model (S$^{2,+}_{(-1)^2,(-2)^3,1^4}$)
we take $\epsilon=+$ and $\theta_D=\pi$.

We evaluate the integrals in the $q\to 0$ limit.
The contributions from the four points of $\mathfrak{M}_{\rm cont}^*$ are
\beqa
(0,0):&&-{1\over 4}(1+y^{1\over 2})^2y^{-{3\over 2}}
(1+7y^{1\over 2}+28y+7y^{3\over 2}+y^2)+O(q)\nn\\
\textstyle{({1\over 2},{1\over 2})}:&&
-{1\over 4}(1-y^{1\over 2})^2y^{-{3\over 2}}
(1-7y^{1\over 2}+28y-7y^{3\over 2}+y^2)+O(q)\nn\\
\textstyle{({\tau\over 2},{\tau\over 2})}:&&
{1\over 4}(y-1)^2y^{-{3\over 2}}(1+y)+O(q)\nn\\
\textstyle{({\tau+1\over 2},{\tau+1\over 2})}:&&
{1\over 4}(y-1)^2y^{-{3\over 2}}(1+y)+O(q),\nn
\eeqa
which sum up to
\beq
-22(y^{-{1\over 2}}+y^{1\over 2})+O(q).
\label{totalEG}
\eeq
The contributions from the six components $\mathfrak{M}_a$
sum up to zero for the choice $\theta_D=\pi$:
The contributions of the first and second $a$'s, 
the third and the fourth $a$'s, and the fifth and the sixth $a$'s
cancel against each other because of $\e^{i\theta_D}=-1$.
Therefore, (\ref{totalEG}) is the total.

\appendix{Topology of $\wt{Y}_S$}\label{app:wtYStop}

We now explain how to obtain the topological data (\ref{wtYStop}) for
the Calabi-Yau manifold $\wt{Y}_S$
that appears in the strongly coupled phase of the model
(S$^{2,+}_{(-1)^2,(-2)^3,1^4}$).\footnote{We would like to thank
 Alexander Kuznetsov and 
Hiromichi Takagi from whom we learned most of what is described here.
We also thank Alexey Bondal for much help.}
Recall that $\wt{Y}_S$ is a double cover of the symmetric determinantal
variety $Y_S=\{{\rm rank} S(p)\leq 3\}$
inside the weighted projective space $\PP^4_{11222}$.
It is a simultaneous unfolding of the $A_1$ singularity of
$Y_S$ along two disjoint
curves, $C_S$ and $\Sigma_S$.
$C_S$ is the locus where rank $S(p)=2$
 and $\Sigma_S$ is the intersection of $Y_S$ and the orbifold locus
 $\mathfrak{S}$ of $\PP^4_{11222}$.

First, let us put
$(y_0,\ldots, y_5)=((p^1)^2,p^1p^2,(p^2)^2,p^3,p^4,p^5)$.
Then, the matrix $S(p)$ can be written as a linear expression $S(y)$.
We also find that $\PP^4_{11222}$ can be regarded as the
degree 2 hypersurface $y_0y_2=y_1^2$ of $\PP^5$.
Therefore, $Y_S$ can be regarded as the intersection of the degree 2
and degree 4 hypersurfaces, $y_0y_2=y_1^2$ and $\det S(y)=0$.
The two hypersurfaces have singularities which are resolved respectively
by inserting $(z_1,z_2)\in \PP^1$ and $(x_1,\ldots, x_4)\in \PP^3$
obeying
\beqa
y_0z_1=y_1z_2,\quad
y_1z_1=y_2z_2,
\label{yz}
\\
\sum_{j=1}^4S(y)^{ij}x_j=0,
\quad i=1,2,3,4.
\label{yx}
\eeqa
Therefore, the singularity of $Y_S\subset \PP^5$ is resolved by
$Z_S\subset \PP^5\times \PP^1\times \PP^3$
defined by (\ref{yz}) and (\ref{yx}):
The curves $\Sigma_S$ and $C_S$ of $A_1$ singularity in $Y_S$ become divisors
$\Delta_S=\{(y,z,x(y))\}$ and $D_S=\{(y,z(y),x)\}$ in $Z_S$,
which are $\PP^1$ fibrations over the curves.
We introduce $\wt{Z}_S$ as the fiber product:
\beq
\begin{array}{ccccc}
\wt{Z}_S&\longrightarrow&Z_S&\subset &\PP^5\times\PP^1\times\PP^3\\
\downarrow&&\downarrow&&\downarrow\\
\wt{Y}_S&\longrightarrow &Y_S&\subset&\PP^5
\end{array}
\eeq
The arrow $\wt{Z}_S\to \wt{Y}_S$ is the blow up of $\wt{Y}_S$
at $C_S$ and $\Sigma_S$. (We denote the preimagaes of
$C_S$ and $\Sigma_S$ under $\wt{Y}_S\to Y_S$ by the same symbols;
we shall do the same for $D_S$ and $\Delta_S$.)  
Therefore, the Hodge numbers of $\wt{Z}_S$ and $\wt{Y}_S$ are related by
\beq
h^{p,q}(\wt{Z}_S)=h^{p,q}(\wt{Y}_S)+h^{p-1,q-1}(C_S)+h^{p-1,q-1}(\Sigma_S).
\label{Hrel}
\eeq
Since the map $\wt{Z}_S\to Z_S$ is one to one on $D_S\cup \Delta_S$
and two to one on the complement, we have the relation between
the Euler numbers,
\beqa
\chi(\wt{Z}_S)&=&\chi(\wt{Z}_S\setminus (D_S\cup \Delta_S))
+\chi(D_S\cup \Delta_S)\nn\\
&=&2\chi(Z_S\setminus (D_S\cup \Delta_S))
+\chi(D_S\cup \Delta_S)\nn\\
&=&2\chi(Z_S)-\chi(D_S\cup \Delta_S).
\label{Erel}
\eeqa

Let $h_1$, $h_2$ and $h_3$ be the hyperplane classes of
$\PP^5$, $\PP^1$ and $\PP^3$ respectively.
The classes of the divisors $\Delta_S$ and $D_S$ are
\beq
[\Delta_S]=(h_1-2h_2)|_{Z_S},\qquad
[D_S]=(3h_1-2h_3)|_{Z_S}.\label{DDS}
\eeq
To show the first, we consider the sequence of embeddings
$\Delta\subset \wh{\PP}^4_{11222}\subset \PP^5\times \PP^1$
where $\wh{\PP}^4_{11222}$ (which we write $\wh{\PP}^4$ for simplicity)
is the resolution (\ref{yz}) of $\PP^4_{11222}$
and $\Delta$ is the preimage $y_0=y_1=y_2$ of the singularity $\mathfrak{S}$.
This induces an exact sequence of the normal bundles
\beq
0\to N_{\Delta,\wh{\PP}^4}\longrightarrow
N_{\Delta,\PP^5\times\PP^1}\longrightarrow
N_{\wh{\PP}^4,\PP^5\times\PP^1}
\to 0.
\eeq
Since $\Delta$ and $\wh{\PP}^4$ are zeroes of the vector bundles
${\mathcal O}(1,0)^{\oplus 3}$ and ${\mathcal O}(1,1)^{\oplus 2}$
over $\PP^5\times\PP^1$, the normal bundles on the middle and on the right
have first Chen-classes $3h_1$ and $2(h_1+h_2)$ respectively.
Therefore, we find
$c_1(N_{\Delta,\wh{\PP}^4})
=3h_1-2(h_1+h_2)=h_1-2h_2$.
Including the $\PP^3$ factor and restricting to
$Z_S$ we find the first equality.
To show the second equality, we consider the following diagrams of
sheaves over $\PP^5\times \PP^1\times \PP^3$,
\beq
\begin{array}{ccccccccc}
0&\to& \!\!\!{\mathcal O}(-h_3)\!\!\!
&\longrightarrow&\!\!\!\!V\otimes {\mathcal O}\,
&\longrightarrow& E&\to& 0\\
&&&&\downarrow S(y)\!\!\!\!\!\!\!\!&&&&\\
0&\to&E^*(h_1)&\longrightarrow&\!\!V^*\!\otimes {\mathcal O}(h_1)\!\!\!\!
&\longrightarrow&\!\!\!{\mathcal O}(h_1+h_3)\!\!\!\!&\to&0
\end{array}
\eeq
where $V\cong \C^4$.
The upper line is the tautological exact sequence on $\PP^3$
and the lower line is its dual tensored with ${\mathcal O}(h_1)$.
Note that $\det E={\mathcal O}(h_3)$.
$Z_S$ is the locus where the map from ${\mathcal O}(-h_3)$ to
$V^*\otimes {\mathcal O}(h_1)$ vanishes or, by the symmetry of $S(y)$,
the locus where the map from $V\otimes {\mathcal O}$
to ${\mathcal O}(h_1+h_3)$ vanishes.
On this locus, there is a map from $E$ to $E^*(h_1)$ and $D_S$ is where this
degenerates. That is, it is the zero of a section of
$\det E^*(h_1)\otimes \det^{-1}E={\mathcal O}(3h_1)\otimes \det^{-2}E
={\mathcal O}(3h_1-2h_3)$. This proves the second equality.

Since $Z_S\subset \PP^5\times \PP^1\times \PP^3$ is the zero of a section of
$F={\mathcal O}(h_1+h_2)^{\oplus 2}\oplus {\mathcal O}(h_1+h_3)^{\oplus 4}$,
it is a Calabi-Yau manifold and
its Euler number is
\beqa
\chi(Z_S)&=&\int\limits_{\PP^5\times \PP^1\times \PP^3}
c_{\rm top}(F)\cdot {c(\PP^5\times \PP^1\times \PP^3)
\over c(F)}\nn\\
&=&\int\limits_{\PP^5\times \PP^1\times \PP^3}
[(h_1+h_2)^2(h_1+h_3)^4]\cdot
{(1+h_1)^6(1+h_2)^2(1+h_3)^4\over (1+h_1+h_2)^2(1+h_1+h_3)^4}
=-88.\nn\\
\eeqa
The equations (\ref{DDS}) show that 
$D_S$ and $\Delta_S$, as submanifolds of $\PP^5\times \PP^1\times \PP^3$,
are zeroes of sections of $F\oplus {\mathcal O}(h_1-2h_2)$
and $F\oplus {\mathcal O}(3h_1-2h_3)$.
By the similar computation as above, we find
\beq
\chi(D_S)=-80,\qquad \chi(\Delta_S)=-8.
\eeq
Since $D_S$ and $\Delta_S$ are $\PP^1$ bundles over $C_S$ and $\Sigma_S$,
this means
\beq
\chi(C_S)=-40,\qquad \chi(\Sigma_S)=-4.
\eeq
Therefore, $C_S$ and $\Sigma_S$ have genus $21$ and $3$ respectively,
meaning that $h^{1,0}(C_S)=21$ and $h^{1,0}(\Sigma_S)=3$.

By (\ref{Erel}), we find
\beq
\chi(\wt{Z}_S)=-2\times 88-(-80-8)=-88,
\eeq
and by (\ref{Hrel}), we find
\beq
h^{2,1}(\wt{Z}_S)=h^{2,1}(\wt{Y}_S)+21+3,\qquad
h^{1,1}(\wt{Z}_S)=h^{1,1}(\wt{Y}_S)+2.
\eeq
We also know from a separate account that
$h^{1,0}(\wt{Z}_S)=h^{2,0}(\wt{Z}_S)=0$ and
$h^{3,0}(\wt{Z}_S)=1$ which implies via (\ref{Hrel})
that $h^{1,0}(\wt{Y}_S)=h^{2,0}(\wt{Y}_S)=0$ and $h^{3,0}(\wt{Y}_S)=1$.
Therefore, we find
\beq
h^{2,1}(\wt{Z}_S)-h^{1,1}(\wt{Z}_S)=44,
\eeq
and
\beq
h^{2,1}(\wt{Y}_S)-h^{1,1}(\wt{Y}_S)=44-22=22.
\label{eqqq}
\eeq

Thus, if we can show either $h^{1,1}(\wt{Y}_S)=1$ or $h^{2,1}(\wt{Y}_S)=23$,
we are done about the Hodge numbers of $\wt{Y}_S$.
We can show the latter by following the argument used in
\cite{hosonotakagi11} for the similar purpose.  It uses the fact from the
deformation theory \cite{Sernesi} that, by $h^{2,0}(\wt{Y}_S)=0$,
any deformation of $\wt{Y}_S$ comes from a deformation of
$Y_S$. It is also known that the number of deformations of
$Y_S$ is equal to the na\"ive count (\ref{numbercountS}), i.e., $23$,
if the Picard number of $Y_S$ is one.
This shows that $h^{2,1}(\wt{Y}_S)\leq 23$. By (\ref{eqqq}) and
$h^{1,1}(\wt{Y}_S)\geq 1$, we find
$h^{1,1}(\wt{Y}_S)=1$ and $h^{2,1}(\wt{Y}_S)=23$.

Finally, we would like to compute the intersection numbers.
Let $M$ be the pull back of $M_{Y_S}={\mathcal O}_{\PP^5}(1)|_{Y_S}$ to 
$\wt{Y}_S$. We have 
\beq
M^3=2M_{Y_S}^3=2\int_{\PP^5}(2h_1)\cdot (4h_1)\cdot h_1^3=16.
\label{MMM}
\eeq
Since $M$ is ample, by Kodaira's vanishing theorem, we have
$\chi({\mathcal O}_{\wt{Y}_S}(M))
=h^0({\mathcal O}_{\wt{Y}_S}(M))$. One can also show that this is equal to
$h^0({\mathcal O}_{Y_S}(M_{Y_S}))=h^0({\mathcal O}_{\PP^5}(1))=6$.
Using Riemann-Roch formula $\chi({\mathcal O}_{\wt{Y}_S}(M))=
{1\over 6}M^3+{1\over 12}c_2(\wt{Y}_S)\cdot M$ and (\ref{MMM}),
we find
\beq
c_2(\wt{Y}_S)\cdot M=40.
\eeq
Let $H_{Y_S}$ be the Weil divisor of $Y_S$ defined by
$p^1=0$. Then, there is a relation $M_{Y_S}=2H_{Y_S}$ in $Y_S$
which pulls back to the relation
\beq
M=2H\quad\mbox{in ~$\wt{Y}_S$,}
\eeq
where now $H$ is a Cartier divisor of $\wt{Y}_S$ which is smooth.
This shows what we wanted: $H^3=2$ and $c_1(\wt{Y}_S)\cdot H=20$.


\appendix{Mirror symmetry checks}
\label{app-mirmet}
We have used the sphere partition function to check whether the hybrid
point in the $r\gg0$ phase is at finite distance in the moduli
space. In all the examples we have discussed in detail the
Picard-Fuchs operator is known. In \cite{Fuji:2010uq} (see also
\cite{Alim:2012ss}) it was shown how to extract the symplectic pairing
of the periods from the Picard-Fuchs, or rather, the associated
Gauss-Manin system. We will demonstrate this procedure for $X_5$. It
works the same way for the other examples, except for $X_7$. The
reason is that this approach requires the knowledge of the flat
coordinate in the given phase. Due to its exceptional behavior it is
not even clear what the flat coordinate is in the pseudo-hybrid phase.

In order to compute the metric
$g_{z\bar{z}}=\partial_z\partial_{\bar{z}}K(z,\bar{z})$ we have to
compute the K\"ahler potential. On the mirror it is given by
$e^{-K}=\Pi^{\dagger}Q\Pi$, where $Q$ is the symplectic form and $\Pi$
is the period vector. We need to determine $Q$ for a given set of
periods which is given by the solutions of the Picard-Fuchs
equations. For this purpose, one defines an antisymmetric bilinear
differential operator which acts on the set solutions of the
Picard-Fuchs equations
\begin{equation}
D_1\wedge D_2(f_1,f_2)=\frac{1}{2}\left(D_1f_1D_2f_2-D_2f_1D_1f_2\right).
\end{equation}
Using this, we make an ansatz for $Q$ as an asymmetric bidifferential
operator, specializing to the one-parameter case:
 \begin{equation}
Q(z)=\sum_{k,l}Q_{k,l}(z)D_{k}(\theta)\wedge D_{l}(\theta),
\end{equation}
where $z$ denotes the complex structure modulus on the mirror.  Since
the symplectic form must be constant over the moduli space, one has to
impose the conditions
\begin{equation}
\theta Q(z)=0.
\end{equation}
This determines the coefficients $Q_{kl}$, which are in general
$z$-dependent. The Picard-Fuchs equation enters into the above
condition.

Let us demonstrate this procedure for $X_5$. The Picard-Fuchs operator
in the Pfaffian phase is \cite{kanazawa10}
\begin{eqnarray}
\mathcal{L}^{r\ll0}&=& \theta^4 + z (2000 \theta^4 + 3904 \theta^3
+ 2708 \theta^2 + 756 \theta + 76) \nonumber\\
&& +  z^2 (63488 \theta^4 + 63488 \theta^3 - 21376 \theta^2 - 18624 \theta
- 2832) \nonumber\\
&&+  z^3 (512000 \theta^4 + 24576 \theta^3 - 37888 \theta^2 + 6144 \theta
+ 3072) \nonumber\\
&&+  z^4 (4096 (2 \theta + 1)^4).
\end{eqnarray}
To get the Picard-Fuchs operator in the $r\gg0$ phase we make the
transformation\footnote{Using this transformation instead of
  $\tilde{z}=\frac{1}{z}$ reveals that this model has maximally
  unipotent monodromy.} $\tilde{z}=\frac{1}{\sqrt{z}}$. The
Picard-Fuchs equation is
$\mathcal{L}\Pi=\mathcal\int(\Omega,\theta\Omega,\theta^2\Omega,
\theta^2\Omega)=0$,
where $\Omega$ is the holomorphic threeform and the vector denotes
elements of $H^3$ of the mirror Calabi-Yau. It is convenient to
rescale the holomorphic three form $\Omega(\tilde{z})\rightarrow
\tilde{z}\Omega(\tilde{z})$. Setting $\tilde{z}\equiv z$, the
Picard-Fuchs operator in the $r\gg0$ phase becomes
\begin{eqnarray}
\mathcal{L}^{r\gg 0}&=& 4096 \theta^4 + z^2(19456 + 96768 \theta
+ 173312 \theta^2 + 124928 \theta^3 +
    32000 \theta^4) \nonumber\\
&&+ z^4(-2832 - 9312 \theta - 5344 \theta^2 + 7936 \theta^3 +
    3968 \theta^4) \nonumber\\
&&+ z^6(12 + 12 \theta - 37 \theta^2 + 12 \theta^3 +
    125 \theta^4) \nonumber\\
&&+ z^8 \left(\frac{1}{16} + \frac{\theta}{4} + \frac{3 \theta^2}{8}
+ \frac{\theta^3}{4} + \frac{\theta^4}{16}\right).
\end{eqnarray}
We make the following ansatz for $Q(z)$ \cite{Masuda:1998eh}:
\begin{equation}
Q(z)=c_1(z)1\wedge\theta^3+c_2(z)\theta\wedge\theta^2+c_3(z)1\wedge\theta^2
+c_4(z)1\wedge\theta
\end{equation}
The condition $\theta Q(z)=0$ leads to a system of differential
equations for the coefficients $c_i(z)$. Furthermore all
$\theta^4$-terms can be eliminated using the Picard-Fuchs
equation. The result is
\begin{equation}
c_1=-c_2=\frac{256 + 1968 z^2 + z^4}{16 + z^2}\quad c_3=
\frac{2 z^2 (31232 + 32 z^2 + z^4)}{(16 + z^2)^2}
\quad c_4=\frac{2 z^2 (24192 - 408 z^2 + z^4)}{(16 + z^2)^2}
\end{equation}
With that we determine the symplectic pairing for a basis of solution
of the Picard-Fuchs equation. To see the behavior of the solutions we
rewrite the Picard-Fuchs equation as a matrix differential equation,
the Gauss-Manin system: $\theta\Pi=M(z)\Pi$ where the period vector
$\Pi$ is as above. The leading behavior of the solutions can be read
off from the Jordan normal form of the connection matrix $M$ evaluated
at $z=0$. If all the eigenvalues are zero, it is also the logarithm of
the monodromy matrix $\mathcal{T}=e^{2\pi i M(0)}$ (see for instance
\cite{coxkatz}). With our choice of coordinates and normalization of
the holomorphic threeform, we get
\begin{equation}
M(0)=\left(\begin{array}{cccc}0&1&0&0\\0&0&1&0\\0&0&0&1\\0&0&0&0 \end{array}
\right).
\end{equation}
This is exactly what one would expect at a large radius
point. Therefore the leading behavior of the period vector is
$\pi\sim(1,\frac{1}{2\pi i}\log z,\frac{1}{(2\pi i)^2}
\log^2z,\frac{1}{(2\pi i)^3}\log^3z)$. The factors $(2\pi i)$
are required for the correct monodromy behavior. The solutions can be
obtained from a power series ansatz and we refrain from giving them
here. Inserting $\pi$ into the expression for $Q(z)$, one finds the
following symplectic matrix for the given basis of solutions:
\begin{equation}
Q=\frac{1}{\pi^3}\left(
\begin{array}{cccc}0&0&0&6i\\0&0&-2i&0\\0&2i&0&0\\-6i&0&0&0\end{array}\right)
\end{equation}
Using this and the period matrix to compute $e^{-K}$, the K\"ahler
metric at $r\gg0$ is
\begin{equation}
g_{z\bar{z}}=\frac{3}{z\bar{z}(\log|z|^2)^2},
\end{equation}
which is exactly the result we got from the sphere partition
function. The same procedure can be repeated for $X_{10}$, $X_{13}$
and the two phases of the $U(1)\times O(2)$ example.

\appendix{Mirror symmetry calculations for $d<3$}
\label{app-mirsmall}
\subsection{Mirror of (A$^2_{5,5}$)}
For this example we can use mirror symmetry to check the result for
the fundamental period we obtained from the sphere partition
function. In the $\ll0$ phase the mirror of the Pfaffian has been
constructed in \cite{boehmthesis} using tropical geometry. The mirror
is given by the rank $2$ locus of the following skew-symmetric matrix
\begin{equation}
\tilde{A}(\tilde{p})=\left(\begin{array}{ccccc}
0&\psi\tilde{p}^5&\tilde{p}^3&\tilde{p}^4&t\tilde{p}^1\\
&0&\psi\tilde{p}^4&\tilde{p}^1&\tilde{p}^2\\
&&0&\psi \tilde{p}^2&\tilde{p}^5\\
&&&0&\psi\tilde{p}^3\\
&&&&0
\end{array}
\right),
\end{equation}
where we denote by $\tilde{p}^i$ the dual coordinates of the $p^i$ and
$\psi$ is the complex structure parameter of the mirror Pfaffian. The
remaining entries of $\tilde{A}(\tilde{p})$ are determined by the
skew-symmetry. Following the prescription in \cite{rodland98} one can
calculate the fundamental period by direct integration. For a Pfaffian
Calabi-Yau there is a globally defined holomorphic threeform
\begin{equation}
\Omega=\frac{(-1)^{\mu}(2\pi i)^3\mathrm{Pf}_{\mu_1\mu_2\mu_3}\Omega_0}
{d\mathrm{Pf}_{\mu_1}\wedge d\mathrm{Pf}_{\mu_2}\wedge d\mathrm{Pf}_{\mu_3}}
=\mathrm{Res}\frac{(-1)^{\mu}\mathrm{Pf}_{\mu_1\mu_2\mu_3}\Omega_0}
{\mathrm{Pf}_{\mu_1}\mathrm{Pf}_{\mu_2}\mathrm{Pf}_{\mu_3}},
\end{equation}
where
\begin{equation}
\Omega_0=\frac{(\tilde{p}^1)^5}{(2\pi i)^4}\bigwedge_{i=2}^5
d\left(\frac{\tilde{p}^i}{\tilde{p}^1}\right).
\end{equation}
The subscripts of the $\mathrm{Pf}$ indicate which rows and columns
are deleted from $\tilde{A}(\tilde{p})$ and $\mu$ is an integer. The
formula above holds for any three $\mu_i\in\{1,\ldots,5\}$. The
fundamental period is than defined by
\begin{equation}
X^0(\psi)=\int_{\gamma_0(\psi)}\Omega(\psi)=\int_{\Gamma}\Psi(\psi).
\end{equation}
In the last equality we used the fact that the period integral can be
rewritten as an integral of a six-form $\Psi(t)$ over a six-cycle
$\Gamma$ around zero. $\Psi$ is defined such that its residue is
$\Omega$:
\begin{equation}
\Psi=\Omega\wedge\frac{d\mathrm{Pf}_{\mu_0}}{2\pi i\mathrm{Pf}_{\mu_0} }\wedge
\frac{d\mathrm{Pf}_{\mu_1}}{2\pi i\mathrm{Pf}_{\mu_1} }\wedge
\frac{d\mathrm{Pf}_{\mu_2}}{2\pi i\mathrm{Pf}_{\mu_2}}=
\frac{(-1)^{\mu}\mathrm{Pf}_{\mu_1\mu_2\mu_3}}
{(2\pi i)^4\mathrm{Pf}_{\mu_1}\mathrm{Pf}_{\mu_2}\mathrm{Pf}_{\mu_3} }
(\tilde{p}^1)^5\bigwedge_{i=2}^5d\left(\frac{\tilde{p}^i}{\tilde{p}^1} \right)
\end{equation}
In order to calculate this integral explicitly we choose
$\mu_1,\mu_2,\mu_3=2,4,5$ and repeat the calculation in
\cite{rodland98} step by step.
\begin{eqnarray}
\frac{(-1)^{\mu}\mathrm{Pf}_{245}}{\mathrm{Pf}_2\mathrm{Pf}_4\mathrm{Pf}_5}&=&
-\frac{\tilde{p}^3}
{(-\tilde{p}^1\tilde{p}^3+\psi(\tilde{p}^4)^2+\psi^2\tilde{p}^2\tilde{p}^5)
(\psi^2\tilde{p}^1\tilde{p}^2+\psi(\tilde{p}^3)^2-\tilde{p}^4\tilde{p}^5)
(-\tilde{p}^2\tilde{p}^3+\psi^2\tilde{p}^1\tilde{p}^4\psi(\tilde{p}^5)^2)}
\nonumber\\
&\stackrel{\tilde{p}^1=1}{=}&
\frac{1}{(\tilde{p}^2\tilde{p}^3\tilde{p}^4\tilde{p}^5)
\left(1-\psi\frac{(\tilde{p}^4)^2}{\tilde{p}^3}
-\psi^2\frac{\tilde{p}^2\tilde{p}^5}{\tilde{p}^3}\right)
\left(1-\psi\frac{(\tilde{p}^3)^2}{\tilde{p}^4\tilde{p}^5}
-\psi^2\frac{\tilde{p}^2}{\tilde{p}^4\tilde{p}^5}\right)
\left(1-\psi\frac{(\tilde{p}^5)^2}{\tilde{p}^2\tilde{p}^3}
-\psi^2\frac{\tilde{p}^4}{\tilde{p}^2\tilde{p}^3}\right)}\nonumber\\
&:=&\frac{1}{(\tilde{p}^2\tilde{p}^3\tilde{p}^4\tilde{p}^5)
(1-v_{11}-v_{12})(1-v_{21}-v_{22})(1-v_{31}-v_{32})}\nonumber\\
&=&\frac{1}{\tilde{p}^2\tilde{p}^3\tilde{p}^4\tilde{p}^5}
\sum_{n=1}^{\infty}(v_{11}+v_{12})^n\sum_{m=1}^{\infty}(v_{21}+v_{22})^m
\sum_{k=1}^{\infty}(v_{31}+v_{32})^k\nonumber\\
&=&\frac{1}{\tilde{p}^2\tilde{p}^3\tilde{p}^4\tilde{p}^5}\sum_{n=1}^{\infty}
\sum_{a=0}^n\left(\begin{array}{c}n\\a\end{array}\right)v_{11}^av_{12}^{n-a}
\sum_{m=1}^{\infty}\sum_{b=0}^m
\left(\begin{array}{c}m\\b\end{array}\right)v_{21}^bv_{22}^{m-b}
\sum_{k=1}^{\infty}\sum_{c=0}^k
\left(\begin{array}{c}k\\c\end{array}\right)v_{31}^cv_{32}^{k-c}\nonumber\\
\end{eqnarray}
There will only be a contribution to the residue integral if
$v_{ij}$-terms are independent of $x_i$. There are the following
relations:
\begin{eqnarray}
v_{11}v_{21}v_{31}v_{22}=\psi^5\equiv z_1\nonumber\\
v_{12}v_{32}v_{21}=\psi^5\equiv z_2
\end{eqnarray}
Thus, only the following monomials will contribute to the residue:
\begin{equation}
z_1^{\alpha}z_2^{\beta}=
v_{11}^{\alpha}v_{12}^{\beta}v_{21}^{\alpha+\beta}v_{22}^{\alpha}
v_{31}^{\alpha}v_{32}^{\beta}
\quad \alpha+\beta=n
\end{equation}
Comparing with the expression above, there are non-zero contributions
if
$a=\alpha,n=\alpha+\beta,b=\alpha,m=2\alpha+\beta,c=\alpha,k=\alpha+\beta$.
Defining $ z=\psi^5$ we find the following expression for the fundamental
period:
\begin{eqnarray}
\varpi_0(z)&=&\sum_{n=0}^{\infty}\sum_{\alpha+\beta=n}
\left(\begin{array}{c}\alpha+\beta\\\alpha\end{array}\right)
\left(\begin{array}{c}2\alpha+\beta\\\alpha\end{array}\right)
\left(\begin{array}{c}\alpha+\beta\\\alpha\end{array}\right) z^n\nonumber\\
&=&\sum_{n=0}^{\infty}\sum_{\alpha=0}^n
\left(\begin{array}{c}n\\\alpha\end{array}\right)^2
\left(\begin{array}{c}n+\alpha\\\alpha\end{array}\right) z^n.
\end{eqnarray}
This is precisely the result we got from the sphere partition function
in both phases.

\subsection{Mirror of (S$^{2,+}_{3,3}$)}
For this example we can check the localization calculation in the
$r\gg0$ phase where the elliptic curve is a free $\mathbb{Z}_2$
quotient of a complete intersection $\tilde{X}$ of three quadric
hypersurfaces $Q_i(u_i,v_i)=0$ of co-dimension $3$ in
$\mathbb{P}^2\times\mathbb{P}^2$. Modding out by the involution
$\sigma=(u_i,v_i)\leftrightarrow(v_i,u_i)$ we obtain the Calabi-Yau
$X\simeq\tilde{X}/\langle\sigma\rangle$. To obtain the mirror we use
the Batyrev-Borisov construction for $\tilde{X}$
\cite{Batyrev:1994pg}. This is the exactly same procedure that was
used to construct the mirror of the Reye congruence in
\cite{hosonotakagi11} which we repeat now for two dimensions less.
One of the $\mathbb{P}^2$ is denoted by the lattice polytope $\Delta$
whose vertices provide a basis of sections
$H^0(\mathbb{P}^2,\mathcal{O}(-K_{\mathbb{P}^2}))$. Its dual polytope
$\Delta^{\ast}$ is given by $\mathrm{Conv.}({e_1,e_2,-e_1-e_2})$ where
$\{e_i\}$ is the canonical basis of
$\mathbb{R}^2$. $\mathbb{P}^2\times\mathbb{P}^2$ is then given by
$\Delta\times\Delta$. The dual polytope $(\Delta\times\Delta)^{\ast}$
has six vertices $\{(e_1,0),\ldots,(0,-e_1-e_2)\}$ which we denote by
$\nu_1,\ldots,\nu_{6}$. They are in one-to-one correspondence with
toric divisors $D_i=D_{\nu_i}$. In order to specify a complete
intersection we also have to give a nef-partition of the
anti-canonical class $-K_{\mathbb{P}_{\Delta\times\Delta}}$. This is
given by $\{D_i,D_{i+3}\}_{i=1,\ldots,3}$. Associated to this
nef-partition we define subpolytopes $\nabla_i$ such that their points
are the sections
$H^0(P_{\Delta\times\Delta},\mathcal{O}_{D_i+D_{i+3}})$. The complete
intersection $\tilde{X}$ is then given by the polynomials
$f_{\nabla_i}=0$. The dual of $\nabla_i$ is given by
$\Delta_i^{\ast}=\mathrm{Conv}(\{0,\nu_i,\nu_{i+3}\})$.  The mirror
$\tilde{X}^{\vee}$ is then given by $f_{\Delta_i^{\ast}}=0$.  These
polynomials have the following form:
\begin{equation}
f_{\Delta_i^{\ast}}=a_i+b_iU_i+c_iV_i\qquad i=1,\ldots,3
\end{equation}
for
$U_{1,\ldots,2},V_{1,\ldots,2}\in(\mathbb{C}^{\ast})^2\times(\mathbb{C^{\ast}})^2$
and $U_1U_2U_3=V_1V_2V_3=1$. Furthermore $a_i,b_i,c_i$ are parameters
which, redundantly, parametrize the complex structure of
$\tilde{X}^{\vee}$. The dual $X^{\vee}$ should be invariant under the
involution $\sigma^{\vee}:U_i\leftrightarrow V_i$. This is realized
when we set $b_i=c_i$. This reduces the number of complex structure
moduli from two to one.

We are going to compute the fundamental period by direct
integration. The period integral for the complete intersection is:
\begin{eqnarray}
\tilde{X}^0&=&\frac{1}{(2\pi i)^{4}}\int_{\gamma}
\frac{a_1a_2a_3}{f_{\Delta_1^{\ast}}f_{\Delta_2^{\ast}}f_{\Delta_3^{\ast}}}
\prod_{i=1}^2\frac{dU_i}{U_i}\frac{dV_i}{V_i}\nonumber\\
&=&\frac{1}{(2\pi i)^4}
\int_{\gamma}\prod_{i=1}^3\frac{dU_i}{U_i}\frac{dV_i}{V_i}
\frac{a_1a_2a_3}
{\prod_{i=1}^2(a_i+b_iU_i+c_iV_i)(a_3+\frac{b_3}{U_1U_2}
+\frac{c_3}{V_1V_2})}\nonumber\\
&=&\frac{1}{(2\pi i)^4}
\int_{\gamma}\prod_{i=1}^2\frac{dU_i}{U_i}\frac{dV_i}{V_i}
\sum_{n=0}^{\infty}(-1)^n\left(\frac{b_1}{a_1}U_1
+\frac{c_1}{a_1}V_1\right)^n
\sum_{m=0}^{\infty}(-1)^m\left(\frac{b_2}{a_2}U_2+\frac{c_2}{a_2}V_2\right)^m
\times\nonumber\\
&&\sum_{k=0}^{\infty}(-1)^k\left(\frac{b_3}{a_3}\frac{1}{U_1U_2}
+\frac{c_3}{a_3}\frac{1}{V_1V_2}\right)^k
\end{eqnarray}
There is only a contribution to the residue if the product of the four
infinite sums is independent of $U_i,V_i$. Defining,
\begin{equation}
x=\frac{b_1b_2b_3}{a_1a_2a_3}\qquad y=\frac{c_1c_2c_3}{a_1a_2a_3}
\end{equation}
we get:
\begin{equation}
\tilde{\Pi}_0=\sum_{a,b=0}^{\infty}
\left(\begin{array}{c}a+b\\a\end{array}\right)^3x^ay^b
\end{equation}
In order to make the transition from $\tilde{X}^{\vee}$ to $X^{\vee}$
we set $x=y\equiv z$. Then the fundamental period of $X^{\vee}$ is,
\begin{equation}
X^0=\sum_{n=0}^{\infty}
\sum_{k=0}^n\left(\begin{array}{c}n\\k\end{array}\right)^3z^n,
\end{equation}
which again coincides with the result from the sphere partition function.

\bibliographystyle{utphys}
\bibliography{bibliography}

\end{document}